\documentclass[manuscript,screen]{acmart}
\AtBeginDocument{%
  }

\setcopyright{acmlicensed}
\copyrightyear{2026}
\acmYear{2026}
\acmDOI{XXXXXXX.XXXXXXX}
\usepackage[utf8]{inputenc}
\usepackage[english]{babel}

\usepackage{graphicx}
\usepackage{transparent}
\usepackage{rotating}

\usepackage{tikz}
\usetikzlibrary{automata, positioning, arrows, calc}
\usepackage{float}
\graphicspath{{Images/}}

\usepackage[font=scriptsize]{subfig}
\usepackage[normalem]{ulem}

\usepackage{amsmath}
\usepackage{empheq} 

\usepackage{tabularx}
\usepackage{longtable}
\usepackage{colortbl}
\usepackage{multirow} 

\usepackage[ruled, vlined, linesnumbered]{algorithm2e}

\usepackage{xcolor}

\usepackage[inline]{enumitem}

\usepackage{ifthen}
\usepackage{orcidlink}
\usepackage{pdfpages}
\usepackage{comment}
\usepackage{tcolorbox}
\usepackage{pgfplots}
\pgfplotsset{compat=1.18}
\usepackage{wrapfig}
\usepackage{subcaption}

\AtBeginDocument{

}

\newtheorem{definition}{Definition}[section]

\newtheorem{lemma}{Lemma}[section]

\newtheorem{proposition}{Proposition}[section]

\theoremstyle{definition}
\newtheorem{example}{Example}[section]

\theoremstyle{definition}
\newtheorem{remark}{Remark}[section]

\usepackage{acronym}
\usepackage[textwidth=3.5cm,textsize=footnotesize]{todonotes}

\newcommand{\TCG}{Timed CLTLoc Game}

\newcommand{\cmax}{c_m}

\newcommand{\act}{\mathit{Act}}

\newcommand{\aut}{\mathcal{A}}

\newcommand{\transta}{\mathit{TS}(\aut)}

\newcommand{\transtap}{\mathit{TS}(\aut')}

\newcommand{\ackc}{X_{\aut}}

\newcommand{\cckc}{X_{\winCond}}

\newcommand{\valZ}{v[Y \leftarrow 0]}

\newcommand{\AP}{\mathit{AP}}

\newcommand{\eval}{\mathit{Eval}_{\aut}}

\newcommand{\frc}{\mathit{fr}}

\newcommand{\lrf}[1]{\lfloor #1 \rfloor}

\newcommand{\rtsstate}{\mathcal{S}}

\newcommand{\Next}{\vcenter{\hbox{$\bigcirc$}}}

\newcommand{\Prev}{\vcenter{\hbox{\scalebox{1.375}{$\ominus$}}}}

\newcommand{\Until}{\mathbf{U}}

\newcommand{\Since}{\mathbf{S}}

\newcommand{\Globally}{\Box}

\newcommand{\Eventually}{\Diamond}

\newcommand{\rtsarrow}{\rightsquigarrow}

\newcommand{\invariant}{\mathit{Inv}}

\newcommand{\outcome}{\Theta_{(\strategyc, \strategye)}}

\newcommand{\cont}{\mathbf{c}}

\newcommand{\env}{\mathbf{e}}

\newcommand{\arena}{\mathcal{A}}

\newcommand{\timedArena}{TAr}

\newcommand{\dce}{\delta}

\newcommand{\winCond}{\mathcal{W}}

\newcommand{\step}{\mathit{step}}

\newcommand{\plays}{\mathit{Plays}}

\newcommand{\tcgshort}{\mathcal{T}}

\newcommand{\strategyc}{\sigma_{\cont}}

\newcommand{\strategye}{\sigma_{\env}}

\newcommand{\playstate}{(q_n, v_n)}

\newcommand{\ckcass}{\nu_{\langle \rho, \nabla \rangle}}

\newcommand{\RTSA}{\mathcal{R}_{\mathcal{A}}}

\newcommand{\stateSatB}{\mathcal{S}_{\varphi}}

\newcommand{\toolname}{\textsc{Tarzan}}

\newcommand{\tiga}{\textsc{Uppaal Tiga}}
\newcommand{\stratego}{\textsc{Uppaal Stratego}}
\newcommand{\prism}{\textsc{Prism-games}}
\newcommand{\synthia}{\textsc{Synthia}}

\newcommand{\strategygraph}{\mathcal{G}_{\tcgshort}}

\newcommand{\reg}{\mathcal{R}}
\newcommand{\lidx}{\boldsymbol{\ell}}
\newcommand{\ridx}{\mathfrak{r}}
\newcommand{\loc}{q}
\newcommand{\rank}{\mathsf{r}}

\newcommand{\vertspace}{~ \vert ~}

\newboolean{compileExtendedVersion}
\setboolean{compileExtendedVersion}{false}
\newcommand{\ite}{\ifthenelse{\boolean{compileExtendedVersion}}}

\newcommand{\Red}[1]{{\color{black}#1}}

\newcommand{\Blue}[1]{{\color{black}#1}}

\newcommand{\nosemic}{\SetEndCharOfAlgoLine{\relax}}
\newcommand{\dosemic}{\SetEndCharOfAlgoLine{\string;}}
\let\oldnl\nl%
\newcommand{\nonl}{\renewcommand{\nl}{\let\nl\oldnl}}

\newcommand{\bea}{\begin{eqnarray}} 

\newcommand{\eea}{\end{eqnarray}}

\acrodef{hri}[HRI]{Human-Robot Interaction}
\acrodef{ta}[TA]{Timed Automaton}
\acrodef{sg}[SG]{Strategy Graph}
\acrodefplural{ta}[TA]{Timed Automata}
\acrodef{tarena}[TAr]{Timed Arena}
\acrodefplural{tarena}[TArs]{Timed Arenas}
\acrodef{tg}[TG]{Timed Game}
\acrodefplural{tg}[TGs]{Timed Games}
\acrodef{ltl}[LTL]{Linear Temporal Logic}
\acrodef{cltloc}[CLTLoc]{Constraint Linear Temporal Logic over clocks}
\acrodef{cltl}[CLTL]{Constraint LTL}
\acrodef{ctl}[CTL]{Computation Tree Logic}
\acrodef{tctl}[TCTL]{Timed Computation Tree Logic}
\acrodef{tol}[TOL]{Timed Obstruction Logic}
\acrodef{tcg}[TCG]{\TCG}
\acrodefplural{tcg}[TCGs]{\TCG{s}}
\acrodef{rts}[RTS]{Region Transition System}
\acrodefplural{rts}[RTSs]{Region Transition Systems}
\acrodef{gba}[GBA]{Generalized Büchi Automaton}
\acrodefplural{gba}[GBA]{Generalized Büchi Automata}
\acrodef{tarzan}[TARZAN]{Timed Automata Region and Zone library for real-time systems ANalysis}

\newcommand{\aprime}{\ac{tarena}-extension}

\begin{document}

\title{Timed Games under Environmental Interference with Real-Time Objectives}

\author{Andrea Manini}
\email{andrea.manini@polimi.it}
\orcid{0009-0009-6103-9244}
\author{Matteo Rossi}
\email{matteo.rossi@polimi.it}
\orcid{0000-0002-9193-9560}
\author{Pierluigi {San Pietro}}
\email{pierluigi.sanpietro@polimi.it}
\orcid{0000-0002-2437-8716}
\affiliation{%
  \institution{Politecnico di Milano}
  \city{Milano}
  \country{Italy}
}
\renewcommand{\shortauthors}{Manini et al.}

\begin{abstract}
  Classical Timed Games formulations may be unsuitable, or require substantial modeling effort, to capture complex interaction patterns between a controller and its surrounding environment, in which non-determinism must be resolved after the controller has chosen which action to perform. 
In this paper, we develop the theory of Timed CLTLoc Games (TCGs), a Timed Game variant designed to facilitate the modeling of such patterns. 
Unlike classical Timed Games formulations, TCGs partition locations rather than transitions, and use Constraint Linear Temporal Logic over clocks (CLTLoc) to specify the controller’s objectives.

We develop algorithms for solving TCGs and synthesizing winning controller strategies for games whose objectives include formal properties such as reachability and safety.
{For reachability objectives, our synthesis algorithm yields winning controller strategies that minimize the number of moves required to reach a goal state.}
Solving reachability TCGs is EXPTIME-complete, matching the complexity of solving reachability Timed Games.
All of our algorithms operate backwards and are implemented in our C\texttt{++} region-based library \toolname{}, leveraging OpenMP for efficient parallelization.
Experimental results demonstrate the {feasibility} of \toolname{}.
\end{abstract}




\maketitle

\section{Introduction}
\label{sec:introduction}
The interactions between a reactive system~\cite{REACTIVESYSTEMS} and its environment are naturally modeled as game-theoretic problems, in which the system represents a controller acting against an adversarial opponent.
Ensuring safety and functional correctness properties is essential, as the environment may introduce interference under which the controller must still guarantee correct behavior.
A canonical domain is robotics, in which autonomous robots act as controllers in dynamic environments shared with other robots and human operators.

When timing constraints are involved, \acp{tg} defined over \acp{ta} address the controller synthesis problem in real-time settings~\cite{PnueliTG}.
Typical \acp{tg} semantics (refer to~\cite{timedgameschapter} for an introduction), as implemented in state-of-the-art tools such as \tiga{}~\cite{tiga}, focus on the concurrency between two players: the controller and its environment. 
Informally, after both players choose an action and a delay, the action corresponding to the shortest delay is executed, breaking ties in favor of the environment.
While effective for modeling concurrent interactions, this semantics may be unsuitable, or require substantial modeling effort, to capture interactions in which the non-determinism over transitions and, in some cases, also delays, must be resolved \emph{after} the controller chooses the action to perform (and possibly also an associated delay).

To facilitate the modeling of such interactions, we propose a novel \ac{tg} variant, called \emph{\acp{tcg}}, in which the controller always chooses an action to perform, and its objectives (hereafter referred to as the \emph{winning condition}) are specified by using \ac{cltloc} formulae~\cite{BRSCLTLocTool,BRSCLTLoc}.
The non-determinism over transitions is then resolved by the environment; non-determinism over delays depends on whether a location belongs to the controller or the environment.
Notably, \acp{tcg} partition locations rather than transitions, better capturing environmental interference that may occur after the controller has chosen an action.
For instance, when the environment consists of human operators, a robot may require human intervention to complete some tasks~\cite{ASKARPOUR2019465,Lestingi2024};
here, \acp{tcg} facilitate the modeling of the following \acl{hri}~\cite{hripaper} phenotypes from Hollnagel’s foundational taxonomy~\cite{hollnagel1998cognitive}: 
\emph{delay}/\emph{premature action}---an action is executed at an unintended time---and 
\emph{insertion}/\emph{intrusion}---an action not included in the original plan is executed.
{These interaction patterns and phenotypes, in which the environment resolves non-determinism after the controller has chosen an action, cannot be faithfully encoded under the standard \acp{tg} semantics, where the controller retains the ability to interfere with the environment's timing and uncontrollable transitions carry no liveness obligation~\cite{surprise,TIGAManual}; 
thus, \acp{tcg} and standard \acp{tg} target different application domains.}

{For instance, consider a manufacturing robot (the controller) cooperating with a human operator (the environment). The robot determines which task the operator must perform, but it controls neither the outcome of the task nor its duration: given the same command, the operator may react in different ways and at a controller-unpredictable time.
Since controllability in standard \acp{tg} is associated with transitions, a single transition cannot simultaneously represent a controller-selected command while leaving both the delay and the resolution of the resulting non-determinism to the environment.
To encode such an interaction, the model must decouple the command and its execution into two consecutive steps: a controllable transition issuing the command and leading to an auxiliary location, followed by uncontrollable transitions exiting from that location, each representing one possible reaction of the operator. 
However, since the environment controls the resolution of uncontrollable transitions, it may choose to never fire any of them when a location has only outgoing uncontrollable transitions, allowing the game to remain indefinitely in the auxiliary location.
Avoiding this behavior requires equipping the auxiliary location with an invariant, thereby imposing an artificial upper bound on the operator’s reaction time.
Besides introducing an auxiliary location for each command, this modification alters the modeled system itself.
In \acp{tcg}, instead, a single environment location is sufficient: the robot selects the action, then the operator determines the delay and subsequently resolves the non-determinism by selecting one of the enabled transitions.
Thus, \acp{tcg} decouple controllability from timing and outcome resolution in a natural way, allowing the original interaction to be modeled without auxiliary locations or artificial timing assumptions.}

\acp{tcg} are further distinguished by \ac{cltloc}'s linear-time nature, which facilitates the specification of constraints over the timeline of the interaction between a controller and its environment. 
Moreover, \ac{cltloc} winning conditions can express timing constraints on clocks external to a given model, enabling the specification of properties without embedding additional structural elements into the model itself. 
As a result, the same model can be reused among different winning conditions, avoiding structural modifications whenever the specification changes.

We implement algorithms for solving \acp{tcg} and synthesizing winning controller strategies in our library \toolname{} (Timed Automata Region and Zone library for real-time systems ANalysis)~\cite{tarzan_forte}, originally developed for forward and backward reachability analysis of \acp{ta}.
Existing tools may integrate \toolname{} to broaden their game-solving capabilities.

\textbf{Related works.}
Existing tools supporting variants of \acp{tg} include \tiga{}; \stratego{}~\cite{STRATEGO}, which targets stochastic priced Timed Games~\cite{Goettmann2025}; \prism{}~\cite{PRISM}, which supports turn-based probabilistic timed multi-player games~\cite{kwiatkowska2019verification}; and \synthia{}~\cite{SYNTHIA}, which implements the \acp{tg} semantics of~\cite{PnueliTG}.
Winning conditions in \tiga{} are specified by using \ac{tctl}~\cite{ALUR19932}, whereas \prism{} adopts Probabilistic Alternating-time
Temporal Logic with Rewards (rPATL)~\cite{rPATL}. 
We recall that \ac{tctl} and rPATL are branching-time logics, whereas \ac{cltloc} is a linear-time logic.
Our implementation in \toolname{} relies on the region abstraction~\cite{ALURDILL}; in contrast, \tiga{}, \stratego{}, and \synthia{} adopt zone-based abstractions~\cite{HENZINGER1994193}; \prism{} implements a digital clocks approach~\cite{kwiatkowska2019verification}.
Existing libraries for \acp{ta} primarily operate on zone abstractions~\cite{PyDBM,UDBM} or more general polyhedral abstractions~\cite{apron,PPL}, and provide generic manipulation operations over these symbolic representations rather than dedicated support for \acp{tg} solving.

\textbf{Contributions.}
The main contributions of this paper are the following\footnote{A preliminary version of this work appeared in~\cite{tase2026}.}: 
(i) we introduce \acp{tcg} and formally define their semantics with respect to \ac{cltloc} winning conditions; 
(ii) we develop backward algorithms for solving \acp{tcg} by adapting the traditional controllable predecessor operator, and extend them to synthesize winning controller strategies; 
(iii) we prove that the problem of solving reachability \acp{tcg} is EXPTIME-complete; 
(iv) we implement the proposed algorithms in \toolname{}, providing a C\texttt{++} library for modeling, solving, and synthesizing winning controller strategies for \acp{tcg}; and 
(v) we empirically evaluate \toolname{}, validating our theoretical results and assessing its performance.

This paper is organized as follows. 
\autoref{sec:theoretical_background} reviews theoretical background. 
\autoref{sec:timed_CLTLoc_games} introduces \acp{tcg} and formalizes their semantics. 
\autoref{sec:tar_extension_and_region_semantics} presents an approach for handling external clocks, followed by a semantic equivalence between \acp{tcg} defined over configurations and state regions.
\autoref{sec:solving_timed_CLTLoc_games} develops algorithms for solving \acp{tcg} by adapting the traditional controllable predecessor operator; it then establishes the computational complexity of solving \acp{tcg}.
\autoref{sec:synthesizing_winning_controller_strategies} extends the solving algorithms to synthesize winning controller strategies.
\autoref{sec:implementation_and_experimental_evaluationn} introduces \toolname{} and presents a detailed experimental evaluation. 
Finally, \autoref{sec:conclusion} concludes the paper and discusses future developments.

\section{Theoretical Background}
\label{sec:theoretical_background}
The timing behavior of reactive systems is naturally modeled by the formalism of \ac{ta}, which is obtained by extending classical finite-state automata with real-valued clocks.
In the following, we use the term \emph{action} to refer to the events that a reactive system can exhibit.
\acp{ta} are defined as follows~\cite{ALURDILL}:

\begin{definition}
\label{def:tadef}
A \emph{\acl{ta}} $\aut$ is a tuple $\aut = (\act, Q, Q_{0}, \ackc, T, \invariant)$, where
$\act$ is a finite set of actions,
$Q$ is a finite set of locations,
$Q_{0} \subseteq Q$ is a set of initial locations,
$\ackc$ is a finite set of clocks,
$T \subseteq Q \times \act \times \Gamma(\ackc) \times 2^{\ackc} \times Q$ is a transition relation, and
$\invariant : Q \rightarrow \Gamma(\ackc)$ is an invariant-assignment function.
\end{definition}

Let $\aut$ be a \ac{ta}.
Clocks are special variables that can only be reset or compared against non-negative integer constants. 
The value of all clocks is initialized to zero and grows with derivative 1 until a reset occurs.
When a clock is reset, its value becomes zero; after a reset, its value starts increasing again.
$\Gamma(\ackc)$ is a set of \emph{clock constraints}, \emph{i.e.}, conditions {over clocks} that must be satisfied for a given transition to fire.
Clock constraints are defined by the following grammar: 
$\gamma := 
x \sim c
\vertspace{}
\gamma \land \gamma$, 
where $x \in \ackc$, $c \in \mathbb{N}$ is a constant, and $\sim \; \in \{ \le, <, =, >, \ge \}$.
Notice that, by definition, clock constraints define convex subsets of clock valuations in $\mathbb{R}^{|\ackc|}$.
Without loss of generality, in the remainder of this work we allow clock constraints to contain disjunctions.
Indeed, a transition whose guard is a disjunction of clock constraints is equivalent to multiple transitions originating from the same location, each labeled with one of the disjuncts as its guard.
The powerset $2^{\ackc}$ 
indicates
that a transition may reset a {subset} of the clocks.
We denote by $\cmax \in \mathbb{N}$ the maximum constant appearing in $\aut$ and assume, without loss of generality, that it is the same for all clocks of $\aut$.

A pair $(q, v)$ is a \emph{configuration} of a \ac{ta} $\aut$, where $q \in Q$ and $v : \ackc \rightarrow \mathbb{R}_{\ge 0}$ is a \emph{clock valuation} function.
We denote by $\eval$ the set of all clock valuations of $\aut$.
Given a subset $Y \subseteq \ackc$ of clocks, we denote by $\valZ$ the clock valuation defined as $\valZ(x) = 0$ if $x \in Y$ and $\valZ(x) = v(x)$ otherwise.
Given a constant $\delta \in \mathbb{R}_{\ge 0}$ and a clock $x \in \ackc$, the value $(v + \delta)(x)$ is defined as $v(x) + \delta.$

The \emph{transition system} $\transta = (S, \act \cup \mathbb{R}_{\ge 0}, I, \hookrightarrow)$ captures the semantics of $\aut$~\cite[Chapter 9]{Baier}, where 
$S = Q \times \eval$ is a set of states, each representing a configuration of $\aut$,
$I = \{ (q, v) \in S \; \vert \; q \in Q_0 \land \forall x \in \ackc : v(x) = 0 \}$ is the set of initial states (all clocks are initialized to zero in an initial state), and 
$\hookrightarrow$ is the transition relation containing:
(i) \emph{discrete} transitions $((q,v), a, (q',v'))$ if there exists a transition $(q, a, \gamma, Y, q') \in T$ in $\aut$ such that $v$ satisfies both $\gamma$ and $\invariant(q)$, and $v'(x) = \valZ(x)$ satisfies $\invariant(q')$, where $a \in \act$, for all $x \in \ackc$, and
(ii) \emph{delay} transitions $((q,v), \delta, (q,v'))$ if there exists a constant $\delta \in \mathbb{R}_{\ge 0}$ such that $v'(x) = (v + \delta)(x)$ satisfies $\invariant(q)$, for all $x \in \ackc$.

Since delays range over a continuous domain, every state in $S$ admits uncountably many outgoing delay transitions.
As a result, $\transta$ has uncountably many states and transitions.
The infinite state space of \acp{ta} can be conveniently analyzed by resorting to a finite abstraction, the \emph{\ac{rts}}~\cite{ALURDILL}.
In the following, for any real number $m \in \mathbb{R}$, we denote its integral and fractional parts by $\lrf{m}$ and $\frc(m)$, respectively.
\begin{definition}[Clock equivalence relation]
\label{def:clock_equivalence}
Let $\aut = (\act, Q, Q_{0}, \ackc, T, \invariant)$ be a \ac{ta}.
Then, two clock valuations $v, v' \in \eval$ are called \emph{clock-equivalent}, denoted by $v \cong v'$, if, and only if,
for all  $x, y \in \ackc$, the following conditions hold:
\emph{(i)} $v(x) > \cmax \Leftrightarrow v'(x) > \cmax$, and 
\emph{(ii)} if  $v(x), v'(x) \le \cmax$, then 
\emph{(ii$_a$)} $\lrf{v(x)} = \lrf{v'(x)}$ and $\frc(v(x)) = 0$ $\Leftrightarrow$ $\frc(v'(x)) = 0$, and
\emph{(ii$_b$)}  if $v(y), v'(y) \le \cmax$, then $\frc(v(x)) \sim \frc(v(y))$ $\Leftrightarrow$ $\frc(v'(x)) \sim \frc(v'(y))$, where $\sim \; \in \{<, =\}$.
\end{definition}
\begin{definition}[Region]
\label{def:region}
Let $\aut$ be a \ac{ta} and let $s = ( q, v )$ be a configuration of $\aut$.
The \emph{clock region} $[v]$ of $v \in \eval$ is the set of all clock valuations equivalent to $v$ under the equivalence relation of \autoref{def:clock_equivalence}.
The set of all clock regions of $\aut$ is denoted by $\eval / \cong$.
The \emph{state region} $[s]$ of $s$ is the pair $[s] = ( q, [v] ) = \{ ( q, v' ) \mid v' \in [v] \}$;
$[s]$ is \emph{initial} if, and only if, $q \in Q_0$ and $\forall v' \in [v], \forall x \in \ackc : v'(x) = 0$.
\end{definition}
\begin{definition}
\label{def:region_transition_system}
Let $\aut$ be a \ac{ta}, let $\transta = (S, \act \cup \mathbb{R}_{\ge 0}, I, \hookrightarrow)$ be the transition system derived from $\aut$, and let $\cong$ be the clock equivalence relation of \autoref{def:clock_equivalence}.
The \emph{\acl{rts}} $\RTSA$ of $\arena$ is a tuple
$\RTSA = 
(\rtsstate, 
\mathit{Act} \cup \{ \tau \}, 
\mathcal{I},
\rtsarrow)$,
where
$\rtsstate = S / \cong \; = 
\{ [s] \; \vert \; s \in S \}$
is the set of all state regions (also called states) of $\aut$,
$\mathcal{I} = \{ [s] \; \vert \; s \in I \}$ is the set of initial states, and
$\rtsarrow \; \subseteq \rtsstate \times (\act \cup \{ \tau \}) \times \rtsstate$ is a transition relation containing discrete and delay transitions.
{We denote by $\rtsarrow^*_{\tau}$ the reflexive and transitive closure of delay transitions in $\RTSA$.}
\end{definition}

{Notice that $\rtsarrow^*_{\tau}$ captures an arbitrary delay---including the zero delay---and, since every delay transition requires the invariant of the current location to hold, the invariant holds continuously throughout the delay.}
A clock region $[v'] \in \eval / \cong$ is the \emph{successor} of a clock region $[v] \in \eval / \cong$, denoted by $[v'] = \mathit{succ}([v])$, if it either holds that:
(i) $[v] = [v']$ and $\forall \bar{v} \in [v], \forall x \in \ackc : \bar{v}(x) > \cmax$, or
(ii) $[v] \neq [v']$ and, for every $\bar{v} \in [v]$, 
there exists $\delta \in \mathbb{R}_{> 0}$ such that $(\bar{v} + \delta) \in [v']$
and 
$\forall \; 0 \le \delta' \le \delta : (\bar{v} + \delta') \in ([v] \cup [v'])$.
{A discrete transition $(s = (q, [v]), a, s') \in \; \rtsarrow$ requires the existence of a transition $(q, a, \gamma, Y, q') \in T$ in $\aut$ such that, for all $\bar{v} \in [v]$, $\bar{v}$ satisfies both $\gamma$ and $\invariant(q)$, and $\bar{v}[Y \leftarrow 0]$ satisfies $\invariant(q')$; 
then, $[s'] = (q',  [v[Y \leftarrow 0]])$.}
A delay transition $(s = (q, [v]), \tau, s') \in \; \rtsarrow$ requires that, for all $\bar{v} \in [v]$, $\bar{v}$ satisfies $\invariant(q)$ and, given $\mathit{succ}([v]) = [v'']$, for all $\tilde{v} \in [v'']$, $\tilde{v}$ satisfies $\invariant(q)$; then, $s' = (q, [v''])$.
Notice that, in \autoref{def:region_transition_system}, $\tau$ is a fictitious action used to denote delay transitions.

In the remainder of this work, to avoid ambiguity, we use the term \emph{state} exclusively for state regions in $\RTSA$, while the elements of $S$ in $\transta$ are called \emph{configurations}.
For visual clarity, we omit square brackets when referring to a \ac{rts} state; that is, we write $s$ instead of $[s]$ to denote a state in $\mathcal{S}$. 

\textbf{Constraint LTL over clocks.}~Let $\AP$ be a finite set of atomic propositions and let $\cckc$ be a finite set of clocks.
\ac{cltloc} formulae are defined over the constraint system $\mathcal{D} = (\mathbb{R}_{\ge 0}, \{ <, = \})$ as a particular case of \ac{cltl}~\cite{DEMRI2007380}.
The first element of $\mathcal{D}$ is the domain of interpretation of clocks, \emph{i.e.},
$\forall x \in \cckc : x \in \mathbb{R}_{\ge 0}$.
Without loss of generality, we assume that \ac{cltloc} clocks are initialized to zero.
The second element of $\mathcal{D}$ is a family of relations on $\mathbb{R}_{\ge 0}$, allowing formulae such as $x \sim c$, where $x \in \cckc$, $c \in \mathbb{N}$, and $\sim \; \in \{ <, = \}$.
Well-formed \ac{cltloc} formulae are defined according to the following grammar:
\begin{equation}
\label{eq:cltloc_grammar}
\varphi := 
\,
p 
\vertspace{}
x \sim c
\vertspace{}
\neg \varphi
\vertspace{}
\varphi \land \varphi 
\vertspace{}
\Next \varphi 
\vertspace{}
\Prev \varphi 
\vertspace{}
\varphi \Until \varphi 
\vertspace{}
\varphi \Since \varphi
\end{equation}
where $p \in \AP$ and $\Next$, $\Prev$, $\Until$, and $\Since$ are the usual \emph{next}, \emph{previous}, \emph{until}, and \emph{since} operators of \ac{ltl}.
The \emph{eventually} ($\Eventually$)
and \emph{globally} ($\Globally$)
temporal operators are defined as usual:
$\Eventually \varphi \equiv \top \Until \varphi$ and
$\Globally \varphi \equiv \neg \Eventually \neg \varphi$, respectively.
\ac{cltloc} formulae that are Boolean combinations of atomic propositions and clock constraints, without temporal operators, are called \emph{pure} \ac{cltloc} formulae.
Since, as shown in~\cite{BRSCLTLoc}, every \ac{cltloc} formula can be translated into an equivalent one using only future operators ($\Next$ and $\Until$), in the remainder of this work we focus exclusively on this fragment of \ac{cltloc}, called \emph{future-only} \ac{cltloc}.
The semantics of \ac{cltloc}, fully described in~\cite{BRSCLTLoc}, will be presented in the context of our new \ac{tg} variant in \autoref{sec:timed_CLTLoc_games}.
%
%
%

\section{\aclp{tcg}}
\label{sec:timed_CLTLoc_games}
We introduce a novel \ac{tg} variant, called \emph{\aclp{tcg}}, in which winning conditions are \ac{cltloc} formulae.
In the remainder of this work we focus on \emph{zero-sum, two-player} games, \emph{i.e.}, games in which exactly two players are involved and a win for one player implies a loss for the other.
We call one player the \emph{controller} (denoted by $\cont$) and the other the \emph{environment} (denoted by $\env$).
The controller's objective is to ensure that a system behaves {according to a given specification, despite the environment acting as an opponent}.

{To formalize the adversarial interaction between players, we propose the notions of \emph{\ac{tarena}} and \emph{move}.
Unlike in traditional \acp{tg}, which typically focus on the concurrency between players, our definitions treat the environment as an external perturbation rather than a concurrent opponent, facilitating the modeling of controllers that must be resilient to environmental interference.}

\begin{definition}
\label{def:timed_arena}
    A \emph{\acl{tarena}} $\arena$ is a \ac{ta} $\aut = (\act, Q, Q_{0}, \ackc, T, \invariant)$, where $Q$ is partitioned into controller $Q_{\cont} \subseteq Q$ and environment $Q_{\env} \subseteq Q$ locations, such that $Q = (Q_{\cont} \cup Q_{\env})$ and $(Q_{\cont} \cap Q_{\env}) = \varnothing$ hold.
    The partition of set $Q$ is reflected on both $\transta$ and $\RTSA$.
    More precisely, 
    the set $\mathcal{S}$ of $\RTSA$ is partitioned into controller states $\mathcal{S}_{\cont} \subseteq \mathcal{S}$ and environment states $\mathcal{S}_{\env} \subseteq \mathcal{S}$, such that $\mathcal{S} = (\mathcal{S}_{\cont} \cup \mathcal{S}_{\env})$ and $(\mathcal{S}_{\cont} \cap \mathcal{S}_{\env}) = \varnothing$ hold.
\end{definition}

A \emph{winning condition} $\winCond$
over a \ac{tarena} $\arena$ is a \ac{cltloc} formula generated by the future-only fragment of the grammar in \autoref{eq:cltloc_grammar}, with atomic propositions given by $\mathit{AP} = Q$, \emph{i.e.}, the atomic propositions of $\winCond$ predicate over the locations of $\arena$.
Notably, the set of clocks $\cckc$ of $\winCond$ may overlap with $\ackc$,
{and clocks in $\cckc \setminus \ackc$ (also referred to as \emph{external} clocks) can still be reset, if needed, to satisfy $\winCond$.}
{The reset behavior of external clocks is determined by the controller, as described in the following paragraphs.}
A \ac{tcg} $\tcgshort$ is defined as a pair $\tcgshort = (\arena, \winCond)$, where $\arena$ is a \ac{tarena} and $\winCond$ is a winning condition.
The way in which $\tcgshort$ evolves is defined according to the following notion of move:

\begin{definition}
\label{def:move}
Let $\arena$ be a \ac{tarena} and let $(q,v)$ be a configuration of $\arena$. 
A \emph{move} from $(q,v)$ is a pair $(a, \dce)$, where $a \in \mathit{Act}$ is an action always chosen by the controller, and $\dce \in \mathbb{R}_{\ge 0}$ is a delay chosen by the controller if $q \in Q_{\cont}$ and by the environment if $q \in Q_{\env}$ (the delay $\dce$ must be chosen after action $a$).
{Then, from the set of all enabled transitions $(q, a, \gamma, Y, q') \in T$, for which $v + \delta$ satisfies $\invariant(q)$ and $\gamma$, the environment resolves the non-determinism by choosing a single transition to fire.}
\end{definition}

Let $\tcgshort = (\arena, \winCond)$ be a \ac{tcg} and let $\transta = (S, \act \cup \mathbb{R}_{\ge 0}, I, \hookrightarrow)$.
The mapping $\step : (S \times ( \act \times \mathbb{R}_{\ge 0} ) ) \rightarrow \wp(S)$, where $\wp$ denotes the powerset operator, returns all configurations $(q', v')$ reachable from $(q, v)$ with action $a$ and delay $\delta$, \emph{i.e.}, it captures every choice the environment can make to resolve the non-determinism over discrete transitions in a move.
Specifically, $\step$ is defined as follows:
\[
\step((q, v), (a, \delta)) = 
\left\{ (q', (v + \dce)[Y \leftarrow 0]) \in S
\; \middle| \;
\begin{array}{@{}l@{}}
\big((q, v), \delta, (q, v + \delta)\big) \in \; \hookrightarrow \\
\land \; \big((q, v + \delta), a, (q', (v + \dce)[Y \leftarrow 0])\big) \in \; \hookrightarrow
\end{array}
\right\}.
\]
Informally, $\step((q, v), (a, \delta))$ requires to compute the delay transition $((q, v), \delta, (q, v + \delta)) \in \; \hookrightarrow$, followed by all discrete transitions $((q, v + \delta), a, (q', (v + \dce)[Y \leftarrow 0])) \in \; \hookrightarrow$ in $\transta$.

{\textbf{Plays and strategies.}~Let $\arena$ be a \ac{tarena} and let $\transta = (S, \act \cup \mathbb{R}_{\ge 0}, I, \hookrightarrow)$.
A \emph{play} 
$\rho$ over $\arena$ is an infinite sequence
\[
\rho = (q_0,v_0) \xrightarrow{(a_0,\delta_0)}(q_1,v_1) \xrightarrow{(a_1,\delta_1)} (q_2, v_2) \cdots
\]
alternating configurations and moves, such that:
$(q_0, v_0) \in I$ and, for all $n \in \mathbb{N}$,
$(q_{n + 1}, v_{n + 1}) \in \step((q_n, v_n), (a_n, \delta_n))$.
Since both players observe the passage of time, the delays occurring along a play determine the values of external clocks from their perspective, allowing them to maintain complete information about the values of external clocks.
The set of all finite prefixes of the plays of $\arena$ is denoted by $\plays$.
Given a play $\rho$, for all $n \in \mathbb{N}$, we denote by $\rho_{\le n}$ its prefix of length $n$, ending in a configuration $\playstate \in S$.}

A \emph{controller strategy}
is a mapping $\strategyc : \mathit{Plays} \rightarrow ((\act \times \mathbb{R}_{\ge 0}) \cup \act)$ defined as:
$\strategyc(\rho_{\le n}) = (a, \delta)$ if $q_n \in Q_{\cont}$, while
$\strategyc(\rho_{\le n}) = a$ if $q_n \in Q_{\env}$, where $a \in \act$ and $\delta \in \mathbb{R}_{\ge 0}$, associating either $(a, \delta)$ or $a$ with a finite prefix $\rho_{\le n} \in \mathit{Plays}$.
Next, let 
$\strategye^1 : \mathit{Plays} \rightarrow \mathbb{R}_{\ge 0}$ be a partial mapping associating a delay in $\mathbb{R}_{\ge 0}$ with a finite prefix in set $\mathit{Plays}$ if $q_n \in Q_{\env}$;
$\strategye^1$ is undefined otherwise.
We also introduce the mapping 
$\strategye^2 : (\mathit{Plays} \times ( \wp(S) \setminus \{ \varnothing \})) \rightarrow S$ which, given a finite prefix of a play and a non-empty set of configurations, returns a configuration in that set.
Then, an \emph{environment strategy} is a
pair $\strategye = (\strategye^1, \strategye^2)$.
Finally, given a pair of strategies $(\strategyc, \strategye)$, their \emph{outcome} $\outcome : \plays \rightarrow S$
is defined as:
\[
\outcome(\rho_{\le n}) = 
\begin{cases}
\strategye^2 \big(\rho_{\le n}, \step((q_n, v_n), \strategyc(\rho_{\le n}))\big) & \text{if $q_n \in Q_{\cont}$} \\
\strategye^2 \big(\rho_{\le n}, \step((q_n, v_n), (\strategyc(\rho_{\le n}), \strategye^1(\rho_{\le n})))\big) & \text{if $q_n \in Q_{\env}$}
\end{cases}.
\]
Informally, an outcome $\outcome$ is a mapping that returns the next configuration obtained after a move is fully applied.
Note that this requires the environment to resolve the non-determinism over multiple enabled discrete transitions.

{To enforce infinite plays, we assume that $\transta$ has no terminal configurations and that, for every play $\rho$ and every $n \in \mathbb{N}$, where $(q_n, v_n) \in S$ denotes the last configuration of $\rho_{\le n}$,
it holds that: 
if $q_n \in Q_{\cont}$, then $\step((q_n, v_n), \strategyc(\rho_{\le n})) \neq \varnothing$; 
if $q_n \in Q_{\env}$, then $\step((q_n, v_n), (\strategyc(\rho_{\le n}), \strategye^1(\rho_{\le n}))) \neq \varnothing$.}

{Since \acp{tcg} are zero-sum, we focus on controller strategies.
A play $\rho$ is \emph{consistent} with a controller strategy $\strategyc$ if there exists an environment strategy $\strategye = (\strategye^1, \strategye^2)$ such that, for every finite prefix $\rho_{\le n}$,
the next configuration reached from $(q_n, v_n)$ in $\rho$ is the outcome of $\strategyc$ and $\strategye$, \emph{i.e.}, $(q_{n + 1}, v_{n + 1 }) = \outcome (\rho_{\le n})$ such that, if $q_n \in Q_{\cont}$, then $\strategyc(\rho_{\le n}) = (a_n, \delta_n)$; if $q_n \in Q_{\env}$, then $\strategyc(\rho_{\le n}) = a_n$ and $\strategye^1(\rho_{\le n}) = \delta_n$.
It follows that every finite prefix of $\rho$ is consistent with $\strategyc$.}

In the remainder of this work,  we will assume that \acp{tcg} do not exhibit \emph{Zeno} behaviors, where an infinite number of moves is applied within a finite time span.
{Indeed, Zeno plays are not physically realizable and could allow either player to win, or to prevent the opponent from winning, by freezing time.
To rule out Zeno behaviors, we adopt the assumption of~\cite{ASARIN1998447} that every considered \ac{tarena} $\aut$ is \emph{strongly non-Zeno}. 
Specifically, every cycle of $\transta$ contains (i) a discrete transition that resets some clock $x \in \ackc$ to zero and (ii) a discrete transition whose guard requires $x \geq 1$.}
Alternative solutions can be adapted to \acp{tcg}, \emph{e.g.}, by punishing the player responsible for Zeno behaviors~\cite{surprise}.

\begin{table}[tb]
\centering
\renewcommand{\arraystretch}{1.4}
\begin{tabular}{rcl}
$(\rho, \ckcass), i \models p$ & 
$\Leftrightarrow$ & 
$p = q_i$, for $p \in \AP$ 
\\
$(\rho, \ckcass), i \models x \sim c$ & 
$\Leftrightarrow$ & 
$\ckcass(i, x) \sim \ckcass(i, c)$, for $\sim \; \in \{ <, = \}$
\\
$(\rho, \ckcass), i \models \neg \varphi$ & 
$\Leftrightarrow$ & 
$(\rho, \ckcass), i \not\models \varphi$ 
\\
$(\rho, \ckcass), i \models \varphi \land \psi$ & 
$\Leftrightarrow$ & 
$(\rho, \ckcass), i \models \varphi$ and $(\rho, \ckcass), i \models \psi$ 
\\
$(\rho, \ckcass), i \models \Next \varphi$ & 
$\Leftrightarrow$ &
$(\rho, \ckcass), i + 1 \models \varphi$
\\
$(\rho, \ckcass), i \models \varphi \Until \psi$ & 
$\Leftrightarrow$ & 
$\exists j \ge i : (\rho, \ckcass), j \models \psi \land \forall n \in [i, j) : (\rho, \ckcass), n \models \varphi$
\end{tabular}

\caption{Semantics of a \ac{tcg} $\tcgshort = (\arena, \winCond)$ over a play $\rho$ at position $i$. Clock values are defined by the clock assignment $\ckcass$.}
\label{tab:semantics}
\end{table}

{\textbf{\acp{tcg} semantics.}~The semantics of a given \ac{tcg} $\tcgshort = (\arena, \winCond)$ over a play $\rho$ is defined with respect to a strict linear order $(\mathbb{N}, <)$, which represents positions in time.
A position $i \in \mathbb{N}$ denotes the configuration $(q_i, v_i)$ of $\rho$.
{Clocks in $\cckc \setminus \ackc$ are reset according to the controller through a \emph{reset policy} $\nabla : \mathit{Plays} \rightarrow \wp(\cckc \setminus \ackc)$, where $\nabla(\rho_{\le i})$ is the set of external clocks  reset on the move from position $i$ to position $i + 1$. 
Like the controller's choices (action and possibly a delay), this set is fixed before the environment chooses the delay (if in an environment location) and resolves the non-determinism over transitions.
Given a play $\rho$ and a reset policy $\nabla$, the \emph{clock assignment} induced by $\rho$ and $\nabla$ is the mapping
$\ckcass : \mathbb{N} \times (\ackc \cup \cckc) \to \mathbb{R}_{\ge 0}$
defined as follows: 
for every $x \in \ackc$ and for all $i \in \mathbb{N}$, $\ckcass(i, x) = v_i(x)$;
for every $x \in (\cckc \setminus \ackc)$, $\ckcass(0, x) = 0$; moreover, for every $i \in \mathbb{N}$, 
if $x \in \nabla(\rho_{\le i})$, then $\ckcass(i+1, x) = 0$; otherwise, $\ckcass(i+1, x) = \ckcass(i, x) + \delta_i$,
where $\delta_i \in \mathbb{R}_{\ge 0}$ is the delay of the move from position $i$ to position $i+1$ in $\rho$.

An \emph{interpretation} of $\winCond$ is a pair $(\rho, \ckcass)$.
The semantics of $\winCond$ at position $i$ under $(\rho, \ckcass)$ is defined in \autoref{tab:semantics},
where $\ckcass(i, c) = c$ whenever $c \in \mathbb{N}$ is a constant value.
We say that $\winCond$ is \emph{satisfiable} if there exists an interpretation $(\rho, \ckcass)$
satisfying $\winCond$ at position $i = 0$---\emph{i.e.}, if $(\rho, \ckcass), 0 \models \winCond$ holds;
in this case,
we say that $(\rho, \ckcass)$ \emph{satisfies} $\winCond$ and succinctly write $(\rho, \ckcass) \models \winCond$.
}

\begin{definition}[Winning strategy]
\label{def:winning_strategy}
{A controller strategy $\strategyc$ is \emph{winning} for a \ac{tcg} $\tcgshort = (\arena, \winCond)$ from an initial configuration $(q_0, v_0) \in I$ if there exists a reset policy $\nabla$ such that, for every play $\rho$ from $(q_0, v_0)$ consistent with $\strategyc$, the interpretation $(\rho, \ckcass)$
satisfies $\winCond$.
\emph{Solving} $\tcgshort$ requires determining if there exists a winning controller strategy for $\tcgshort$.}
\end{definition}
\begin{figure}[b]
    \centering
    \includegraphics[width=0.715\linewidth]{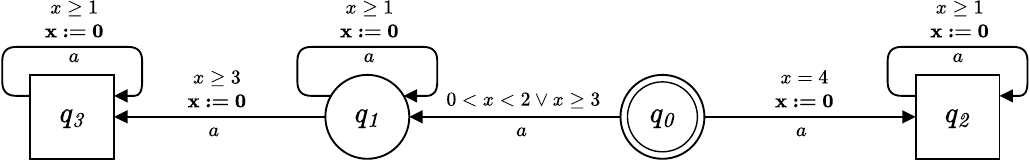}
    \caption{A \timedArena{} $\aut$, where $\mathit{Act} = \{ a \}$, $Q = \{ q_0, q_1, q_2, q_3 \}$ is partitioned into controller $\{ q_0, q_1 \} \subset Q_{\cont}$ and environment $\{ q_2, q_3 \} \subset Q_{\env}$ locations, $Q_0 = \{q_0\}$, and $\ackc = \{ x \}$.
Clock resets are represented in bold, while clock guards are represented in italic.}
\Description{Example \timedArena{}}
\label{fig:move_example}
\end{figure}

\begin{example}
Consider the \ac{tarena} $\arena$ depicted in \autoref{fig:move_example}.
The initial configuration of $\arena$ is $(q_0, \{ x = 0 \})$, where $q_0 \in Q_{\cont}$ has two outgoing transitions, both labeled with the same action $a \in \act$.
To define a move, the controller must first choose an action labeling a transition exiting from $q_0$; here, the choice is forced to be $a$.
Then, since $q_0 \in Q_{\cont}$, the controller chooses a delay $\delta \in \mathbb{R}_{\ge 0}$.
The resulting move is $( a, \delta )$.
{Since $\arena$ contains no external clocks, the reset policy $\nabla$ always returns the empty set.}
If $\delta$ is such that $(0 < \delta < 2 \lor \delta \ge 3)$ and $\delta \neq 4$ holds, then the only transition that can fire, after a total of $\delta$ time units, is the one from $q_0$ to $q_1$; the new configuration of $\arena$ is $\step((q_0, \{x = 0 \}), (a, \delta)) = \{ (q_1, \{ x =\delta \}) \}$.
If instead $\delta = 4$ holds, then the guards of both transitions exiting from $q_0$ are satisfied; thus, the environment
must choose only one transition between these two to fire, meaning that $\step((q_0, \{x = 0 \}), (a, \delta = 4)) = \{ (q_1, \{ x = 4 \}), (q_2, \{ x = 0 \}) \}$.

If the controller chooses $\delta > 4$ as a delay, then this must fully elapse, \emph{i.e.}, neither player is allowed to make further decisions until a new configuration is reached upon applying $\step$.
This means that, in \autoref{fig:move_example}, the environment cannot choose the transition from $q_0$ to $q_2$ as the one to fire when clock $x$ is equal to 4 while this clock is increasing to reach $\delta > 4$.
This implies that the controller has a winning strategy for the winning condition $\winCond_0 = \Next q_1$, which consists in avoiding a delay $\delta = 4$ in the initial configuration.
It is also easy to verify that, in this example, there exists no winning controller strategy for the winning condition $\winCond_1 = \Diamond q_3$.
\end{example}

\label{rmk:external}
Specifying winning conditions with constraints on clocks external 
to a \ac{tarena} $\arena$ enhances the specification of controller behaviors.
Directly encoding such constraints into $\arena$ would require an exhaustive analysis of its structure and a careful assessment of the effects introduced by modifying its locations and transitions.
This process is highly error-prone if performed manually in Timed Arenas with a large number of locations and transitions, and also computationally expensive if automated as a preprocessing step, thus introducing unnecessary modeling effort and semantic complexity.
In contrast, specifying these constraints declaratively in a winning condition shifts the focus from low-level structural encodings to high-level requirements specification, leading to cleaner models and more maintainable \ac{tcg} formulations.
Hence, a given \ac{tarena} can be reused across multiple controller specifications.
The next example captures this point.

\begin{example}
Consider the \ac{tarena} $\arena$ of \autoref{fig:move_example} and suppose that the controller is required to visit $q_1$ at least twice, and that the elapsed time between two such visits is exactly three time units.
However, the guard $x \ge 1$ on the self loop of $q_1$ does not enforce the aforementioned timing constraints between consecutive moves, nor does it provide any means of determining how many times $q_1$ has been visited.
The introduction of two external clocks $y_1,y_2 \in \cckc$ such that $(\cckc \cap \ackc) = \varnothing$ resolves this issue in a natural way.
The resulting winning condition is the following:
\[
\winCond :=
\Next \big(
  (q_1 \land y_1 = 0)
  \Until
  (q_1 \land y_1 = 3 \land y_2 > 3) \big).
\]
In particular, the constraint $y_2 > 3$ is required to ensure that $q_1$ is visited at least twice; without this constraint, $\winCond$ could be trivially satisfied by moving in $q_1$ after three time units from the initial configuration.
The constraint $y_1 = 0$ ensures that $y_1$ is reset at least once while in $q_1$. 
Satisfying $\winCond$ requires $y_1$ to be reset on every traversal of the self loop in $q_1$, until a move with a delay of exactly three time units is performed, resulting in a configuration that is again in $q_1$.
In this case, a winning controller strategy does not exist, since the environment can resolve the non-determinism over transitions forcing the game to reach $q_3$ once in $q_1$ {if the chosen delay is greater than or equal to three.}

Without the support of clocks external to $\arena$, encoding the per-move timing constraints would require unrolling the self-loop on $q_1$ into a chain of distinct locations and transitions, each with specific guards and resets. 
Modifying the specification, \emph{e.g.}, by changing the required number of moves or the time constraints between consecutive moves, would require a different unrolling.
In contrast, with \acp{tcg} only the winning condition changes, while the \ac{tarena} is untouched.
\end{example}

\section{\aclp{tcg} over Region Transition Systems}
\label{sec:tar_extension_and_region_semantics}
This section first presents an approach for handling external clocks by directly embedding them into a \ac{tarena} (\autoref{subsec:tar_extension}). 
It then adapts the semantics of \acp{tcg} to state regions (\autoref{subsec:using_regions_for_tcgs}).

\subsection{Semantic Equivalence for Handling External Clocks}
\label{subsec:tar_extension}
The algorithms presented in \autoref{subsec:solvingalg}
and \autoref{sec:synthesizing_winning_controller_strategies} for solving \acp{tcg} and synthesizing winning controller strategies, respectively, are defined under the assumption that all clocks belong to the \ac{tarena}. 
To justify this assumption, we first describe a construction that embeds clocks appearing exclusively in a winning condition (also referred to as external clocks) into a \ac{tarena}.
We then prove that solving a \ac{tcg} on the resulting extended \ac{tarena} is equivalent to solving the original \ac{tcg} in which some clocks belong only to the winning condition.
We present the construction using the transition system semantics of \ac{ta}; recall that, given a \ac{tarena} $\aut$ and a play $\rho$ over $\aut$, we refer to a single pair $(q,v)$ occurring in $\rho$ as \emph{configuration}.
The same construction applies when considering $\RTSA$; recall that the \ac{rts} abstraction will be used as the semantic representation on which the algorithms introduced later in this section operate.

\begin{definition}
\label{def:tar-extension}
    Let $\tcgshort = (\aut, \winCond)$ be a \ac{tcg}, where $\aut = (\act, Q, Q_{0}, \ackc, T, \invariant)$.
    The \emph{\aprime{}} of $\tcgshort$ is a \ac{tcg} $\tcgshort' = (\aut', \winCond)$, where
    $\aut' = (\act', Q, Q_{0}, \ackc', T', \invariant)$ is such that:
    \begin{itemize}[nosep]
        \item $\act' = \{ a_{\mathcal{X}} \; | \; a \in \act \land \mathcal{X} \in 2^{(\cckc \setminus \ackc)} \}$ is the extended set of actions;
        \item $\ackc' = (\cckc \cup \ackc)$ is the extended set of clocks;
        \item For every $t = (q, a, \gamma, Y, q') \in T$, let 
    $\mathbf{T}_t = \{ (q, a_{\mathcal{X}}, \gamma, Y \cup \mathcal{X}, q') \in (Q \times \act' \times \Gamma(\ackc) \times 2^{\ackc'} \times Q) \; | \; \mathcal{X} \in 2^{(\cckc \setminus \ackc)} \}$; 
    then, $T' = \bigcup_{t \in T} \mathbf{T}_t$ is the extended set of transitions.
    \end{itemize}
\end{definition}

Informally, the \ac{tarena}-extension of a \ac{tcg} $\tcgshort = (\aut, \winCond)$ extends the set of actions $\act$ by introducing, for every action in $\aut$, a corresponding action for each possible subset of external clocks that can be reset over a transition. 
The set $\ackc$ is then extended with clocks in $\cckc$; therefore, the resulting \ac{tarena}-extension contains no external clocks.
Finally, for every transition in $T$, a new transition is introduced for every extended action $a_{\mathcal{X}} \in \act'$. 
Clocks in $\mathcal{X}$, whose reset is specified by the action subscript, are also explicitly added to the set $Y$ of the corresponding original transition. 
This construction reflects the possibility of external clocks being arbitrarily reset (by the controller) over a transition in $T$.

{Notably, since $\cckc \subseteq \ackc'$ in the \ac{tarena}-extension $\tcgshort' = (\aut', \winCond)$ of $\tcgshort$, every reset policy always returns the empty set. 
Hence, we simply denote the clock assignment associated with a play $\rho'$ over $\transtap$ by $\nu'$, where $\nu'$ assigns values only to clocks in $\ackc'$ (see \autoref{sec:timed_CLTLoc_games} for clock assignments).}
Thus, for every $i \in \mathbb{N}$, given a configuration $(q'_i, v'_i)$ for $\tcgshort'$, it holds that: 
$\forall x \in \ackc' : \nu'(i, x) = v'_i(x)$.
It follows that $\nu'$ depends only on $\rho'$.
In the remainder of this subsection, the concepts of plays and clock assignments 
are general and are not necessarily related to winning controller strategies.

\begin{definition}[$\nu$-equivalent plays]
\label{def:nu-equivalence}
    {Let $\tcgshort = (\arena, \winCond)$ be a \ac{tcg}, let $\tcgshort' = (\arena', \winCond)$ be its
    \ac{tarena}-extension, let $\nabla$ be a reset policy for $\tcgshort$, and let
    $\rho = (q_0, v_0) \xrightarrow{(a_0, \delta_0)} (q_1, v_1) \cdots$ and
    $\rho' = (q'_0, v'_0) \xrightarrow{(a'_0, \delta'_0)} (q'_1, v'_1) \cdots$
    be plays over $\aut$ and $\aut'$, respectively.
    Then, we say that $\rho$ and $\rho'$ are \emph{$\nu$-equivalent under $\nabla$}, written $\rho \equiv_{\nu, {\nabla}} \rho'$, if:
    \emph{(i)} for every position $i \in \mathbb{N}$, $q'_i = q_i$, $v'_i(x) = v_i(x)$ for all $x \in \ackc$, $\delta'_i = \delta_i$, and
    $a'_i = (a_i)_{\mathcal{X}_i}$, where $\mathcal{X}_i = \nabla(\rho_{\le i})$, and 
    \emph{(ii)} for every position $i \in \mathbb{N}$ and for all $x \in (\cckc \setminus \ackc)$, $\nu'(i, x) = \ckcass(i, x)$.
    Whenever the context makes clear that a clock assignment $\ckcass$ is induced by a play $\rho$ and a reset policy $\nabla$, we denote it simply by $\nu$ for brevity and say that $\rho$ and $\rho'$ are \emph{$\nu$-equivalent}, denoted by $\rho \equiv_{\nu} \rho'$.}
    %
\end{definition}

Informally, two plays are $\nu$-equivalent if they visit the same locations at the same positions in time, where the values of all clocks also coincide.
{Furthermore, the two plays consist of the same moves, with respect to both actions, delays, and reset external clocks.}
Notice that, although $\nu$-equivalence is defined over infinite plays, it extends in a natural way also to finite prefixes of such plays.
Indeed, any pair of finite prefixes of equal length extracted from $\nu$-equivalent plays is itself $\nu$-equivalent according to the conditions of \autoref{def:nu-equivalence}, restricted to such prefixes.
Thus, given a pair of plays $\rho$ and $\rho'$ such that $\rho \equiv_{\nu} \rho'$, we denote the corresponding $\nu$-equivalent finite prefixes by $\rho_{\le i} \equiv_{\nu} \rho'_{\le i}$, for every $i \in \mathbb{N}$.

We now prove that, given a \ac{tcg} $\tcgshort$, a controller strategy for $\tcgshort$, {and a reset policy for $\tcgshort$}, there always exists a corresponding controller strategy for the \ac{tarena}-extension of $\tcgshort$ that induces $\nu$-equivalent plays, and vice versa.

\begin{lemma}
\label{lem:ttp_eq}
    {Let $\tcgshort = (\aut, \winCond)$ be a \ac{tcg} and let $\tcgshort' = (\aut', \winCond)$ be its \aprime{}.
    For every controller strategy $\strategyc$ for $\tcgshort$ and every reset policy $\nabla$ for $\tcgshort$, there exists a controller strategy $\strategyc'$ for $\tcgshort'$ such that, for every play $\rho'$ starting from an initial configuration $(q'_0, v'_0)$ consistent with $\strategyc'$,
    there exists a play $\rho$ starting from an initial configuration $(q_0, v_0)$ consistent with $\strategyc$ such that $\rho \equiv_{\nu} \rho'$.}
\end{lemma}

{\begin{proof}
    Given $\strategyc$ and $\nabla$, we define a controller  strategy $\strategyc'$ for $\tcgshort'$.
    Let $\mathsf{proj}$ denote the projection from plays over $\aut'$ to plays over $\aut$, obtained by removing the external clocks and the action subscripts, while preserving locations, delays, underlying actions, and the values of the clocks in $\ackc$.
    The same notation is used for finite play prefixes.
    For every $i \in \mathbb{N}$ and every play prefix $\rho'_{\le i}$ 
    over $\aut'$,  we define $\strategyc'(\rho'_{\le i})$ from $\strategyc(\mathsf{proj}(\rho'_{\le i}))$ as follows:
    \[
    \strategyc'(\rho'_{\le i}) =
    \begin{cases}
        (a_{\mathcal X},\delta)
        & \text{if $q_i\in Q_{\cont}$ and
            $\strategyc(\mathsf{proj}(\rho'_{\le i}))=(a,\delta)$}\\
        a_{\mathcal X}
        & \text{if $q_i\in Q_{\env}$ and
            $\strategyc(\mathsf{proj}(\rho'_{\le i}))=a$}
    \end{cases},
    \qquad 
    \text{where}
    \qquad
    \mathcal X=\nabla(\mathsf{proj}(\rho'_{\le i})).
    \]
    
    Let $\rho'$ be any play consistent with $\strategyc'$. 
    By \autoref{def:tar-extension}, every transition of $\aut'$ corresponds to a transition of $\aut$ with the same source and target locations, guard, and resets of the clocks in $\ackc$.
    Moreover, both plays $\mathsf{proj}(\rho')$ and $\rho'$ have the same delays; hence, $\mathsf{proj}(\rho')$ is consistent with $\strategyc$.
    
    By construction of $\strategyc'$, external clocks reset at each position of $\rho'$ are exactly those specified by $\nabla$ on the corresponding prefix of $\mathsf{proj}(\rho')$.
    Since these clocks are initially zero and evolve according to the same delays, their values in $\rho'$ coincide with those given by $\nu_{\langle\mathsf{proj}(\rho'),\nabla\rangle}$.
    Thus, in each position, the two plays have the same locations, delays, and values of all clocks, and their actions and external-clock resets correspond.
    Hence, $\mathsf{proj}(\rho')\equiv_\nu\rho'$, where $\nu=\nu_{\langle\mathsf{proj}(\rho'),\nabla\rangle}$.  
\end{proof}}

\begin{lemma}
\label{lem:tpt_eq}
    {Let $\tcgshort = (\aut, \winCond)$ be a \ac{tcg} and let $\tcgshort' = (\aut', \winCond)$ be its \aprime{}.
    For every controller strategy $\strategyc'$ for $\tcgshort'$, there exists a controller strategy $\strategyc$ for $\tcgshort$ and a reset policy $\nabla$ for $\tcgshort$ such that, for every play $\rho$ from $(q_0, v_0)$ consistent with $\strategyc$, there exists a play $\rho'$ from $(q'_0, v'_0)$ consistent with $\strategyc'$ such that $\rho \equiv_{\nu} \rho'$.}
\end{lemma}

{\begin{proof}
    Given $\strategyc'$, we construct $\strategyc$ and $\nabla$ by induction on the length of finite play prefixes of $\aut$.
    
    For every initial prefix $\rho_{\le 0}=(q_0,v_0)$, let $\rho'_{\le 0}$ be the corresponding initial prefix of $\aut'$, where all clocks are zero.
    
    Let $i\ge 0$ and suppose, by the induction hypothesis, that $\rho'_{\le i}$ has been defined for a corresponding prefix $\rho_{\le i}$ ending in $(q_i,v_i)$.
    We define $\strategyc$ and $\nabla$ on $\rho'_{\le i}$ according to $\strategyc'(\rho'_{\le i})$ as follows:
    \[
    \strategyc(\rho_{\le i}) =
    \begin{cases}
        (a,\delta)
        & \text{if $q_i\in Q_{\cont}$ and
            $\strategyc'(\rho'_{\le i})
            =(a_{\mathcal X},\delta)$}\\
        a
        & \text{if $q_i\in Q_{\env}$ and
            $\strategyc'(\rho'_{\le i})
            =a_{\mathcal X}$}
    \end{cases},
    \qquad
    \text{where}
    \qquad
    \nabla(\rho_{\le i})=\mathcal X.
    \]
    
    For 
    every one-step extension $\rho_{\le i+1}$ of $\rho_{\le i}$, we then define $\rho'_{\le i+1}$ by extending $\rho'_{\le i}$ with the corresponding transition 
    of $\aut'$, preserving the original action and delay and resetting the additional clocks in $\mathcal X$.
    By \autoref{def:tar-extension}, such a transition    always exists. This completes the inductive construction of $\strategyc$, $\nabla$, and
    the extended prefixes.
    
    Let $\rho$ be any play consistent with $\strategyc$.
    By construction, the extended prefixes $\rho'_{\le i}$ of $\rho$ form an infinite play $\rho'$ consistent with $\strategyc'$.
    Since the external clocks are reset according to $\nabla$ and since, in each position, all other clocks, locations, and delays
    are preserved, it follows that $\rho\equiv_\nu\rho'$, where $\nu=\nu_{\langle\rho,\nabla\rangle}$.
    \end{proof}}

The following proposition shows that, when considering two $\nu$-equivalent plays, one from a \ac{tcg} and one from its \ac{tarena}-extension, the satisfaction of the winning condition is preserved.

\begin{proposition}
\label{prop:wincond_satisfaction}
    {Let $\tcgshort = (\aut, \winCond)$ be a \ac{tcg}, let $\tcgshort' = (\aut', \winCond)$ be its \aprime{}, let $\nabla$ be a reset policy for $\tcgshort$, and let $\rho$ and $\rho'$ be plays over $\aut$ and $\aut'$, respectively.
    If $\rho \equiv_{\nu} \rho'$, then $(\rho, \nu) \models \winCond \Leftrightarrow (\rho', \nu') \models \winCond$, i.e., $\winCond$ is satisfied at position 0 in both interpretations.}
\end{proposition}

\begin{proof}
    The proposition follows directly from \autoref{def:tar-extension}.
    Indeed, since $\rho \equiv_{\nu} \rho'$ holds, it follows that, for every position $i \in \mathbb{N}$, the configurations $(q_i, v_i)$ and $(q'_i, v'_i)$ of $\rho$ and $\rho'$, respectively, are such that $q'_i = q_i$ (the location is the same) and $\forall x \in (\ackc \cup \cckc) : \nu'(i,x) = \nu(i,x)$ (the value of every clock is the same for both configurations).
    Hence, if $(\rho, \nu)$ satisfies $\winCond$ as detailed in \autoref{tab:semantics}, then $(\rho', \nu')$ must also satisfy $\winCond$, and vice versa.
\end{proof}

\begin{theorem}
\label{thr:semantic_equivalence}
    {Let $\tcgshort = (\aut, \winCond)$ be a \ac{tcg} and let
    $\tcgshort' = (\aut', \winCond)$ be its \aprime{}.
    Then, there exists a winning controller strategy for $\tcgshort$ starting from an initial configuration $(q_0, v_0)$ if, and only if,
    there exists a winning controller strategy for $\tcgshort'$ starting from the same initial configuration.}
\end{theorem}

\begin{proof}
    By \autoref{lem:ttp_eq} and \autoref{prop:wincond_satisfaction}, if there exists a winning controller strategy  from $(q_0, v_0)$ in $\tcgshort$, then there exists a winning controller strategy  from $(q_0, v_0)$ in $\tcgshort'$.
    Conversely, by \autoref{lem:tpt_eq} and \autoref{prop:wincond_satisfaction}, if there exists a winning controller strategy  from $(q_0, v_0)$ in $\tcgshort'$, then there exists a winning controller strategy  from $(q_0, v_0)$ in $\tcgshort$.
    %
\end{proof}

Thanks to \autoref{thr:semantic_equivalence},
solving a \ac{tcg} $\tcgshort = (\aut, \winCond)$ is equivalent to solving its \aprime{} $\tcgshort' = (\aut', \winCond)$; hence, in the remainder of this work, we consider only \ac{tarena}-extensions of \acp{tcg}.
Accordingly, whenever the notation of a \ac{tcg} is used, it implicitly refers to its corresponding \aprime{} defined in \autoref{def:tar-extension}, {\emph{e.g.}, subscripts are omitted from action names to improve readability}.
{Notably, every reset policy $\nabla$ now always returns the empty set, since the \ac{tarena}-extension has no external clocks.}

\subsection{Adapting \texorpdfstring{\aclp{tcg}}{TCGs} to Region Transition Systems}
\label{subsec:using_regions_for_tcgs}

{ 
Let $\aut$ be a \ac{tarena}.
Recall that the \ac{rts} $\RTSA$ is a finite abstraction of $\transta$.
Although a controller strategy over $\transta$ may distinguish
configurations whose clock valuations belong to the same clock region,
the bisimulation between $\transta$ and $\RTSA$ allows us to relate
plays generated in the two semantics.
In the following, we adapt moves, plays, and strategies to state
regions and then show that the existence of winning controller
strategies is preserved.

{Since we consider \ac{tarena}-extensions, $\aut$ also contains external clocks, and the maximum constant $\cmax$ must account for them as well. 
Hence, $\cmax$ is the maximum constant against which any clock is compared, either in $\aut$ or in $\winCond$. 
Recall the assumption made in \autoref{sec:theoretical_background} that the same $\cmax$ is used for every clock of a \ac{ta}, which here means every clock in $\ackc$.}

Let $\RTSA = (\rtsstate, \mathit{Act} \cup \{ \tau \}, \mathcal{I}, \rtsarrow)$ be the \ac{rts} of $\arena$.
When considering regions, an exact delay is abstracted by the clock region
reached after such a delay has elapsed and before a discrete transition
is taken.
Accordingly, a \emph{region move} is a pair $(a,[w])$, where
$a \in \act$ and $[w] \in \eval/\cong$.
The mapping $\step_{\mathcal{R}} : (\rtsstate \times ( \act \times \eval / \cong ) ) \rightarrow \wp(\rtsstate)$ returns all states $s'$ that are reachable from $s$ by letting time
elapse until the clock region $[w]$ is reached and then taking discrete transitions labeled with action $a$.
Specifically, $\step_{\mathcal{R}}$ is defined as follows:
\begin{equation}
    \label{eq:step_over_states}
    \step_{\mathcal{R}}(s, (a, [w])) = 
    \left\{ s' = (q', [v']) \in \rtsstate
    \; \middle| \;
    (s = (q, [v]), \tau, s_1 = (q, [w])) \in \; \rtsarrow^*_{\tau} \land \; (s_1, a, s') \in \; \rtsarrow
    \right\}.
\end{equation}
The clock region $[v']$ of $s'$ may differ from $[w]$ because some clocks may be reset over a discrete transition.
A \emph{(region) play} $\rho_{\mathcal{R}}$ over $\RTSA$ is an infinite sequence
\[
\rho_{\mathcal R}=s_0\xrightarrow{(a_0,[w_0])}s_1\xrightarrow{(a_1,[w_1])}s_2\cdots
\]
alternating states and region moves, where
$s_0 \in \mathcal I$ and, for all $n \in \mathbb N$,
$s_n=(q_n,[v_n]) \in \rtsstate$ and
$
s_{n+1} \in \step_{\mathcal R}(s_n,(a_n,[w_n]))
$.
We recall the assumption that an infinite play can always be defined in \acp{tcg} (see \autoref{sec:timed_CLTLoc_games}).
The set of all finite prefixes $\rho_{\mathcal{R} \le n}$ of region plays is denoted by $\mathit{Plays}_{\mathcal{R}}$.
Given a play
$
\rho = (q_0,v_0) \xrightarrow{(a_0,\delta_0)}(q_1,v_1)\xrightarrow{(a_1,\delta_1)}(q_2,v_2)\cdots
$
over $\arena$, its \emph{region abstraction}, denoted by
$\mathsf{reg}(\rho)$, is defined as:
\begin{equation}
\label{eq:region_abstraction}
\mathsf{reg}(\rho)
=
(q_0,[v_0]) \xrightarrow{(a_0,[v_0+\delta_0])}(q_1,[v_1])\xrightarrow{(a_1,[v_1+\delta_1])}(q_2,[v_2])\cdots
\end{equation}
Similarly, the region abstraction of a finite prefix
$\rho_{\le n}$ is denoted by $\mathsf{reg}(\rho_{\le n})$.
By construction of $\RTSA$ and $\step_{\mathcal{R}}$, if $\rho$ is a
play over $\arena$, then $\mathsf{reg}(\rho)$ is a play over $\RTSA$.
Notice that $\mathsf{reg(\rho)}$ is not injective: infinitely many plays over configurations may share the same region abstraction.}

A \emph{(region) controller strategy} $\strategyc^{\mathcal{R}}$ is a mapping 
$\strategyc^{\mathcal{R}} : \mathit{Plays}_{\mathcal{R}} \rightarrow ((\act \times \eval / \cong) \cup \act)$; a \emph{(region) environment strategy} is a pair 
$\strategye^{\mathcal{R}} = (\strategye^{1,\mathcal{R}}, \strategye^{2,\mathcal{R}})$, 
where 
$\strategye^{1,\mathcal{R}} : \mathit{Plays}_{\mathcal{R}} \rightarrow \eval / \cong$ is a partial mapping (the environment chooses a clock region only at  locations in $Q_{\env}$)
and 
$\strategye^{2, \mathcal{R}} : (\mathit{Plays}_{\mathcal{R}} \times ( \wp(\mathcal{S}) \setminus \{ \varnothing \})) \rightarrow \mathcal{S}$.
The choices made by the players in their strategies depend on the current location (see \autoref{sec:timed_CLTLoc_games}).
The outcome $\Theta_{(\strategyc^{\mathcal{R}}, \strategye^{\mathcal{R}})}^{\mathcal{R}} : \plays_{\mathcal{R}} \rightarrow \rtsstate$ is defined as follows:
\[
\Theta_{(\strategyc^{\mathcal{R}}, \strategye^{\mathcal{R}})}^{\mathcal{R}}(\rho_{\mathcal{R} \le n}) = 
\begin{cases}
\strategye^{2, \mathcal{R}} \big(\rho_{\mathcal{R} \le n}, \step_{\mathcal{R}}(s_n = (q_n, [v_n]), \strategyc^{\mathcal{R}}(\rho_{\mathcal{R} \le n}))\big) & \text{if $q_n \in Q_{\cont}$} \\

\strategye^{2,\mathcal{R}} \big(\rho_{\mathcal{R} \le n}, \step_{\mathcal{R}}(s_n = (q_n, [v_n]), (\strategyc^{\mathcal{R}}(\rho_{\mathcal{R} \le n}), \strategye^{1, \mathcal{R}}(\rho_{\mathcal{R} \le n})))\big) & \text{if $q_n \in Q_{\env}$}
\end{cases}.
\]

{Consistency of a region play with a controller strategy over $\RTSA$ is defined
analogously to the configuration-based case.
In particular, a play
$
\rho_{\mathcal R}=
s_0 \xrightarrow{(a_0,[w_0])} s_1\xrightarrow{(a_1,[w_1])} s_2 \cdots
$
is consistent with $\strategyc^{\mathcal R}$ if there exists an
environment strategy $\strategye^{\mathcal R}$ such that, for every
$n\in\mathbb N$,
$
s_{n+1} = \Theta^{\mathcal R}_{(\strategyc^{\mathcal R},  \strategye^{\mathcal R})}
          (\rho_{\mathcal R\le n})
$
and the move $(a_n,[w_n])$ agrees with the choices of both players, that is,
$\strategyc^{\mathcal R}(\rho_{\mathcal R\le n})
=(a_n,[w_n])$ if $q_n\in Q_{\cont}$, whereas
$\strategyc^{\mathcal R}(\rho_{\mathcal R\le n})=a_n$ and
$\strategye^{1,\mathcal R}(\rho_{\mathcal R\le n})=[w_n]$
if $q_n\in Q_{\env}$.
It follows that every finite prefix of $\rho_{\mathcal R}$ is consistent with $\strategyc^{\mathcal R}$.}

\begin{remark}
\label{rem:region_winning_condition}
The semantics of \acp{tcg} is the same as in \autoref{tab:semantics}, except that plays over states are used instead of plays over configurations.
A state $(q, [v'])$ satisfies a pure \ac{cltloc} formula $\varphi$ if there exists a configuration $(q, v)$ satisfying $\varphi$ such that $v \in [v']$.
Since \ac{cltloc} formulae predicate over locations and clock constraints (\autoref{eq:cltloc_grammar}) their satisfaction depends only on the location and clock region, rather than on a particular clock valuation within the clock region. 
Hence, if a configuration $(q, v)$ with $v \in [v']$ satisfies $\varphi$, then every configuration $(q, \bar{v})$, with $\bar{v} \in [v']$, also satisfies $\varphi$.

Since the domain of reset policies is empty in the \ac{tarena}-extension, clock assignments are induced exclusively by plays. 
Accordingly, we omit clock assignments from the satisfaction notation introduced in \autoref{tab:semantics} for both the configuration and region semantics. 
For instance, we write $\rho, i \models \varphi$ to denote that formula $\varphi$ is satisfied by play $\rho$ at position $i$.
\end{remark}

\begin{proposition}
\label{prop:region_semantics_satisfaction}
    {Let $\tcgshort = (\aut, \winCond)$ be a \ac{tcg} and let $\rho$ be a play over $\aut$.
    It follows that, for every position $i \in \mathbb{N}$, $\rho,i \models \winCond$ if, and only if, $\mathsf{reg}(\rho),i \models \winCond$.
    In particular, $\rho \models \winCond$ if, and only if, $\mathsf{reg}(\rho) \models \winCond$.}
\end{proposition}

\begin{proof}
    {The proof proceeds by structural induction on subformulae $\phi$ of $\winCond$, using the future-only fragment of the \ac{cltloc} grammar in \autoref{eq:cltloc_grammar} and the \acp{tcg} semantics in \autoref{tab:semantics}.

    \begin{itemize}[nosep]
    \item 
    \emph{Case 1 (atomic proposition): $\phi = p$.} 
    This follows directly from \autoref{rem:region_winning_condition}.

    \item 
    \emph{Case 2 (clock constraint): $\phi = x \sim c$.}
    This follows directly from \autoref{rem:region_winning_condition}.

    \item 
    \emph{Case 3 (negation): $\phi = \neg \varphi$.}
    By \autoref{tab:semantics}, $\rho, i \models \neg \varphi$ if, and only if, $\rho, i \not\models \varphi$, which by the inductive hypothesis holds if, and only if, $\mathsf{reg}(\rho), i \not\models \varphi$, \emph{i.e.}, $\mathsf{reg}(\rho), i \models \neg \varphi$.

    \item 
    \emph{Case 4 (conjunction): $\phi = \varphi \land \psi$.}
    By \autoref{tab:semantics}, $\rho, i \models \varphi \land \psi$ if, and only if, $\rho, i \models \varphi$ and $\rho, i \models \psi$, which by the inductive hypothesis holds if, and only if, $\mathsf{reg}(\rho), i \models \varphi$ and $\mathsf{reg}(\rho), i \models \psi$, \emph{i.e.}, $\mathsf{reg}(\rho), i \models \varphi \land \psi$.

    \item 
    \emph{Case 5 (next): $\phi = \Next \varphi$}.
    By \autoref{tab:semantics}, $\rho, i \models \Next \varphi$ if, and only if, $\rho, i + 1 \models \varphi$, which by the inductive hypothesis (applied at position $i + 1$) holds if, and only if, $\mathsf{reg}(\rho), i + 1 \models \varphi$, \emph{i.e.}, $\mathsf{reg}(\rho), i \models \Next \varphi$.

    \item 
    \emph{Case 6 (until): $\phi = \varphi \Until \psi$.} By \autoref{tab:semantics}, $\rho, i \models \varphi \Until \psi$ if, and only if, there exists $j \ge i$ such that $\rho, j \models \psi$ and $\rho, n \models \varphi$ for every $n \in [i, j)$. 
    By the inductive hypothesis (applied to $\varphi$ at every position $n \in [i, j)$ and to $\psi$ at position $j$) this holds if, and only if, there exists $j \ge i$ such that $\mathsf{reg}(\rho), j \models \psi$ and $\mathsf{reg}(\rho), n \models \varphi$ for every $n \in [i, j)$, \emph{i.e.}, $\mathsf{reg}(\rho), i \models \varphi \Until \psi$.\qedhere
    \end{itemize}}
\end{proof}

{Accordingly, a controller strategy $\strategyc^{\mathcal{R}}$ is winning for a \ac{tcg} $\tcgshort = (\aut, \winCond)$ from an initial state $s_0 \in \mathcal{I}$ if, for every play ${\rho_{\mathcal{R}}}$ from $s_0$ and consistent with $\strategyc^{\mathcal{R}}$, $\rho_{\mathcal R}$
satisfies $\winCond$ (see \autoref{prop:region_semantics_satisfaction}).

We now show that solving a \ac{tcg} using configurations is equivalent to solving it using states.}

{\begin{lemma}
    \label{lem:conf_to_state_strat}
    Let $\tcgshort=(\aut,\winCond)$ be a \ac{tcg}.
    For every controller strategy $\strategyc$ starting from an initial configuration $(q_0, v_0)$, there exists a controller
    strategy $\strategyc^{\mathcal R}$ starting from an initial state $(q_0, [v_0])$ such that, for every play
    $\rho_{\mathcal R}$ consistent with $\strategyc^{\mathcal R}$,
    there exists a play $\rho$ consistent with $\strategyc$ such that
    $
    \rho_{\mathcal R}=\mathsf{reg}(\rho).
    $
\end{lemma}

{\begin{proof}
    Given $\strategyc$, we construct $\strategyc^{\mathcal R}$ by induction on finite region-play prefixes.
    At the same time, we fix for every prefix $\rho_{\mathcal R\le i}$ consistent with $\strategyc^{\mathcal R}$ a concrete representative  $\rho_{\le i}$ consistent with $\strategyc$, such that $\rho_{\mathcal R\le i}=\mathsf{reg}(\rho_{\le i})$.
    Each representative is chosen to extend the representative already fixed for its preceding prefix.
    
    \emph{Base case.}
    For $i=0$, let $\rho_{\mathcal R\le 0}=(q_0,[v_0])$.
    Since all clocks are initially zero, we fix $\rho_{\le 0}=(q_0,v_0)$ as its representative.
    
    \emph{Inductive step.}
    For $i\ge 1$, suppose that a representative $\rho_{\le i-1}$ has already been fixed for a prefix $\rho_{\mathcal R\le i-1}$ consistent with $\strategyc^{\mathcal R}$.
    Let $(q_{i-1},v_{i-1})$ be its last configuration.
    We define $\strategyc^{\mathcal R}$ as follows:
    \[
    \strategyc^{\mathcal R}(\rho_{\mathcal R\le i-1})=
    \begin{cases}
        (a,[v_{i-1}+\delta])
        &
        \text{if $q_{i-1}\in Q_{\cont}$ and
            $\strategyc(\rho_{\le i-1})=(a,\delta)$}\\
        a
        &
        \text{if $q_{i-1}\in Q_{\env}$ and
            $\strategyc(\rho_{\le i-1})=a$}
    \end{cases}.
    \]
    
    Consider any one-step extension $\rho_{\mathcal R\le i}$ consistent with $\strategyc^{\mathcal R}$, and let $[w]$ be the clock region reached before its discrete transition.
    If $q_{i-1}\in Q_{\cont}$, then $[w]=[v_{i-1}+\delta]$, where $\delta$ is the delay chosen by $\strategyc$.
    If $q_{i-1}\in Q_{\env}$, then the environment chooses $[w]$, and the properties of the \ac{rts} guarantee an admissible delay $\delta\ge 0$ such that $(v_{i-1}+\delta)\in[w]$.
    In both cases, the discrete transition of $\rho_{\mathcal R\le i}$ is enabled from $v_{i-1}+\delta$, since guards are preserved by clock-region equivalence.
    Applying the same transition and resets therefore yields a concrete successor in the corresponding target region.
    Thus, $\rho_{\le i-1}$ can be extended to a prefix $\rho_{\le i}$ consistent with $\strategyc$ such that $\rho_{\mathcal R\le i}=\mathsf{reg}(\rho_{\le i})$.
    We fix this extension as the representative of $\rho_{\mathcal R\le i}$.
    Finally, let $\rho_{\mathcal R}$ be any infinite
    play consistent with $\strategyc^{\mathcal R}$.
    By construction, the representatives of its successive prefixes extend one another,
    thus determining an infinite concrete play $\rho$ consistent with $\strategyc$ such that $\rho_{\mathcal R}=\mathsf{reg}(\rho)$.
\end{proof}
}

\begin{lemma}
    \label{lem:state_to_conf_strat}
    Let $\tcgshort=(\aut,\winCond)$ be a \ac{tcg}.
    For every controller strategy $\strategyc^{\mathcal R}$ starting from an initial state $(q_0, [v_0])$, there exists
    a controller strategy $\strategyc$ starting from an initial configuration $(q_0, v_0)$ such that, for every play $\rho$
    consistent with $\strategyc$, there exists a play
    $\rho_{\mathcal R}$ consistent with $\strategyc^{\mathcal R}$ such that
    $
    \rho_{\mathcal R}=\mathsf{reg}(\rho).
    $
\end{lemma}

{\begin{proof}
    Given $\strategyc^{\mathcal R}$, we define $\strategyc$ on every concrete play prefix $\rho_{\le i}$ ending
    in $(q_i,v_i)$, with $\rho_{\mathcal R\le i}=\mathsf{reg}(\rho_{\le i})$,  as follows:
    \[
    \strategyc(\rho_{\le i})=
    \begin{cases}
        (a,\delta)
        \text{ such that }(v_i+\delta)\in[w]
        &
        \text{if $q_i\in Q_{\cont}$ and
            $\strategyc^{\mathcal R}(\rho_{\mathcal R\le i})
            =(a,[w])$}\\
        a
        &
        \text{if $q_i\in Q_{\env}$ and
            $\strategyc^{\mathcal R}(\rho_{\mathcal R\le i})=a$}
    \end{cases}.
    \]
   In the first case, the existence of such a delay $\delta$ follows from the properties of the \ac{rts}, since $[w]$ is reachable from every clock valuation in $[v_i]$.  
     
    Let $\rho$ be any play consistent with $\strategyc$. We show that $\mathsf{reg}(\rho)$ is consistent with
    $\strategyc^{\mathcal R}$.
    At every position $i$, if $q_i\in Q_{\cont}$, then the controller chooses an action and a delay reaching
    the clock region prescribed by $\strategyc^{\mathcal R}$.
    The discrete transition selected by the environment has a corresponding transition in the \ac{rts}.
    If $q_i\in Q_{\env}$, then both strategies prescribe the same action, while the environment can select the clock region reached by the concrete delay and the corresponding discrete successor state.
    Hence, in each position, every move of $\rho$ corresponds to a move 
    allowed by $\strategyc^{\mathcal R}$, and therefore $\rho_{\mathcal R}=\mathsf{reg}(\rho)$ is a play consistent with $\strategyc^{\mathcal R}$.
\end{proof}}

\begin{theorem}
    {Let $\tcgshort = (\aut, \winCond)$ be a \ac{tcg}.
    There exists a winning controller strategy $\strategyc$ from an initial configuration $(q_0, v_0)$ for $\tcgshort$ if, and only if, there exists a winning controller strategy $\strategyc^{\mathcal{R}}$ from an initial state $(q_0, [v_0])$ for $\tcgshort$.}
\end{theorem}

\begin{proof}
    {By \autoref{lem:conf_to_state_strat}, by \autoref{prop:region_semantics_satisfaction}, and by the fact that $\strategyc$ is winning from $(q_0, v_0)$, it follows that $\strategyc^{\mathcal{R}}$ is winning from $(q_0, [v_0])$.
    Conversely, by \autoref{lem:state_to_conf_strat}, by \autoref{prop:region_semantics_satisfaction}, and by the fact that $\strategyc^{\mathcal{R}}$ is winning from $(q_0, [v_0])$, it follows that $\strategyc$ is winning from $(q_0, v_0)$.}
\end{proof}

Since the underlying notions of play and strategy, as well as the \acp{tcg} semantics, remain unchanged, in the remainder of this work we use $\rho$, $\strategyc$, and $\strategye$ instead of $\rho_{\mathcal{R}}$, $\strategyc^{\mathcal{R}}$, and $\strategye^{\mathcal{R}}$, respectively, to avoid visual clutter.

\section{Solving \aclp{tcg}}
\label{sec:solving_timed_CLTLoc_games}
This section introduces the controllable predecessor operator for \acp{tcg} (\autoref{subsec:controllable_pred_op}) and presents backward algorithms for solving \acp{tcg} with \emph{reachability}, \emph{safety}, and \emph{$\Next$-conjunction} winning conditions (\autoref{subsec:solvingalg}). 
Then, it introduces additional winning conditions (\autoref{subsec:additional_winning_conditions}) and proves that solving reachability \acp{tcg} is EXPTIME-complete (\autoref{subsec:exptime_completeness}).

Without loss of generality, the theorems presented in \autoref{subsec:solvingalg} and \autoref{sec:synthesizing_winning_controller_strategies} are formulated with respect to initial state regions since plays (defined over state regions rather than configurations, as detailed in \autoref{subsec:using_regions_for_tcgs}), including those consistent with winning controller strategies, start from initial state regions.

\subsection{Controllable Predecessor Operator}
\label{subsec:controllable_pred_op}
Let $\arena$ be a \ac{tarena} and let $\RTSA = (\rtsstate, \mathit{Act} \cup \{ \tau \}, \mathcal{I}, \rtsarrow)$ be the \ac{rts} of $\arena$.
In the following, $G \subseteq \mathcal{S}$ denotes a subset of states of $\RTSA$.
Recall that, by \autoref{def:timed_arena}, $\mathcal{S} = (\mathcal{S}_{\cont} \cup \mathcal{S}_{\env})$ and $(\mathcal{S}_{\cont} \cap \mathcal{S}_{\env}) = \varnothing$.
We introduce the \emph{controllable predecessor operator} $\pi : \wp(\mathcal{S}) \rightarrow \wp(\mathcal{S})$, used to solve \acp{tcg}.
We split $\pi$ into the operators $\pi_{\cont} : \wp(\mathcal{S}) \rightarrow \wp(\mathcal{S}_{\cont})$ and $\pi_{\env} : \wp(\mathcal{S}) \rightarrow \wp(\mathcal{S}_{\env})$, by considering controller and environment states, respectively. 
Then,
$\pi= (\pi_{\cont}\cup\pi_{\env})$,
where:
\begin{equation}
\label{eq:controllable_predecessor_operator_c}
\pi_{\cont}(G) = 
\left\{
s \in \mathcal{S}_{\cont} 
\; \middle| \;
\begin{array}{@{}l@{}}
\exists a \in \act : \exists s_1 \in \mathcal{S}, \exists s_2 \in G : (s, \tau, s_1) \in \; \rtsarrow^*_{\tau} \land \; (s_1, a, s_2) \in \; \rtsarrow \\
\land \; \neg \exists s_3 \in (\mathcal{S} \setminus G) : (s_1, a, s_3) \in \; \rtsarrow
\end{array}
\right\},
\end{equation}
\begin{equation}
\label{eq:controllable_predecessor_operator_e}
\pi_{\env}(G) = 
\left\{
s \in \mathcal{S}_{\env} 
\; \middle| \;
\begin{array}{@{}l@{}}
\exists a \in \act : \exists s_1 \in \mathcal{S}, \exists s_2 \in G : (s, \tau, s_1) \in \; \rtsarrow^*_{\tau} \land \;  (s_1, a, s_2) \in \; \rtsarrow \\
\land \; \neg ( \exists s_3 \in \mathcal{S}, \exists s_4 \in (\mathcal{S} \setminus G) : (s, \tau, s_3) \in \; \rtsarrow^*_{\tau} \land \; (s_3, a, s_4) \in \; \rtsarrow )
\end{array}
\right\}.
\end{equation}

Intuitively, operator $\pi$ returns all the predecessor states of $G$ from which the controller is guaranteed to reach a state in $G$ after the application of exactly one move.
In particular, both \autoref{eq:controllable_predecessor_operator_c} and \autoref{eq:controllable_predecessor_operator_e} require the existence of such a move.
This corresponds to the first line of \autoref{eq:controllable_predecessor_operator_c} and \autoref{eq:controllable_predecessor_operator_e}, 
{where $(s,\tau,s_1) \in \; \rtsarrow^*_{\tau}$ represents the passage of an
arbitrary amount of time, possibly zero, abstracted by the clock
region of $s_1$,}
while $(s_1, a, s_2) \in \; \rtsarrow$ is a discrete transition with action $a \in \act$.

Furthermore, since the controller must be guaranteed to reach a state in $G$, the following conditions must be satisfied.
If $s \in \mathcal{S}_{\cont}$, then the controller is in charge of choosing a clock region.
Hence, by the second line of \autoref{eq:controllable_predecessor_operator_c}, the environment cannot resolve the non-determinism over multiple enabled discrete transitions (that can be taken after the controller’s chosen clock region has been reached) by selecting a discrete transition ending in a state not contained in $G$.
If $s \in \mathcal{S}_{\env}$, then the environment is in charge of choosing a clock region.
Hence, by the second line of \autoref{eq:controllable_predecessor_operator_e} it must hold that, after action $a$ has been chosen by the controller, the environment cannot choose a clock region and resolve the non-determinism over multiple enabled discrete transitions (that can be taken after the environment’s chosen clock region has been reached) by selecting a discrete transition ending in a state not contained in $G$.

\begin{proposition}
\label{prop:pi_complexity}
Let $\tcgshort=(\arena,\winCond)$ be a \ac{tcg}, let $n=|\mathcal{S}|$ be the number of states of
$\RTSA$, let $\act$ be its set of actions,
and let $G\subseteq\mathcal{S}$.
The 
set $\pi(G)$ 
can be computed in $\mathcal{O}(n^2 \cdot |\act|)$ time.
\end{proposition}
\begin{proof}
For every state $s \in \mathcal S$ and action $a \in \act$, we first determine whether $s$ has an outgoing discrete transition labeled with $a$ to a state in $G$, and whether it has one to a state outside $G$.
This requires $\mathcal{O}(n^2 \cdot |\act|)$ time.

We then consider separately controller and environment states.

For each $s\in\mathcal{S}_{\cont}$, we examine the states reachable from $s$ by delay transitions.
By \autoref{eq:controllable_predecessor_operator_c}, $s\in\pi_{\cont}(G)$ if there exists a state
$s_1$ and an action $a$ having at least one successor in $G$ and no successors outside $G$.

For each $s\in\mathcal{S}_{\env}$, we similarly examine all states reachable from $s$ by delay transitions.
By \autoref{eq:controllable_predecessor_operator_e}, $s\in\pi_{\env}(G)$ if there exists an action $a$ having at least one successor in $G$ from some delay-reachable state, and no successor outside $G$ from any delay-reachable state.

{By the definition of the \ac{rts}, each state has at most one outgoing delay transition, leading to its successor clock region.
Thus, all delay-reachable states from a given state can be enumerated in $\mathcal{O}(n)$ time by following successive delay transitions.
Consequently, the delay-reachability relation can be computed in $\mathcal{O}(n^2)$ time.}

Since the conditions above {for controller and environment states} can be checked using the precomputed information, each state
can be processed in $\mathcal{O}(n \cdot |\act|)$ time. 
Therefore, $\pi(G)$ can be computed in $\mathcal{O}(n^2 \cdot |\act|)$ time.
\end{proof}

\subsection{Reachability, Safety, and \texorpdfstring{$\Next$-conjunction}{Next-conjunction} Winning Conditions}
\label{subsec:solvingalg}

We now show how to solve \acp{tcg} when the winning condition follows certain classical patterns expressing reachability or safety properties.
We then present the $\Next$-conjunction winning condition, which is a particular case of a reachability property where the states to be traversed can be explicitly specified move by move.

\textbf{Reachability.}~A \emph{reachability}
winning condition has the form
$
\winCond := \varphi \Until \psi
$,
where $\varphi$ and $\psi$ are pure \ac{cltloc} formulae (see \autoref{sec:theoretical_background} for reference).
The formula $\psi$ identifies the states of $\RTSA$ (also called \emph{goal} states) that the controller must reach to win in a \ac{tcg} $\tcgshort = (\arena, \winCond)$.
Recall that, if $\varphi \equiv \top$ holds, then $\varphi \Until \psi \equiv \Diamond \psi$.
%

\begin{figure}[tb]
\centering
\noindent
\begin{minipage}{0.48\textwidth}
\begin{algorithm}[H]
    \caption{Reachability}
    \label{alg:solve_tcgs}
    
    \KwData{$\tcgshort = (\arena, \winCond := \varphi \Until \psi)$: a \ac{tcg} where $\varphi, \psi$ are pure CLTLoc formulae.}
    \KwResult{\texttt{Reach}: controller-winning states in $\tcgshort$.}
    
    \Begin() {
        $\stateSatB \gets \text{states satisfying $\varphi$}$\; \label{ln:init_sPhi_reach}
        $G \gets \text{states satisfying $\psi$}$\; \label{ln:init_G}

        
        \Repeat{$G = \overline{G}$}  { \label{ln:start_reach_loop}
            $\overline{G} \gets G$\; \label{ln:copy_G}
            $G \gets G \cup (\pi(G) \cap \stateSatB)$\; \label{ln:update_G}
        } \label{ln:check_fixpoint}

        
        \If{$(\mathcal{I} \cap G) \neq \varnothing$}{ \label{ln:check_if_winning}
            $\texttt{Reach} \gets G$\;
            \Return{\texttt{Reach}}\;
            }
            $\texttt{Reach} \gets \varnothing$\;
        \Return{\texttt{Reach}}\;
    }
\end{algorithm}
\end{minipage}
\hfill
\begin{minipage}{0.48\textwidth}
\begin{algorithm}[H]
    \caption{Safety}
    \label{alg:safety} 
    \KwData{$\tcgshort = (\arena, \winCond := \Box \varphi)$: a \ac{tcg} where $\varphi$ is a pure CLTLoc formula.}
    \KwResult{\texttt{Safe}: controller-winning states in $\tcgshort$.}
    
    \Begin() {
    \nosemic \nonl \ \;
    
        \dosemic $G \gets \text{states satisfying $\varphi$}$\; \label{ln:init}

     
        \Repeat{$G = \overline{G}$} { \label{ln:begin_safety_loop}
        $\overline{G} \gets G$\;   \label{ln:updateG} 
       
        $G \gets (\pi(G) \cap G) $\; \label{ln:newG}
            
        }  \label{ln:until}

        
        \If{$(\mathcal{I} \cap G) \neq \varnothing$}  { \label{ln:hasinitial}
         $\texttt{Safe} \gets G$\;
                \Return{\texttt{Safe}}\;
        }
        $\texttt{Safe} \gets \varnothing$\;
        \Return{\texttt{Safe}}\;
    }
\end{algorithm}
\end{minipage}
\Description{Pseudocode of the algorithms used to solve reachability and safety \acp{tcg}.}
    \label{fig:comparison_of_algorithms}
\end{figure}

\autoref{alg:solve_tcgs} solves a \ac{tcg} having a reachability winning condition, by determining whether an initial state can be reached backwards from the set of goal states.
In particular, Line~\ref{ln:init_sPhi_reach} initializes a set $\stateSatB$ containing the states that the controller must traverse before reaching a goal state. Notice that such a set coincides with $\rtsstate$ if $\varphi \equiv \top$.
Since $\varphi$ and $\psi$ are pure, the states satisfying $\varphi$ and $\psi$ can be computed in time linear in the size of $\rtsstate$.
A set $G$ is initialized in Line~\ref{ln:init_G} with the set of goal states.
The controllable predecessor operator $\pi$ is then iteratively applied to $G$ in the loop starting at Line~\ref{ln:start_reach_loop} until no more states can be added to $G$.
Line~\ref{ln:copy_G} creates a copy $\widebar{G}$ of $G$, which is used in Line~\ref{ln:check_fixpoint} to check whether the current iteration has added any states to $G$.
Line~\ref{ln:update_G} updates $G$ by adding the states returned by $\pi(G)$, restricted to those that also belong to $\stateSatB$.
If no additional states can be added to $G$, then the fixed point is reached, and Line~\ref{ln:check_if_winning} checks whether at least one initial state belongs to $G$.
If this holds, then a winning controller strategy exists.

{We extend the notion of winning controller strategy (see \autoref{def:winning_strategy}) to an arbitrary state $s \in \mathcal S$ in the natural way: a controller strategy $\strategyc$ is winning for a \ac{tcg} $\tcgshort = (\aut, \winCond)$ from a state $s \in \mathcal{S}$ if, for every play $\rho$ from $s$ consistent with $\strategyc$, the play $\rho$ satisfies $\winCond$ (recall that we now omit clock assignments from the notation, see \autoref{rem:region_winning_condition}).}

{\begin{definition}[Rank]
\label{def:rank}
Let $\tcgshort = (\arena, \winCond)$ be a \ac{tcg}, where $\winCond := \varphi \Until \psi$, let $\stateSatB$ be the set of states satisfying $\varphi$, let $G_0$ be the set of states satisfying $\psi$ and, for every $n > 0$, let $G_n = \bigl(G_{n-1} \cup (\pi(G_{n-1}) \cap \stateSatB)\bigr)$; hence, $G^* = \bigcup_{n \in \mathbb{N}} G_n$ is the fixed point computed by the loop of \autoref{alg:solve_tcgs}. 
For every state $s \in \mathcal{S}$, the \emph{rank} of $s$ is
$
  \rank(s) = \min \{ n \in \mathbb{N} \mid s \in G_n \}
$,
with the convention $\min \varnothing = \infty$; hence, $\rank(s) < \infty$ if, and only if, $s \in G^*$.
\end{definition}}

{
\begin{lemma}
\label{lem:reach_properties}
Let $\tcgshort=(\arena,\winCond)$ be a \ac{tcg}, where
$\winCond:=\varphi\Until\psi$, and let $G^*$ be the fixed point
computed by the loop of \autoref{alg:solve_tcgs}.
Then, for every state $s\in\mathcal S$, there exists a winning
controller strategy for $\tcgshort$ from $s$ if, and only if,
$s\in G^*$.
\end{lemma}
\begin{proof}
    Let $G_0$ be the set of states satisfying $\psi$ and, for every
    $n \geq 0$, let
    $
    G_{n+1}
    =
    \big(G_n \cup (\pi(G_n)\cap \stateSatB)\big)
    $,
    where $\stateSatB$ is the set of states satisfying $\varphi$.
    Since $\mathcal S$ is finite, this sequence {$G_0, G_1, \dots$ of sets of states} eventually reaches the
    fixed point $G^*$ computed by the loop of
    \autoref{alg:solve_tcgs}.
    
    Suppose first that $s\in G^*$.
    If $\rank(s) = 0$, then $s$ satisfies $\psi$, and therefore
    $\varphi\Until\psi$ is satisfied from $s$.
    Otherwise, $\rank(s) = n>0$. By construction of $G_n$, it holds that
    $s$ satisfies $\varphi$ and $s\in\pi(G_{n-1}).$
    Hence, by
    \autoref{eq:controllable_predecessor_operator_c} and
    \autoref{eq:controllable_predecessor_operator_e}, the controller
    has a choice such that, regardless of the choices of the
    environment, the next state (whose rank is at most $n-1$) belongs to $G_{n-1}$.
    Applying the same argument recursively, the controller can ensure
    that the rank of the visited states strictly decreases after every
    move, until a state in $G_0$ is reached.
    All states visited before reaching $G_0$ satisfy $\varphi$, while
    the state reached in $G_0$ satisfies $\psi$.
    Therefore, there exists a winning controller strategy from $s$.
    
Conversely, suppose that $s\notin G^*$.
Since every state satisfying $\psi$ belongs to $G_0\subseteq G^*$,
$s$ does not satisfy $\psi$.
If $s$ does not satisfy $\varphi$, then
$\varphi\Until\psi$ is already false at $s$, so there is no
winning controller strategy from $s$.
Assume now that $s$ satisfies $\varphi$.
The fixed point satisfies $
    G^*
    =
    \big(G^* \cup (\pi(G^*)\cap \stateSatB)\big)
    $
and is constructed backwards from the goal states.
Hence, since $s\notin G^*$ and $s\in\stateSatB$, it follows that
$s\notin\pi(G^*)$.
By the definition of the controllable predecessor operator, the environment can force a successor of $s$, denoted by $s'$,
that is still outside $G^*$.
If $s'$ also satisfies $\varphi$, the same argument applies again.
Otherwise, since $s'\notin G^*$ also implies that
$s'$ does not satisfy $\psi$, the formula
$\varphi\Until\psi$ is violated.
Hence, either a state not satisfying $\varphi$ is eventually
reached, or the environment can keep the play forever outside $G^*$, thus not allowing reaching a state satisfying $\psi$ (since they belong to $G_0\subseteq G^*$).
In either case, $\varphi\Until\psi$ is not satisfied.
\end{proof}
}

The following theorem proves the correctness of the algorithm.

\begin{theorem}
    Let $\tcgshort = (\arena, \winCond)$ be a \ac{tcg}, where $\winCond := \varphi \Until \psi$.
    Then, there exists at least one winning controller strategy for $\tcgshort$ {from some initial state}
    if, and only if, the set \texttt{Reach} returned by \autoref{alg:solve_tcgs} with input $\tcgshort$
    is not empty.
\end{theorem}
{\begin{proof}
    By \autoref{lem:reach_properties}, there exists a winning  controller strategy for $\tcgshort$ if, and only if,
    $
        (\mathcal I \cap G^*) \neq \varnothing
    $.
    By \autoref{alg:solve_tcgs}, if this condition holds, then
    $\texttt{Reach}=G^*$ is returned; otherwise,
    $\texttt{Reach}=\varnothing$ is returned.
    Hence, a winning controller strategy for $\tcgshort$ exists if,
    and only if, $\texttt{Reach}\neq\varnothing$.
\end{proof}
}

Let $\tcgshort = (\arena, \winCond)$ be a \ac{tcg}.
Since $G$ grows monotonically in \autoref{alg:solve_tcgs}, the worst-case complexity occurs when the states of $\RTSA$ are added to $G$ individually.
Under the assumption that $\lvert G \rvert = 1$ after initialization in Line~\ref{ln:init_G}, the loop of \autoref{alg:solve_tcgs} executes at most $\lvert \mathcal{S} \rvert - 1$ times.
Since, by \autoref{prop:pi_complexity}, the complexity of $\pi$ is $\mathcal{O}(\lvert \mathcal{S} \rvert^2 \cdot |\act|)$, the overall complexity of \autoref{alg:solve_tcgs} is:
$\sum_{i = 1}^{\lvert \mathcal{S} \rvert - 1} \mathcal{O} (\lvert \mathcal{S} \rvert^2 \cdot |\act|)
=
\mathcal{O} ( \lvert \mathcal{S} \rvert ^3 \cdot |\act|)
$, \emph{i.e.}, polynomial in the size of $\RTSA$ and exponential in the size of the \ac{tarena} $\arena$~\cite[Chapter 9]{Baier}.

\textbf{Safety.} A \emph{safety} winning condition
has the form $\winCond := \Box \varphi$,
where $\varphi$ is a pure CLTLoc formula.
In this case, the controller is required to always stay within the set of states identified by $\varphi$, regardless of the choices of the environment.

\autoref{alg:safety}, whose complexity is $\mathcal{O}(\lvert \mathcal{S} \rvert ^3 \cdot |\act|)$, solves a \ac{tcg} having a safety winning condition, by determining whether the controller can ensure that $\varphi$ remains true in a play consistent with a winning controller strategy.
In particular, Line~\ref{ln:init} initializes a set $G$ containing all states satisfying $\varphi$.
The controllable predecessor operator $\pi$ is then iteratively applied to $G$ in the loop starting
at Line~\ref{ln:begin_safety_loop} until no more states can be removed from $G$.
Line~\ref{ln:updateG} creates a copy $\widebar{G}$ of $G$, which is used in Line~\ref{ln:until} to check whether the current iteration has removed any states from $G$.
Line~\ref{ln:newG} updates $G$ by retaining only its states returned by $\pi(G)$.
If no states can be removed from $G$, then the fixed point is reached, and Line~\ref{ln:hasinitial} checks whether at least one initial state belongs to $G$. 
If this holds, then a winning controller strategy exists.

{
\begin{lemma}
\label{lem:safe_properties}
Let $\tcgshort=(\arena,\winCond)$ be a \ac{tcg}, 
where $\winCond := \Box \varphi$, and let $G^*$ be the fixed point
computed by the loop of \autoref{alg:safety}.
Then, for every state $s\in\mathcal S$, there exists a winning
controller strategy for $\tcgshort$ from $s$ if, and only if,
$s\in G^*$.
\end{lemma}
\begin{proof}
    Let $G_0$ be the set of states satisfying $\varphi$ and, for every
    $n\geq 0$, let $G_{n+1}=(G_n\cap\pi(G_n))$.
        Since $\mathcal S$ is finite, this decreasing sequence eventually
    reaches the fixed point $G^*$ computed by
    \autoref{alg:safety}.
    
    Suppose first that $s\in G^*$.
    At the fixed point, $G^*=(G^*\cap\pi(G^*))$,
    therefore $G^*\subseteq\pi(G^*)$.
    Hence, from every state in $G^*$, the controller can choose a move
    such that, regardless of the choices of the environment, the next
    state also belongs to $G^*$.
    Choosing such a move at every state in $G^*$ defines a controller
    strategy under which every consistent play remains forever in $G^*$.
    Since $G^*\subseteq G_0$, every state visited by such a play satisfies
    $\varphi$.
    Therefore, $\Box\varphi$ is satisfied, and the controller strategy is
    winning from $s$.
    
    Conversely, suppose that $s\notin G^*$.
    If $s\notin G_0$, then $s$ does not satisfy $\varphi$, and therefore
    $\Box\varphi$ is already false at $s$.
    Assume now that $s\in G_0$.
    Since $s\notin G^*$, there exists an index $n\geq 0$ such that
    $
    s\in G_n
    $ and $
    s\notin G_{n+1}
    $.
    Since $G_{n+1}=(G_n\cap\pi(G_n))$, it follows that
    $s\notin\pi(G_n)$.
    Thus, whatever choice the controller makes, the environment can force
    a successor $s'$ of $s$
    outside $G_n$.
    If $s'\notin G_0$, then $s'$ does not satisfy $\varphi$, and $\Box \varphi$ is violated in $s$.
    Otherwise, since the sequence
    $G_0\supseteq G_1\supseteq\cdots$ is decreasing and
    $s'\notin G_n$, the state $s'$ must have been removed at some earlier
    iteration.
    The same argument can therefore be applied again with a strictly
    smaller removal index.
    Consequently, after finitely many such steps, the environment can
    force the play to reach a state outside $G_0$, \emph{i.e.}, a state not
    satisfying $\varphi$.
    Hence, no controller strategy from $s$ can guarantee
    $\Box\varphi$.
\end{proof}
}

The following theorem proves the correctness of the algorithm:

\begin{theorem}
\label{thm:safety_correctness}
    Let $\tcgshort = (\arena, \winCond)$ be a \ac{tcg}, where $\winCond := \Box \varphi$.
    Then, there exists at least one winning controller strategy for $\tcgshort$ {from some initial state}
    if, and only if, the set \texttt{Safe} returned by \autoref{alg:safety} with input $\tcgshort$
    is not empty.
\end{theorem}

{\begin{proof}
    By \autoref{lem:safe_properties}, there exists a winning
    controller strategy for $\tcgshort$ if, and only if, $(\mathcal I \cap G^*) \neq \varnothing$.
    By \autoref{alg:safety}, if this condition holds, then
    $\texttt{Safe}=G^*$ is returned; otherwise,
    $\texttt{Safe}=\varnothing$ is returned.
    Hence, a winning controller strategy for $\tcgshort$ exists if,
    and only if, $\texttt{Safe}\neq\varnothing$.
\end{proof}
}

\textbf{$\Next$-conjunction.}~In the following, we use $\Next^n$ to denote the repeated application of the $\Next$ operator $n \ge 0$ times;
if $n = 0$, then $\Next^0(\varphi) \equiv \varphi$.
Since the semantics of the $\Next$ operator (see \autoref{tab:semantics} for reference)
captures the next state reached within a play after a move is fully applied, the controller's behavior can be formally specified using a \emph{$\Next$-conjunction} winning condition of the form
$\winCond := \bigwedge_{i = 0}^k \Next^{d_i} \varphi_i$, where $k \in \mathbb{N}$,
each $\varphi_i$ is a pure \ac{cltloc} formula characterizing a set of states, and $d_i < d_{i + 1}$ holds for every $0 \le i < k$.
The controller’s objective is to reach a state in the set of goal states characterized by $\varphi_k$. Notice that $\varphi_k$ must be reached after exactly $d_k$ moves.

\begin{example}
    Consider the \ac{tarena} $\arena$ of \autoref{fig:move_example}.
    Suppose that the controller is required to visit $q_1$ after seven time units from the initial state, then cycle two times in $q_1$, with the elapsed time between consecutive visits being one and then two time units, respectively.
    This requirement can be effectively captured by the following $\Next$-conjunction winning condition, where we introduce one external clock $y$ such that $\ackc = \{ x \}$ and $\cckc = \{ x, y \}$:
    \[
    \winCond := q_0 \land 
               \Next (q_1 \land x = 7 \land y = 0) \land 
               \Next^2(q_1 \land y = 1) \land
               \Next^3(q_1 \land y = 3).
    \]
    In this case, a winning controller strategy {does not exist since $x = 7$ the first time $q_1$ is reached; indeed, the environment can force the transition from $q_1$ to $q_3$ in the next move}.
    The use of $y$ here is necessary to handle the timing between consecutive moves in $q_1$, since $x$ is {always} reset on the self loop at $q_1$.
    A different formulation of the same requirement would be
    $\winCond := \Next (q_1 \land y = 7 ) \land 
               \Next^2(q_1 \land y = 8) \land
               \Next^3(q_1 \land y = 10)$, in which case $(\cckc \cap \ackc) = \varnothing$.
    \end{example}

\autoref{alg:next_conjunction} solves a \ac{tcg} having a $\Next$-conjunction winning condition, by determining whether an initial state can be reached backwards from the set of goal states.
Differently from reachability and safety \acp{tcg}, solving a $\Next$-conjunction \ac{tcg} does not require a fixed-point algorithm.
In particular, from Line~\ref{ln:init_array_next} to Line~\ref{ln:init_next}, a vector $\stateSatB$ is initialized to contain, at each position $i$, the states satisfying $\varphi_i$.
A set $G$ is subsequently initialized to the set of goal states in Line~\ref{ln:init_G_next}; since the controller must reach one of these states after exactly $d_k$ moves, the backward analysis starts from this set.
The algorithm then iterates over each conjunct in $\winCond$ from $k$ down to $0$ (Line~\ref{ln:main_for_loop_next}).
During the $i$-th iteration, if $i > 0$ (Line~\ref{ln:number_of_pi_app_if_next}), then the controllable predecessor  operator $\pi$ is applied $\Delta = d_i - d_{i-1}$ (Line~\ref{ln:get_number_of_pi_applications}) times to $G$, each time setting the new value of $G$ to $\pi(G)$ (lines \ref{ln:inner_for_loop_next} and \ref{ln:update_G_next}); if $i = 0$, then $\pi$ is applied $\Delta = d_0$ (Line~\ref{ln:get_number_of_pi_applications_zero}) times to $G$. 
After all $\Delta$ applications
of $\pi$ and before starting a new iteration, if $i > 0$, then $G$ is intersected with $\stateSatB[i-1]$, enforcing that the states reachable at that point must also satisfy $\varphi_{i-1}$ (lines \ref{ln:intersect_G_next_cond} and \ref{ln:intersect_G_next}).
If a winning controller strategy exists (Line~\ref{ln:hasinitial_next}), then $\top$ is returned in Line~\ref{ln:return_top_next}; $\bot$ otherwise.
It follows that the complexity of \autoref{alg:next_conjunction} is 
$\mathcal{O} (d_k \cdot | \mathcal{S} |^2 \cdot | \act |)$.

\begin{figure}[tb]
\centering
\noindent
\begin{minipage}{0.68\textwidth}
\begin{algorithm}[H]
    \caption{$\Next$-conjunction}
    \label{alg:next_conjunction} 
    \KwData{$\tcgshort = (\arena, \winCond := \bigwedge_{i = 0}^k \Next^{d_i} \varphi_i)$: a \ac{tcg} where each $\varphi_i$ is a pure CLTLoc formula, for $0 \le i \le k$, and $d_i < d_{i + 1}$, for $0 \le i < k$.}
    \KwResult{$\top$: if at least one winning controller strategy exists; $\bot$ otherwise.}
    
    \Begin() {
        $\stateSatB[0, \dots, k] \gets (\varnothing, \dots, \varnothing)$\; \label{ln:init_array_next}

        \For{$i \gets 0$ \KwTo $k$}{ \label{ln:init_G_for_loop_next}
            $\stateSatB[i] \gets \text{states satisfying $\varphi_i$}$\; \label{ln:init_next}
        }

        $G \gets \stateSatB[k]$\; \label{ln:init_G_next}

        \SetKw{KwDownTo}{downto}
        \For{$i \gets k$ \KwDownTo $0$}{ \label{ln:main_for_loop_next}
            \uIf{$i > 0$}{ \label{ln:number_of_pi_app_if_next}
                $\Delta \gets d_i - d_{i - 1}$\; \label{ln:get_number_of_pi_applications}
            }
            \Else{
                $\Delta \gets d_0$\; \label{ln:get_number_of_pi_applications_zero}
            }

            \For{$j \gets 1$ \KwTo $\Delta$}{ \label{ln:inner_for_loop_next}
                $G \gets \pi(G)$\; \label{ln:update_G_next}
            }

            \If{$i > 0$}{ \label{ln:intersect_G_next_cond}
                $G \gets G \cap \stateSatB[i - 1]$\; \label{ln:intersect_G_next}
            }
        }

        \If{$(\mathcal{I} \cap G) \neq \varnothing$}  { \label{ln:hasinitial_next}
            \Return{$\top$}\; \label{ln:return_top_next}
        }
        \Return{$\bot$}\;
    }
\end{algorithm}
\end{minipage}
\Description{Pseudocode of the algorithms used to solve $\Next$-conjunction \acp{tcg}.}
    \label{fig:next_conjunction_fig}
\end{figure}

The following theorem proves the correctness of the algorithm:

\begin{theorem}
\label{thm:next_conjunction_solving}
    Let $\tcgshort = (\arena, \winCond)$ be a \ac{tcg}, where $\winCond := \bigwedge_{i = 0}^k \Next^{d_i} \varphi_i$ is such that each $\varphi_i$ is a pure \ac{cltloc} formula and $d_i < d_{i + 1}$ holds for $0 \le i < k$.
    Then, there exists an initial state $s_0 \in \mathcal{I}$ 
    and a winning controller strategy for
    $\tcgshort$ from $s_0$ if, and only if, \autoref{alg:next_conjunction} with input $\tcgshort$ returns $\top$.
\end{theorem}
{\begin{proof}
        For every $0 \le m \le d_k$, let $G_m$ be the set defined backwards as follows: 
        \[ G_{d_k} = \stateSatB[k]  \qquad \text{and, for $0 \le m < d_k$,} \qquad  G_m = \begin{cases} \pi(G_{m+1}) \cap \stateSatB[i] & \text{if } m = d_i \text{ for some } 0\le i < k\\ \pi(G_{m+1}) & \text{otherwise} \end{cases}. \] 
        These are exactly the sets computed by \autoref{alg:next_conjunction}, which returns $\top$ if, and only if, $(G_0 \cap \mathcal{I}) \neq \varnothing$.
        
        We prove, by backward induction on $m\le d_k$, that $s \in G_m$ if, and only if, considering $s$ as the state at position $m$ of a play consistent with a controller strategy,
        the controller can guarantee that $\varphi_i$ holds at position $d_i$ of the play,
        for every $i$ such that $d_i \ge m$.
        For $m = d_k$, the claim follows immediately from $G_{d_k} = \stateSatB[k]$.
        Now, let $m < d_k$ and suppose that the claim holds for $m+1$. 
        By the definition of $\pi$ (\autoref{eq:controllable_predecessor_operator_c} and \autoref{eq:controllable_predecessor_operator_e}), $s \in \pi(G_{m+1})$ if, and only if, the controller has a choice such that every possible successor consistent with the controller's choice
        belongs to $G_{m+1}$. 
        By the induction hypothesis, from each such successor the controller can guarantee all the conditions required at positions $d_i \ge m+1$. Thus, $s \in \pi(G_{m+1})$ if, and only if, the controller can guarantee all these conditions starting from $s$ at position $m$.
        If $m = d_i$ for some $i < k$, then the additional intersection with $\stateSatB[i]$ ensures that $\varphi_i$ also holds at position $m$. Hence, the claim holds for $m$.
        In particular, $G_0$ contains exactly the states from which the controller can guarantee $\winCond$. Therefore, there exists an initial state $s_0 \in \mathcal{I}$ and a winning controller strategy from $s_0$ if, and only if, $(G_0 \cap \mathcal{I}) \neq \varnothing$.
\end{proof}
}

Remarkably, the $\Next$ operator enables a declarative, move-by-move specification of the controller's behavior directly within a winning condition, without introducing structural modifications into a given \ac{tarena}.
For instance, it can exclude initial states from evaluation, preventing trivial falsification due to initial clock values.
Conversely, \acp{tg} would require augmenting the arena with specific guards and resets, or formulating complex winning conditions, to capture the same behavior. 
This decoupling between specification and arena significantly reduces modeling effort and  design errors.

\subsection{Additional Winning Conditions}
\label{subsec:additional_winning_conditions}
The algorithms presented in this section can be adapted, or combined, to solve \acp{tcg} with additional winning conditions and to synthesize corresponding winning controller strategies.
For instance, the ability to solve a safety \ac{tcg}
implies the ability to solve a \ac{tcg} in which the winning
condition is $\neg(\varphi \Until \psi)$, since
$
\neg (\varphi \Until \psi) \equiv ( (\varphi \land \neg \psi) \Until (\neg \varphi \land \neg \psi) ) \lor \Box (\varphi \land \neg \psi)
$.
If the controller wins for either of the two disjuncts, then it also wins for $\neg(\varphi \Until \psi)$.

\textbf{$\Next$-based winning conditions.}~\autoref{alg:solve_tcgs} and \autoref{alg:safety} can be adapted to solve a \ac{tcg} 
$\tcgshort = (\arena, \winCond)$ where $\winCond := \Next^n (\varphi \Until \psi)$ or $\winCond := \Next^n (\Box \varphi)$, respectively, \emph{i.e.}, a reachability or a safety formula holds after $n$ moves are applied from an initial state.
The adaptation is identical for both algorithms. Once the fixed point over set $G$ has been reached, a vector of size $n+1$ is initialized, with each element at indices $1, \dots, n$ being an empty set of states and the element at index $0$ set to $G$. 
The operator $\pi$ is then applied iteratively: for each $i = 0,\ldots,n-1$, $\pi$ is applied to the set stored at index $i$, and the resulting set is stored at index $i+1$. 
This vector is needed to keep track of the number of moves associated with the states collected during the backward traversal. 
In particular, index $i$ contains the states from which the reachability or safety property can be guaranteed after exactly $i$ moves. 
Consequently, the check in the if-statement in both algorithms is performed on the set of states at index $n$ of the vector.

Similarly, winning conditions can be defined recursively by nesting the $\Next$ and $\Until$ temporal operators.
For instance,
$\winCond_0 := \Next^{n_0} ( \varphi_0 \Until \psi_0 )$;
$\winCond_k := \Next^{n_k} ( \varphi_k \Until \winCond_{k - 1} )$, for $k \in \mathbb{N}_{> 0}$.
The previous adaptation of \autoref{alg:solve_tcgs} can be extended to handle nesting.
In particular, the innermost formula $\winCond_0$ is handled as detailed above, resulting in a set $G_0$ of states, which defines the goal states for the formula $\winCond_1$ at the upper nesting level, \emph{i.e.}, $\winCond_1 = \Next^{n_1} ( \varphi_1 \Until \winCond_0 )$.
Although $\winCond_0$ is not a pure \ac{cltloc} formula, the set of states it specifies is already computed.
Hence, $\winCond_1$ is handled as detailed above, resulting in a set $G_1$ of states.
This bottom-up evaluation is iterated up to the outermost formula $\winCond_k$.

\subsection{Computational Complexity of Solving Reachability \aclp{tcg}}
\label{subsec:exptime_completeness}
We now prove that the problem of solving reachability \acp{tcg} is EXPTIME-complete by reducing the EXPTIME-complete problem of solving \emph{countdown games}~\cite{countdownGames} to it.
A countdown game $\mathcal{C}$ between Player 1 and Player 2 is a pair $\mathcal{C} = (\Xi, \Lambda)$, where $\Xi$ is a set of states and $\Lambda \subseteq \Xi \times \mathbb{N}_{> 0} \times \Xi$ is a transition relation.
A \emph{configuration} of a countdown game is a pair $(\xi, B) \in (\Xi \times \mathbb{N})$.
The way in which a countdown game evolves from $(\xi, B)$ is defined according to the following notion of \emph{move}:
first, Player 1 chooses an integer $d$ (also known in the literature as \emph{duration}) such that $0 < d \le B$ and there exists a transition $(\xi, d, \xi') \in \Lambda$; then, Player~2 chooses a transition $(\xi, d, \xi'') \in \Lambda$.
The resulting configuration is $(\xi'', B - d)$.
A configuration $(\xi, 0)$ is \emph{winning} for Player 1, while a configuration $(\xi, \bar{B})$ such that $\forall (\xi, d, \xi') \in \Lambda : d > \bar{B}$ is \emph{winning} for Player 2.
Notice that a winning configuration (either for Player 1 or for Player 2) is \emph{terminal}, \emph{i.e.}, the game stops since no moves are available.
A sequence of moves leading to a terminal configuration is called a \emph{$\mathcal{C}$-play}.
The problem of solving a countdown game consists in determining whether Player 1 can reach a winning configuration from a given initial configuration $(\xi_0, B_0)$.

\begin{theorem}
\label{thm:complexity}
    The problem of solving reachability \acp{tcg} is EXPTIME-complete on Timed Arenas with at least two clocks.
\end{theorem}

\begin{proof}
\Blue{We first prove membership in EXPTIME.
Let $\aut$ be a \ac{tarena}.
The number of states of $\RTSA$ is at most exponential in the size of the input, by the standard bound on the number of clock regions.
Since \autoref{alg:solve_tcgs} runs in time polynomial in the size of $\RTSA$, which can be constructed in exponential time, solving reachability \acp{tcg} belongs to EXPTIME.
}

    To prove EXPTIME-hardness of solving reachability \acp{tcg}, we reduce the EXPTIME-complete problem of solving countdown games to it.
    Our reduction is similar to that of~\cite{exptimereachabilitygames}, but the \acp{tcg} semantics allows for a simpler construction.
    
    Let $\mathcal{C} = (\Xi, \Lambda)$ be a countdown game.
    We define a reachability \ac{tcg} $\tcgshort = (\aut, \winCond)$ in such a way that Player 1 wins in $\mathcal{C}$, starting from an initial configuration $(\xi_0, B_0)$, if, and only if, the controller wins in $\tcgshort$.
    Let $\aut = (\act, Q, Q_{0}, \ackc, T, \invariant)$ be a \ac{tarena} (recall that $T \subseteq Q \times \act \times \Gamma(\ackc) \times 2^{\ackc} \times Q$) defined as follows (we take all location invariants to be $\top$):

    \begin{itemize}[nosep]
        \item $\act = \big( \{ a \} \cup \{ a_d \; | \; (\xi, d, \xi') \in \Lambda \} \big)$;

        \item $Q = (Q_{\cont} \cup Q_{\env})$, where $Q_{\cont} = ( \Xi \cup \{ q_{\mathit{goal}} \} )$, $Q_{\env} = \varnothing$, and $Q_0 = \{ \xi_0 \}$;

        \item $\ackc = \{x, y \}$, where clock $x$ is reset by every transition, while clock $y$ is never reset by any transition;

        \item $T = \big(\{ (\xi, a_d, x = d, \{x\}, \xi') \; | \; (\xi, d, \xi') \in \Lambda \} \cup 
                   \{ (\xi, a, x = 0 \land y = B_0, \{x\}, q_{\mathit{goal}}) \; | \; \xi \in \Xi \} \cup
                   \{ (q_{\mathit{goal}}, a, x \ge 1,$ $ \{x\}, q_{\mathit{goal}}) \} \big) \cup
                   \{ (\xi, a, x \ge 1 \land y > B_0, \{x\}, \xi) \mid \xi\in\Xi\}.
                   $
    \end{itemize}
    This definition of $\aut$ ensures that the reduction can be performed in polynomial time in the size of $\mathcal{C}$.
    Indeed, $\aut$ has $|\Xi| + 1$ locations, at most $|\Lambda| + 1$ actions, $|\Lambda| + 2|\Xi| + 1$ transitions, and two clocks.
    \Blue{
    Each guard is a conjunction of at most two  clock constraints, using either the fixed constants $0$ and $1$ or constants $d$ and $B_0$ from $\mathcal{C}$. Since constants are encoded in binary in both $\mathcal{C}$ and $\aut$, their representation does not increase the complexity of the reduction beyond polynomial time.
    
   
   Since every countdown transition has a strictly positive duration
   $d$ and resets $x$, and all self-loops are guarded by $x\ge 1$ and also reset $x$, every cycle requires positive elapsed time. Hence, the constructed \ac{tarena} satisfies the strongly non-Zeno assumption introduced in \autoref{sec:timed_CLTLoc_games}.
    The additional self-loops on locations $\xi\in\Xi$ ensure that an admissible move is always available, even when the corresponding countdown configuration has no legal moves. However, taking such a self-loop makes $q_{\mathit{goal}}$ permanently unreachable, since $y>B_0$ and $y$ is never reset.
    We recall that the countdown in $\mathcal{C}$ is initialized to $B_0$ and decreases after each move. 
    Each countdown move of duration $d$ corresponds to a delay of $d$ time units in $\aut$, followed by a transition resetting $x$ but not $y$.}
    
    Next, we define the following reachability winning condition: $\winCond := \Diamond (q_{\mathit{goal}})$.
    
    We now prove that Player 1 wins in $\mathcal{C}$ if, and only if, the controller wins in $\tcgshort$.

    \Blue{A countdown configuration $(\xi,B)$, with $0 \le B \le B_0$, corresponds to a configuration of $\tcgshort$ at location $\xi$, with $v(x)=0$ and $v(y)=B_0-B$. 
    This correspondence holds initially, since both clocks are initialized to zero.

    Suppose $B>0$. For every legal countdown move of duration $d \le B$, the controller can choose an action $a_d$ and 
    a delay of $d$ time units. 
    The environment then chooses among the enabled transitions $(\xi,a_d,x=d,\{x\},\xi')$, exactly as Player~2 chooses the successor $\xi'$ in $\mathcal{C}$.
    Since $x$ is reset and $y$ is not, the resulting configuration satisfies 
    \[
    v'(x)=0,\qquad
    v'(y)=B_0-B+d=B_0-(B-d),
    \]
    thus preserving the correspondence.
    
    Conversely, consider a configuration $(\xi,v)$ of $\tcgshort$ corresponding to $(\xi,B)$ with $B>0$. 
    The controller cannot reach $q_{\mathit{goal}}$ by choosing action $a$, since the guard $x=0\land y=B_0$ cannot be satisfied after any delay.
    The same action $a$ can enable the additional self-loop, but only when $y>B_0$, making $q_{\mathit{goal}}$ permanently unreachable.
    Any other admissible choice must use an action $a_d$ after a delay equal to $d$.
    If $d>B$, then the resulting value of $y$ exceeds $B_0$ and, since $y$ is never reset, $q_{\mathit{goal}}$ becomes unreachable.
    Hence, every choice compatible with a winning controller strategy corresponds to a legal countdown move.
    In particular, suppose that $(\xi,B)$ is a terminal configuration of $\mathcal{C}$ with $B>0$. If $\xi$ has outgoing transitions in $\mathcal{C}$, all their durations satisfy $d>B$. 
    The corresponding actions $a_d$ are still admissible in $\tcgshort$, but executing any of them makes $q_{\mathit{goal}}$ permanently unreachable.
    If, instead, $\xi$ has no outgoing transitions, no action $a_d$ is admissible.
    In both cases, the controller can choose action $a$, which can only enable the additional self-loop, again making $q_{\mathit{goal}}$ unreachable. 
    Thus, no winning controller choice is possible. 
    
    On the other hand, when $B=0$, the controller can reach $q_{\mathit{goal}}$ by choosing action $a$ with zero delay.
    The additional self-loop on $\xi$ is not enabled by this choice, since its guard requires $x\ge 1$.
    Thus, the environment cannot prevent the transition to $q_{\mathit{goal}}$.
    The self-loop on $q_{\mathit{goal}}$ allows the play to continue indefinitely.
    
    Therefore, a winning strategy for Player~1 in $\mathcal{C}$ induces a winning controller strategy in $\tcgshort$, by simulating each countdown move and reaching $q_{\mathit{goal}}$ when the counter becomes zero.
    Conversely, every winning controller strategy must select legal countdown moves until $q_{\mathit{goal}}$ is reached.
    Since every such move strictly decreases the non-negative counter, the corresponding countdown play must terminate, and it cannot terminate with a positive counter.
    Thus, it reaches counter value zero, yielding a winning strategy for Player~1 in $\mathcal{C}$.}
\end{proof}

\section{Synthesizing Winning Controller Strategies}
\label{sec:synthesizing_winning_controller_strategies}
The algorithms presented in \autoref{subsec:solvingalg} solve \acp{tcg}.
We now extend such algorithms to synthesize winning controller strategies.
To do this, we introduce a graph-based data structure that captures all the necessary information for synthesizing winning controller strategies.

\begin{definition}
\label{def:strategy_graph}
    Let $\tcgshort = (\arena, \winCond)$ be a \ac{tcg} and let $\RTSA = (\rtsstate, \mathit{Act} \cup \{ \tau \}, \mathcal{I}, \rtsarrow)$.
    A \emph{\ac{sg}} $\strategygraph$ of $\tcgshort$ is a triple 
    $\strategygraph = (\mathcal{V}, \mathcal{E}, \lambda)$, where 
    $\mathcal{V} \subseteq \mathcal{S}$ is a subset of states, 
    $\mathcal{E} \subseteq \mathcal{V} \times \mathcal{V}$ is a set of edges,
    and 
    $\lambda : \mathcal{E} \rightarrow (\wp(\mathcal{S} \times \act) \setminus \{ \varnothing \})$ is a labeling function.
    With an abuse of notation, we denote by $(s_1, \langle s_2, a \rangle, s_3)$ an edge $(s_1,s_3)\in\mathcal{E}$ together with one of its associated labels, where $\langle s_2,a\rangle \in \lambda((s_1, s_3))$ is a label of $(s_1,s_3)$, and write $(s_1, \langle s_2, a \rangle, s_3) \in \mathcal{E}$. 
    Thus, the same edge may be denoted by multiple such triples, one for each of its associated labels.
\end{definition}

Let $\tcgshort = (\arena, \winCond)$ be a \ac{tcg}, let $\strategygraph = (\mathcal{V}, \mathcal{E}, \lambda)$ be a corresponding \ac{sg},
and let $(s_1, \langle s_2, a \rangle, s_3) \in \mathcal{E}$.
The set $\mathcal{E}$ captures the move semantics of \autoref{subsec:using_regions_for_tcgs}.
In particular, $s_1 = (q, [v_1])$ is the \emph{source} state of a move $(a, [v_2])$, where the delay transition from $s_1$ leads to $s_2  = (q, [v_2])$, and the subsequent discrete transition labeled by $a$ leads from $s_2$ to the \emph{target} state $s_3$.
Notably, the delay transition from $s_1$ to $s_2$ captures a set of admissible {concrete} delays.
%

\textbf{Controller strategy synthesis.}~A controller strategy $\strategyc$ from an initial state $s_0 = (q_0, [v_0])$ can be \emph{synthesized} from $\strategygraph$ as follows (we denote by $\rho$ a play consistent with $\strategyc$). 
{At the initial position $i = 0$, where $\rho_{\le 0} = s_0$, the choices of both players are defined according to $s_0$: if $s_0 \in \mathcal{S}_{\cont}$, then $\strategyc(\rho_{\le 0}) = (a, [v])$; otherwise, if $s_0 \in \mathcal{S}_{\env}$, then $\strategyc(\rho_{\le 0}) = a$, leaving the choice of the clock region $[v]$ to the environment.
In particular, the move $(a, [v])$ must be such that at least one edge with source $s_0$ and labeled $\langle (q_0, [v]), a \rangle$ exists.
Then, among all edges $(s_0, \langle (q_0, [v]), a \rangle, s_1) \in \mathcal{E}$, the environment chooses one such edge.
This process is iterated for every position $i \in \mathbb{N}_{> 0}$, ensuring that the source state $s_i$ coincides with the target state of the edge selected at position $i-1$ and that the move at position $i$ is chosen according to the labels of the edges with source $s_i$.
Recall that, since the \ac{tarena}-extension is used, actions are of the form $a_{\mathcal{X}}$, where the subscript $\mathcal{X}$ denotes the clocks that must be reset when the corresponding discrete transition is taken. 
To correctly synthesize a winning controller strategy for the original \ac{tarena}, these subscripts must be removed from the actions, while the controller reset policy must be defined according to the subscript of the action selected at each position.}

{A controller strategy $\strategyc$ is obtained only on the states visited along its induced plays: at every position $i$, the label selected at $s_i$ prescribes the choice of $\strategyc$ on the prefix leading to it, while on the prefixes that are not induced by $\strategygraph$ from the chosen initial state, $\strategyc$ is defined arbitrarily among the admissible moves.
This is irrelevant for winning, since whether $\strategyc$ is winning from an initial state $s_0$ depends only on the plays starting from $s_0$ and consistent with $\strategyc$.
{Recall that, in general, a controller strategy is a function of finite play prefixes.}
Hence, a controller strategy can be synthesized from $\strategygraph$ by selecting, after each relevant prefix of a play consistent with $\strategyc$, a controller choice represented by one of the labels on the outgoing edges of its last state.

A particularly simple technique, used in the remainder of this work, consists in selecting one controller choice for each state of $\strategygraph$ and using it whenever that state is visited.
This yields \emph{positional}~\cite{introductionchapter} strategies for reachability and safety \acp{tcg}, \emph{i.e.}, strategies whose choices depend only on the current state and not on the preceding play prefix.}
This restriction is particularly convenient for implementing strategy synthesis algorithms,
{since a strategy can then be obtained by \ac{sg} traversal.}


However, since plays retain complete information about the evolution of a given \ac{tcg}, 
{the controller may also select different choices after different prefixes ending in the same state.}
Such {non-positional} strategies can still be synthesized from the corresponding \ac{sg}, but {their implementation may require} additional memory structures.
{In particular, for reachability \acp{tcg}, we construct an acyclic \ac{sg} from which cycle-free winning controller strategies can be synthesized, ensuring that no state is visited twice before the first goal state is reached.}

\textbf{Extending the controllable predecessor operator.}~Let $\arena$ be a \ac{tarena} and let $\RTSA = (\rtsstate, \mathit{Act} \cup \{ \tau \}, \mathcal{I}, \rtsarrow)$ be the \ac{rts} of $\arena$.
In the following, $G \subseteq \mathcal{S}$ denotes a subset of states of $\RTSA$.
Recall that, by \autoref{def:timed_arena}, $\mathcal{S} = (\mathcal{S}_{\cont} \cup \mathcal{S}_{\env})$ and $(\mathcal{S}_{\cont} \cap \mathcal{S}_{\env}) = \varnothing$.
To facilitate the construction of \ac{sg}, we introduce the operator
$\mu : \wp(\mathcal{S}) \rightarrow \wp(\mathcal{S} \times \mathcal{S} \times \act \times \mathcal{S})$, \emph{i.e.}, an extension of the operator $\pi$.
We split $\mu$ into the operators 
$\mu_{\cont} : \wp(\mathcal{S}) \rightarrow \wp(\mathcal{S}_{\cont} \times \mathcal{S} \times \act \times \mathcal{S})$ and 
$\mu_{\env} : \wp(\mathcal{S}) \rightarrow \wp(\mathcal{S}_{\env} \times \mathcal{S} \times \act \times \mathcal{S})$, by considering controller and environment states, respectively. 
Then, $\mu = ( \mu_{\cont} \cup \mu_{\env} )$, where:
\begin{equation}
\label{eq:mu_c}
\mu_{\cont}(G) = 
\left\{
(s_1, s_2, a, s_3) \in (\mathcal{S}_{\cont} \times \mathcal{S} \times \act \times \mathcal{S})
\; \middle| \;
\begin{array}{@{}l@{}}
s_3 \in G \land (s_1, \tau, s_2) \in \; \rtsarrow^*_{\tau} \land \;  (s_2, a, s_3) \in \; \rtsarrow \\
\land \; \neg \exists s_4 \in (\mathcal{S} \setminus G) : (s_2, a, s_4) \in \; \rtsarrow
\end{array}
\right\},
\end{equation}
\begin{equation}
\label{eq:mu_e}
\mu_{\env}(G) = 
\left\{
(s_1, s_2, a, s_3) \in (\mathcal{S}_{\env} \times \mathcal{S} \times \act \times \mathcal{S})
\; \middle| \;
\begin{array}{@{}l@{}}
s_3 \in G \land (s_1, \tau, s_2)  \in \; \rtsarrow^*_{\tau} \land \;  (s_2, a, s_3) \in \; \rtsarrow \\
\land \; \neg ( \exists s_4 \in \mathcal{S}, \exists s_5 \in (\mathcal{S} \setminus G) : (s_1, \tau, s_4)  \in \; \rtsarrow^*_{\tau} \land \;  (s_4, a, s_5) \in \; \rtsarrow )
\end{array}
\right\}.
\end{equation}

Unlike $\pi(G)$, which only returns a set of states, $\mu(G)$ explicitly enumerates tuples $(s_1,s_2,a,s_3)$. Since each of the three states can take at most $n=|\mathcal{S}|$ values, $\mu(G)$ can contain $\mathcal{O}(n^3)$ tuples, assuming a fixed action alphabet. Using the same precomputed successor information as for $\pi$ (see \autoref{prop:pi_complexity}), these tuples can be generated in $\mathcal{O}(n^3)$ time.

In the remainder of this section, we extend the algorithms solving \acp{tcg} with operator $\mu$ to enable strategy synthesis.

\begin{remark}
\label{rek:mu_remark}
\autoref{eq:mu_c} and \autoref{eq:mu_e} extend \autoref{eq:controllable_predecessor_operator_c} and \autoref{eq:controllable_predecessor_operator_e}, respectively.
Intuitively, operator $\mu$ returns all tuples $(s_1, s_2, a, s_3)$ representing a delay transition from $s_1$ to $s_2$, followed by a discrete transition with action $a$ from $s_2$ to $s_3$, applying the same conditions as the controllable predecessor operator $\pi$, but explicitly accounting for the delays and actions involved in the moves.
In other words, the operator $\mu$ captures \autoref{eq:step_over_states} (which defines the mapping $\step$ over states) by retaining only the states and actions that guarantee that the controller reaches $G$.
{In particular, 
$\pi_{\cont}(G)=\{s_1\mid\exists s_2,a,s_3:(s_1,s_2,a,s_3)\in\mu_{\cont}(G)\}$ and
$\pi_{\env}(G)=\{s_1\mid\exists s_2,a,s_3:(s_1,s_2,a,s_3)\in\mu_{\env}(G)\}$.}
\end{remark}

\begin{figure}[tb]
\centering
\noindent
\begin{minipage}{0.68\textwidth}
\begin{algorithm}[H]
    \caption{Reachability for strategy synthesis}
    \label{alg:reachability_strategy_strat}
    
    \KwData{$\tcgshort = (\arena, \winCond := \varphi \Until \psi)$: a \ac{tcg} where $\varphi, \psi$ are pure CLTLoc formulae.}
    \KwResult{$\strategygraph$: an acyclic \ac{sg} for $\mathcal{T}$.}
    
    \Begin() {
        $\stateSatB \gets \text{states satisfying $\varphi$}$\;
        $G \gets \text{states satisfying $\psi$}$\; \label{ln:init_G_strat}
        $\mathcal{E} \gets \varnothing$; $\mathbf{M} \gets \varnothing$\; \label{ln:init_V_E_strat}

        
        \Repeat{$G = \overline{G}$}  {
            \ForEach{$(s_1, s_2, a, s_3) \in \mu(G)$}{ \label{ln:foreach_cond_strat}
                \If{$s_1 \notin G \land s_1 \in \stateSatB$} { \label{ln:s_not_in_G_strat}
                    $\mathcal{E} \gets \mathcal{E} \cup \{ (s_1, \langle s_2, a \rangle, s_3) \}$\; \label{ln:update_E_strat}
                    $\mathbf{M} \gets \mathbf{M} \cup \{s_1 \}$\; \label{ln:collect_sources_in_M}
                }
            }

            $\overline{G} \gets G$\; \label{ln:copy_G_strat}
            $G \gets G \cup \mathbf{M}$; $\mathbf{M} \gets \varnothing$\; \label{ln:update_G_strat}
            
        } \label{ln:check_fixpoint_strat}


        \tcp{Derive labeling function $\lambda$ from $\mathcal{E}$}
        
        \If{$(\mathcal{I} \cap G) \neq \varnothing$}{ \label{ln:check_if_winning_strat}
            \Return{$\strategygraph \gets (\mathcal{V} \gets G, \mathcal{E}, \lambda)$}\; \label{ln:return_SG_strat}
            }
        \Return{$\strategygraph \gets (\varnothing, \varnothing, \lambda)$}\; \label{ln:return_empty_sg_strat}
    }
\end{algorithm}
\end{minipage}
\Description{Pseudocode for solving reachability and safety timed CLTLoc games while constructing a \ac{sg}.}
    \label{fig:comparison_of_algorithms_strat}
\end{figure}

\subsection{Strategy Graph for Reachability \texorpdfstring{\aclp{tcg}}{TCGs}}
In reachability \acp{tcg}, where the controller's objective is to reach a set of goal states, cycle-free winning controller strategies can be synthesized over an \emph{acyclic} \ac{sg}.
{Notably, since plays are infinite, in the remainder of this work we restrict them to be acyclic until the first goal state is reached.}
Indeed, recall that reachability winning conditions only require that a goal state be eventually reached.

To correctly construct an acyclic \ac{sg}, \autoref{alg:solve_tcgs} has been extended as detailed in \autoref{alg:reachability_strategy_strat}.
In particular, Line~\ref{ln:init_V_E_strat} was added to initialize the set of edges $\mathcal{E}$ and the auxiliary set $\mathbf{M}$.
Then, the loop of \autoref{alg:solve_tcgs} computing the fixed point over set $G$ has been extended as follows:
for every tuple $(s_1, s_2, a, s_3)$ returned by $\mu(G)$ in Line~\ref{ln:foreach_cond_strat}, if $s_1$ does not belong to $G$ and is contained in $\stateSatB$ (Line~\ref{ln:s_not_in_G_strat}), then 
the corresponding labeled edge is added to $\mathcal{E}$ in Line~\ref{ln:update_E_strat}.
The condition $s_1 \notin G$ in Line~\ref{ln:s_not_in_G_strat} ensures that the returned \ac{sg} is acyclic.
Since $G$ is initialized to the set of states satisfying $\psi$ (Line~\ref{ln:init_G_strat}), by \autoref{eq:mu_c} and \autoref{eq:mu_e} it is sufficient to update $G$ (Line~\ref{ln:update_G_strat}) by adding the source $s_1$ (collected in Line~\ref{ln:collect_sources_in_M}) of each tuple returned by $\mu(G)$ (Line~\ref{ln:foreach_cond_strat}), thereby ensuring that the sources and targets of all edges belong to $\mathcal{V}$ (returned in Line~\ref{ln:return_SG_strat}).
Once the fixed point over $G$ has been reached, if at least one cycle-free winning controller strategy exists (Line~\ref{ln:check_if_winning_strat}), then the \ac{sg} from which such a strategy (and possibly other cycle-free winning controller strategies) can be synthesized is returned in Line~\ref{ln:return_SG_strat}; otherwise, an empty \ac{sg} is returned in Line~\ref{ln:return_empty_sg_strat}.
Since the complexity of $\mu$ is $\mathcal{O}(|\mathcal{S}|^3)$, 
the overall complexity of \autoref{alg:reachability_strategy_strat} 
is $\mathcal{O}(|\mathcal{S}|^4)$.
{This higher complexity, compared with the $\mathcal{O}(|\mathcal{S}|^3 \cdot | \act |)$ complexity of the corresponding solving algorithm, reflects the additional cost of explicitly constructing the \ac{sg}, rather than merely computing the set of winning states.}

\Blue{Strategy synthesis for reachability \acp{tcg} is also EXPTIME-hard, since solving a \ac{tcg} can be reduced to synthesizing its \ac{sg} and checking whether the resulting \ac{sg} is nonempty.
Moreover, the synthesis algorithm runs in exponential time in the size of the input \ac{tcg}, as its complexity is polynomial
in the size of the \ac{rts}.}

{\begin{lemma}
    \label{lem:sg_in_reach_strictly_deacreases_rank}
    Let $\tcgshort = (\arena, \winCond)$ be a \ac{tcg}, where $\winCond := \varphi \Until \psi$, and let $\strategygraph = (\mathcal{V}, \mathcal{E}, \lambda)$ be the \ac{sg} returned by \autoref{alg:reachability_strategy_strat} with input $\tcgshort$. 
    Then, for every edge $(s_1, \langle s_2,a \rangle, s_3) \in \mathcal{E}$, it holds that $\rank(s_3) < \rank(s_1) < \infty$.
\end{lemma}

\begin{proof}
    An edge $(s_1, \langle s_2, a \rangle, s_3) \in \mathcal{E}$ with source $s_1$ is added during an iteration of the loop at Line~\ref{ln:foreach_cond_strat} in which $\mu$ is applied to the current value of $G$, say $G_n$, and only if $s_1 \notin G_n \land s_1 \in \stateSatB$ holds (Line~\ref{ln:s_not_in_G_strat}). 
    By \autoref{eq:mu_c} and \autoref{eq:mu_e}, its targets belong to $G_n$.
    Since $\mathsf{r}(s_3)$ is the least index $m$ such that $s_3 \in G_{m}$ (\autoref{def:rank}), and since $s_3 \in G_n$, it follows that $\rank(s_3) \le n$. 
    Moreover, $s_1$ is added to $G$ at the end of that iteration, so $s_1 \in ( G_{n + 1} \setminus G_n )$, that is, $\rank(s_1) = n + 1$. 
    Therefore, it holds that $\rank(s_3) \le n < n + 1 = \rank(s_1) < \infty$.
\end{proof}}

\begin{proposition}
\label{prop:no_cycles_in_reachability_tcgs}
    Let $\tcgshort = (\arena, \winCond)$ be a \ac{tcg}, where $\winCond := \varphi \Until \psi$, and let $s_0 \in \mathcal{I}$ be an initial state.
    If there exists a winning controller strategy from $s_0$, then there exists a cycle-free winning controller strategy from $s_0$.
\end{proposition}

{\begin{proof}
    By \autoref{lem:reach_properties}, if a winning controller strategy
	exists from $s_0$, then $s_0\in G^*$, {where $G_0, G_1, \dots$, and $G^*$ are as in \autoref{def:rank}.}
	It is immediate to see that, by construction, if $s_0\in G_n$, then there exists a controller strategy $\strategyc$
	that guarantees reaching {$G_0$} in at most $n$ moves.
	If $\rank(s_0)=n>0$, then
	$s_0\in\pi(G_{n-1})$, since $s_0\notin G_{n-1}$ by minimality of $\rank(s_0)$.
	Hence, fix a controller choice witnessing this membership.
	For every play consistent with $\strategyc$, if the current state has
	rank $n>0$, by \autoref{lem:sg_in_reach_strictly_deacreases_rank} every possible successor chosen by the environment
	belongs to $G_{n-1}$, \emph{i.e.}, it has rank strictly smaller than $n$.
	Thus, rank strictly decreases at every move until a state in $G_0$ is
	reached. 
    Since all states in $G_0$ satisfy $\psi$, $\strategyc$ is winning.
	Moreover, no state can occur twice before reaching $G_0$, since equal
	states have equal rank, whereas rank strictly decreases along the play.
	Therefore, every play is acyclic up to the first goal state, and $\strategyc$ is cycle-free.
\end{proof}}


Interestingly, the construction of the \ac{sg} in \autoref{alg:reachability_strategy_strat} yields an acyclic \ac{sg} that enables the synthesis of controller strategies minimizing the number of moves required to reach the set of goal states.

{\begin{definition}
\label{def:distance}
    Let $\tcgshort = (\aut, \winCond)$ be a \ac{tcg}, where $\winCond := \varphi \Until \psi$, let $s_0 \in \mathcal{I}$ be an initial state, and let $\strategyc$ be a controller strategy. 
    We denote by $\plays(\strategyc, s_0)$ the set of plays from $s_0$ consistent with $\strategyc$.
    Then, the \emph{distance} of $\strategyc$ from the set of goal states is defined as
    $
    \mathsf{d}(\strategyc, s_0) = \max_{\rho \in \plays(\strategyc, s_0)} \min \{ n \in \mathbb{N} \mid \rho, n \models \psi \}
    $,
    with the convention $\min \varnothing = \infty$.
    Therefore, $\mathsf{d}(\strategyc, s_0) \in (\mathbb{N} \cup \{\infty\})$, and $\mathsf{d}(\strategyc, s_0) < \infty$ if, and only if, $\strategyc$ guarantees that a goal state is reached from $s_0$.
\end{definition}}

{Recall that a controller strategy is a function, which nonetheless induces a set of plays, since the environment resolves the non-determinism over transitions and, in environment locations, also chooses the delay.
This motivated the definition of distance as in \autoref{def:distance}: in each play, the inner $\min$ selects the first position at which a goal state is reached, while the outer $\max$ accounts for the worst case over the plays that may arise from the environment's choices.}

\begin{definition}[Minimal winning controller strategy]
    A winning controller strategy $\strategyc$ from an initial state $s_0 \in \mathcal{I}$ is \emph{minimal} for a reachability \ac{tcg} $\tcgshort = (\arena, \winCond)$, where $\winCond := \varphi \Until \psi$, if, {for every winning controller strategy $\strategyc'$ from $s_0$, it holds that 
    $\mathsf{d}(\strategyc, s_0) \le \mathsf{d}(\strategyc', s_0)$.}
\end{definition}

{\begin{remark}
\label{rem:same_sequence}
The rank (see \autoref{def:rank}) is defined through the sequence $G_0, G_1, \dots$, which is computed by the loop of \autoref{alg:solve_tcgs}. 
\autoref{alg:reachability_strategy_strat} computes the same sequence: it differs from \autoref{alg:solve_tcgs} only in that, at each iteration, it also collects the states and the labeled edges of the \ac{sg}, while the states added to $G$ are those of $\mu(G)$ that are sources of moves and satisfy $\varphi$ (see \autoref{rek:mu_remark}). 
Hence, after $n$ iterations, in either algorithm $G$ takes the value $G_n$
and the rank of a state $s$ is the index of the iteration in which $s$ is added to $G$.
\end{remark}}

{Given a state $s \in \mathcal{S}$, a play from $s$ is an infinite sequence defined as in \autoref{subsec:using_regions_for_tcgs}, except that it starts from $s$ rather than from an initial state; for
$s \in \mathcal{I}$, the two notions coincide.}

The following theorem proves the correctness of the algorithm:

\begin{theorem}
{Let $\tcgshort = (\arena, \winCond)$ be a \ac{tcg}, where $\winCond := \varphi \Until \psi$, and let $\strategygraph = (\mathcal{V}, \mathcal{E}, \lambda)$ be the \ac{sg} returned by \autoref{alg:reachability_strategy_strat} with input $\tcgshort$. 
Then,
\emph{(i)} every controller strategy synthesized from $\strategygraph$ from an initial state $s_0 \in (\mathcal{I} \cap \mathcal{V})$ is winning and minimal for $\tcgshort$; and
\emph{(ii)} if there exists a winning controller strategy for $\tcgshort$ from some initial state, then
$(\mathcal{I} \cap \mathcal{V}) \neq \varnothing$ and at least one winning and minimal controller
strategy can be synthesized from $\strategygraph$.}
\end{theorem}
\begin{proof}
    We prove Property (i), \emph{i.e.}, every controller strategy 
    synthesized from $\strategygraph$ from an initial state $s_0 \in (\mathcal{I} \cap \mathcal{V})$ is winning and minimal.
    {If $\mathsf{r}(s_0) = 0$, then Property (i) trivially follows, since $s_0$ satisfies $\psi$ and $\winCond$ holds at position 0 of every play from $s_0$, so every controller strategy from $s_0$ is winning and minimal.
    Hence, let $\mathsf{r}(s_0) > 0$.}
    By \autoref{lem:sg_in_reach_strictly_deacreases_rank}, every edge of $\strategygraph$ leads from a state of rank $n + 1$ to states of rank at most $n$. 
    Moreover, by construction of \autoref{alg:reachability_strategy_strat}, every state of $\mathcal{V}$ with positive rank is the source of at least one edge. 
    Hence, along every play consistent with a controller strategy synthesized from $\strategygraph$, the rank strictly decreases at every move, until a state of $G_0$ (where $G_0, G_1, \dots$ are as in \autoref{def:rank})---\emph{i.e.}, a goal state---is reached after at most $\rank(s_0)$ moves; 
    moreover, every state visited before belongs to $\stateSatB$, since sources of edges satisfy $\varphi$. 
    Therefore, the strategy is winning.
    
    Next, let $\strategyc$ be a controller strategy synthesized from $\strategygraph$ from $s_0$. 
    By the argument above, $\mathsf{d}(\strategyc, s_0) \le \rank(s_0)$.
    We begin by proving that, for every state $s \in \mathcal{S}$ and every $n \in \mathbb{N}$, if there exists a winning controller strategy $\strategyc'$ from $s$ with $\mathsf{d}(\strategyc', s) \le n$, then $s \in G_n$. 
    The proof is by induction on $n$. 
    If $n = 0$, then every play from $s$ consistent with $\strategyc'$ satisfies $\psi$ at position $0$, hence $s \in G_0$.
    If $n > 0$ and $s$ satisfies $\psi$, then $s \in G_0 \subseteq G_n$.
    If $n > 0$ and $s$ does not satisfy $\psi$, then $s$ satisfies $\varphi$, since otherwise $\varphi \Until \psi$ would be trivially violated at $s$.
    Consider the choice of $\strategyc'$ from $s$: an action and, if $s \in \mathcal{S}_{\cont}$, also a clock region. 
    Every state $s'$ that the environment can reach from $s$ under this choice is such that every play from $s'$ consistent with $\strategyc'$, prefixed by the move obtained from $\strategyc'$ and by the environment's choices, is a play starting from $s$ consistent with $\strategyc'$; hence, a goal state $s_{\mathit{goal}}$ is reached from $s'$ within $n - 1$ moves, and by the inductive hypothesis $s' \in G_{n-1}$ (the states visited before reaching $s_{\mathit{goal}}$ satisfy $\varphi$).
    Since \autoref{eq:mu_c} and \autoref{eq:mu_e} are an extension of \autoref{eq:controllable_predecessor_operator_c} and \autoref{eq:controllable_predecessor_operator_e}, respectively (see \autoref{rek:mu_remark}), and since we only need the source of the tuples returned by $\mu$, we can use $\pi$ for conciseness. 
    It follows that $s \in \pi(G_{n-1})$, and therefore $s \in (\pi(G_{n-1}) \cap \stateSatB) \subseteq G_n$.
    This intersection corresponds to the condition in Line~\ref{ln:s_not_in_G_strat}.
    Now, let $\strategyc'$ be any winning controller strategy from $s_0$ and let $n = \mathsf{d}(\strategyc', s_0)$; by the claim above, $s_0 \in G_n$; by \autoref{def:rank}, $\rank(s_0) \le n$. 
    It follows that $\mathsf{d}(\strategyc, s_0) \le \rank(s_0) \le \mathsf{d}(\strategyc', s_0)$, \emph{i.e.}, $\strategyc$ is minimal.
    
    We now prove Property (ii), \emph{i.e.}, if there exists a winning controller strategy for $\tcgshort$ from some initial state, then
    $(\mathcal{I} \cap \mathcal{V}) \neq \varnothing$ and at least one winning and minimal controller
    strategy can be synthesized from $\strategygraph$.
    Suppose that there exists a winning controller strategy for $\tcgshort$ from an initial state $s_0$. 
    Let $G^*$ be as in \autoref{def:rank}.
    By \autoref{lem:reach_properties}, {which characterizes $G^*$ as the set of states from which a winning controller strategy exists}, $s_0 \in G^*$ and hence $\rank(s_0) < \infty$ holds.
    If $\rank(s_0) = 0$, then $s_0$ satisfies $\psi$ and $\winCond$ holds at position $0$ of every play from $s_0$, so every controller strategy from $s_0$ is winning and minimal. 
    Assume then $\rank(s_0) = n + 1 > 0$. 
    By \autoref{def:rank}, $s_0 \in (\pi(G_n) \cap \stateSatB)$ and, by minimality of $\rank(s_0)$, $s_0 \notin G_n$. 
    Since $\rank(s_0) < \infty$, the loop of \autoref{alg:reachability_strategy_strat} performs at least $n + 1$ iterations and, by \autoref{rem:same_sequence}, $G = G_n$ at the $(n+1)$-th iteration. 
    During that iteration, since $\mu$ is an extension of $\pi$ (see \autoref{rek:mu_remark}), $\mu(G_n)$ contains a tuple $(s_0, s_2, a, s_3)$ with $s_3 \in G_n$; moreover, the condition of Line~\ref{ln:s_not_in_G_strat} is satisfied for $s_0$, as $s_0 \notin G_n$ and $s_0 \in \stateSatB$. 
    {Hence, Line~\ref{ln:update_E_strat} adds the corresponding labeled edge to $\mathcal{E}$, while Line~\ref{ln:update_G_strat} adds $s_0$ to $G$ (recall that $G$ is returned as $\mathcal{V}$ in Line~\ref{ln:return_SG_strat}).}
    Therefore, $s_0 \in \mathcal{V}$ and at least one edge outgoing from $s_0$ belongs to $\mathcal{E}$, that is, $s_0 \in (\mathcal{I} \cap \mathcal{V})$ and a controller strategy can be synthesized from $\strategygraph$ from $s_0$. 
    By Property (i), every such strategy is winning and minimal for $\tcgshort$.
\end{proof}

\subsection{Strategy Graph for Safety \texorpdfstring{\aclp{tcg}}{TCGs}}
\begin{figure}[tb]
\centering
\noindent
\begin{minipage}{0.68\textwidth}
\begin{algorithm}[H]
    \caption{Safety for strategy synthesis}
    \label{alg:safety_strategy} 
    \KwData{$\tcgshort = (\arena, \winCond := \Box \varphi)$: a \ac{tcg} where $\varphi$ is a pure CLTLoc formula.}
    \KwResult{$\strategygraph$: a \ac{sg} for $\mathcal{T}$.}

    \Begin(){
        $G \gets \text{states satisfying $\varphi$}$\; \label{ln:init_safety_strat}
        $\mathcal{V} \gets \varnothing$; $\mathcal{E} \gets \varnothing$\; \label{ln:init_V_E_safety_strat}


     \tcp{The loop is as in \autoref{alg:safety}}
        {\color{gray}
        \Repeat{$G = \overline{G}$} { \label{ln:begin_fixpoint_loop_safety_strat}
        $\overline{G} \gets G$\;   \label{ln:safety_strat_updateG} 
       
        $G \gets (\pi(G) \cap G) $\; \label{ln:safety_strat_newG}
            
        }  \label{ln:until_safety_strat}
        }

        \ForEach{$(s_1, s_2, a, s_3) \in \mu(G)$}{ \label{ln:foreach_cond_safety_strat}
        \If{$s_1 \in G$}{
               \label{ln:s_in_G_safety_strat}
                    $\mathcal{V} \gets \mathcal{V} \cup \{ s_1, s_3 \}$\; \label{ln:update_V_safety_strat}
                    $\mathcal{E} \gets \mathcal{E} \cup \{ (s_1, \langle s_2, a \rangle, s_3) \}$\; \label{ln:update_E_safety_strat}
            }
            }


        \tcp{Derive labeling function $\lambda$ from $\mathcal{E}$}
        \If{$(\mathcal{I} \cap G) \neq \varnothing$}  { \label{ln:hasinitial_safety_strat}
                \Return{$\strategygraph \gets (\mathcal{V}, \mathcal{E}, \lambda)$}\; \label{ln:return_sg_safety_strat}
        }
        \Return{$\strategygraph \gets (\varnothing, \varnothing, \lambda)$}\; \label{ln:return_empty_sg_safety_strat}
        }
\end{algorithm}
\end{minipage}
\Description{Pseudocode for solving safety timed CLTLoc games while constructing a \ac{sg}.}
    \label{fig:comparison_of_algorithms_strat_safety}
\end{figure}
To correctly construct a \ac{sg} for safety \acp{tcg}, \autoref{alg:safety} has been extended as detailed in \autoref{alg:safety_strategy}.
Notice that lines~\ref{ln:begin_fixpoint_loop_safety_strat}--\ref{ln:until_safety_strat} are identical to lines~\ref{ln:begin_safety_loop}--\ref{ln:until} in \autoref{alg:safety}; hence, these lines are represented in gray.
In particular, Line~\ref{ln:init_V_E_safety_strat} was added to initialize the components of the returned \ac{sg}.
Once the fixed point over set $G$ is reached, the application of $\mu(G)$ in Line~\ref{ln:foreach_cond_safety_strat} returns only tuples that guarantee the victory of the controller.
If the source state of a move as returned by $\mu(G)$ is in $G$ itself (Line~\ref{ln:s_in_G_safety_strat}), then lines \ref{ln:update_V_safety_strat} and \ref{ln:update_E_safety_strat} update the components of the returned \ac{sg}.
Finally, if at least one  
winning controller strategy exists (Line~\ref{ln:hasinitial_safety_strat}), then the \ac{sg} from which such a strategy (and possibly other winning controller strategies) can be synthesized is returned in Line~\ref{ln:return_sg_safety_strat}; otherwise, an empty \ac{sg} is returned in Line~\ref{ln:return_empty_sg_safety_strat}.

The overall complexity of \autoref{alg:safety_strategy} {matches that of \autoref{alg:safety}: the fixed-point computation requires at most $\mathcal{O}(|\mathcal{S}|)$ applications of $\pi$, each taking $\mathcal{O}(|\mathcal{S}|^2)$ time; $\mu$ is then applied only once, after the fixed point is reached.
Thus, the overall complexity 
of \autoref{alg:safety_strategy} is $\mathcal{O}(|\mathcal{S}|^3)$. Unlike reachability synthesis, where complexity is worse than reachability solving, safety synthesis can postpone constructing $\strategygraph$ until the winning-state computation is complete.}

The following theorem proves the correctness of the algorithm:
\begin{theorem}
    {Let $\tcgshort = (\arena, \winCond)$ be a \ac{tcg}, where $\winCond := \Box \varphi$, and let $\strategygraph  = (\mathcal{V}, \mathcal{E}, \lambda)$ be the \ac{sg} returned by \autoref{alg:safety_strategy} with input $\tcgshort$.
    Then, \emph{(i)} every controller strategy synthesized from $\strategygraph$ from an initial state in $\mathcal{I} \cap \mathcal{V}$ is winning; and \emph{(ii)} if there exists a 
    winning controller strategy $\strategyc$ for $\tcgshort$ from some initial state $s_0$, then $s_0 \in (\mathcal{I} \cap \mathcal{V})$ and $\strategyc$ can be synthesized from $\strategygraph$.}
\end{theorem}

\begin{proof}
    {We first prove that every controller strategy synthesized from $\strategygraph$ from an initial state in $\mathcal{I} \cap \mathcal{V}$ is winning. 
    Since the loop at Line~\ref{ln:begin_fixpoint_loop_safety_strat} is the same as in \autoref{alg:safety}, let $G^*$ denote the value of $G$ upon termination of such a loop. 
    Since $G$ is initialized with the states satisfying $\varphi$  (Line~\ref{ln:init_safety_strat}) and the loop only removes states from it, every state in $G^*$ satisfies $\varphi$; moreover, at the fixed point $G^* = (\pi(G^*) \cap G^*)$, hence $G^* \subseteq \pi(G^*)$.
    We now show that every edge $(s_1, \langle s_2, a \rangle, s_3) \in \mathcal{E}$ is such that $s_1, s_3 \in G^*$, and that every state of $G^*$ is the source of at least one edge. 
    The first claim holds since the condition of Line~\ref{ln:s_in_G_safety_strat} requires $s_1 \in G^*$, while $s_3 \in G^*$ follows by \autoref{eq:mu_c} and \autoref{eq:mu_e}. 
    For the second claim, let $s \in G^*$; since $G^* \subseteq \pi(G^*)$, by \autoref{eq:controllable_predecessor_operator_c} and \autoref{eq:controllable_predecessor_operator_e} there is a choice of the controller in $s$ such that, regardless of the choices of the environment, the next state belongs to $G^*$. 
    Since \autoref{eq:mu_c} and \autoref{eq:mu_e} are an extension of \autoref{eq:controllable_predecessor_operator_c} and \autoref{eq:controllable_predecessor_operator_e}, respectively (see \autoref{rek:mu_remark}), $\mu(G^*)$ contains a tuple with source $s$, which the loop at Line~\ref{ln:foreach_cond_safety_strat} collects, adding the corresponding labeled edge to $\mathcal{E}$.
    {To show that the two claims above ensure that a synthesized strategy is winning, we now follow the edges of $\mathcal{E}$, which by construction correspond to the moves collected by $\mu$ from $G^*$.}
    Then, let $\strategyc$ be a controller strategy synthesized from $\strategygraph$ from an initial state $s_0 \in (\mathcal{I} \cap \mathcal{V})$ and let $\rho$ be a play consistent with $\strategyc$. 
    By induction on the positions of $\rho$, every state visited in $\rho$ belongs to $G^*$: this holds at position $0$, since $s_0 \in \mathcal{V} \subseteq G^*$, and it is preserved at every move} {because, by the definition of $\mu(G^*)$, every possible successor under a controller choice represented in $\strategygraph$ belongs to $G^*$.
    Moreover, every state in $G^*$ admits at least one such choice, as shown above.}
    Since every state in $G^*$ satisfies $\varphi$, every state visited in $\rho$ satisfies $\varphi$, \emph{i.e.}, $\rho \models \Box \varphi$. 
    Therefore, $\strategyc$ is winning.
 
    We now prove that, if there exists a 
    winning controller strategy $\strategyc$ for $\tcgshort$ from an initial state $s_0$, then $s_0 \in (\mathcal{I} \cap \mathcal{V})$ and $\strategyc$ can be synthesized from $\strategygraph$. 
    By \autoref{lem:safe_properties} and by the consideration above on the loop at Line~\ref{ln:begin_fixpoint_loop_safety_strat}, it follows that $s_0 \in G^*$.
    %
    {Let $\rho$ be an arbitrary play from $s_0$ consistent with $\strategyc$, let $i \in \mathbb{N}$, and let $s$ be the last state of $\rho_{\le i}$. We show that $s \in G^*$.
    Suppose, by contradiction, that $s \notin G^*$. By \autoref{lem:safe_properties}, no controller strategy from $s$ can guarantee $\Box\varphi$.
    Hence, even if the controller follows the choices prescribed by $\strategyc$ after $\rho_{\le i}$, 
    then $\rho_{\le i}$ can be extended to a play from $s_0$ consistent with $\strategyc$ in which $\varphi$ is false at some position $m \ge i$.
    This is a contradiction with the assumption that $\strategyc$ is winning, hence $s \in G^*$.
        Moreover, 
    every state that the environment can reach from $s$ under the controller choice $\strategyc(\rho_{\le i})$ is the last state of a 
    {one-step extension} of
    $\rho_{\le i}$ consistent with $\strategyc$,}
    and hence belongs to $G^*$ by the same argument. 
    Thus, {the choice $\strategyc(\rho_{\le i})$ satisfies 
    \autoref{eq:controllable_predecessor_operator_c} or \autoref{eq:controllable_predecessor_operator_e} with respect to $G^*$. Since \autoref{eq:mu_c} and \autoref{eq:mu_e} extend {these operators}
    (see \autoref{rek:mu_remark}), $\mu(G^*)$ contains the corresponding tuples with source $s$; as $s \in G^*$, the condition of Line~\ref{ln:s_in_G_safety_strat} is satisfied and the loop at Line~\ref{ln:foreach_cond_safety_strat} adds the corresponding labeled edges to $\mathcal{E}$, together with $s$ in $\mathcal{V}$.
    In particular, $s_0 \in (\mathcal{I} \cap \mathcal{V})$. 
    Since the argument applies to every prefix $\rho_{\le i}$ consistent with $\strategyc$, the choice $\strategyc(\rho_{\le i})$ is always represented in $\strategygraph$. Hence, 
     $\strategyc$ can be synthesized from $\strategygraph$ by selecting its prescribed choice after each such prefix.
    }
\end{proof}

\Red{\autoref{alg:reachability_strategy_strat} and \autoref{alg:safety_strategy} can be adapted as detailed in \autoref{subsec:additional_winning_conditions} to synthesize
winning controller strategies for a given \ac{tcg}. 
In this case, the \ac{sg} must also be extended to account for the additional applications of $\pi$.
\autoref{alg:reachability_strategy_strat} can be adapted to handle nested winning conditions such as $\winCond_0 := \Next^{n_0} ( \varphi_0 \Until \psi_0 )$;
$\winCond_k := \Next^{n_k} ( \varphi_k \Until \winCond_{k - 1} )$, for $k \in \mathbb{N}_{> 0}$.}

\subsection{Strategy Graph for $\Next$-conjunction \texorpdfstring{\aclp{tcg}}{TCGs}}
In $\Next$-conjunction \acp{tcg}, in which the controller’s objective is to reach a set of goal states by traversing specific sets of states,
memory structures are needed to guarantee that states are visited as encoded in the winning condition; for instance, cycles of states may be traversed a given number of times before exiting them.

Memory structures are directly embedded in the \ac{sg} by adding indices to states, resulting in an \emph{unrolled} (thus, acyclic) \ac{sg}.
To define \ac{sg} for $\Next$-conjunction \acp{tcg}, we extend \autoref{def:strategy_graph} as 
$\strategygraph^{\bigcirc} = (\mathcal{V}^{\bigcirc}, \mathcal{E}^{\bigcirc}, \lambda^{\bigcirc})$, where 
$\mathcal{V}^{\bigcirc} \subseteq \mathcal{S} \times \mathbb{N}$ is an extended set of states,
$\mathcal{E}^{\bigcirc} \subseteq \mathcal{V}^{\bigcirc} \times \mathcal{V}^{\bigcirc}$ is an extended set of edges, and 
{$\lambda^{\bigcirc} : \mathcal{E}^{\bigcirc} \rightarrow (\wp((\mathcal{S} \times \mathbb{N}) \times \act) \setminus \{ \varnothing \})$} is an extended labeling function.
We use the same notation with angular brackets for edges and labels as introduced in \autoref{def:strategy_graph} and omit the adjective \emph{extended} whenever it is clear from the context.
The addition of indices to \ac{sg} states makes it possible to determine when a state is visited multiple times in a play; this also guarantees that the extended \ac{sg} remains acyclic.
Furthermore, since indexing causes multiple visits to the same 
{state in the \ac{rts}} to be represented by distinct extended states, the resulting \ac{sg} is effectively unrolled.


\begin{example}
To illustrate the unrolling procedure, consider the left-hand side of \autoref{fig:unrolling_example}, which represents a fragment of an \ac{rts}.
For simplicity, only states and transitions are shown, without distinguishing between delay and discrete transitions. 
All states are represented by circles, regardless of whether they belong to the controller or the environment. 
Since the figure depicts only an \ac{rts} fragment, states and transitions ensuring infinite plays are omitted.
Suppose that, starting from $s_0$, the controller must reach $s_5$ while visiting the cycle from $s_1$ to $s_4$ exactly once. 
Hence, the states visited in a play $\rho$ (up to $s_5$) are, in order: $s_0, s_1, s_2, s_3, s_4, s_1, s_2, s_3, s_4, s_5$.
Without memory structures, determining whether the cycle has already been completed would require strategies to use the states visited in $\rho$. 
By associating an index with each state, the corresponding unrolled \ac{rts} is obtained as shown on the right-hand side of \autoref{fig:unrolling_example}. 
The index records the progress through the cycle, so that the current state together with its index uniquely determines the corresponding stage of the unrolling. 
Consequently, players can determine whether the cycle has been completed by considering only the current extended state, and the resulting \ac{rts} is acyclic.

\begin{figure}
    \centering
    \includegraphics[width=\linewidth]{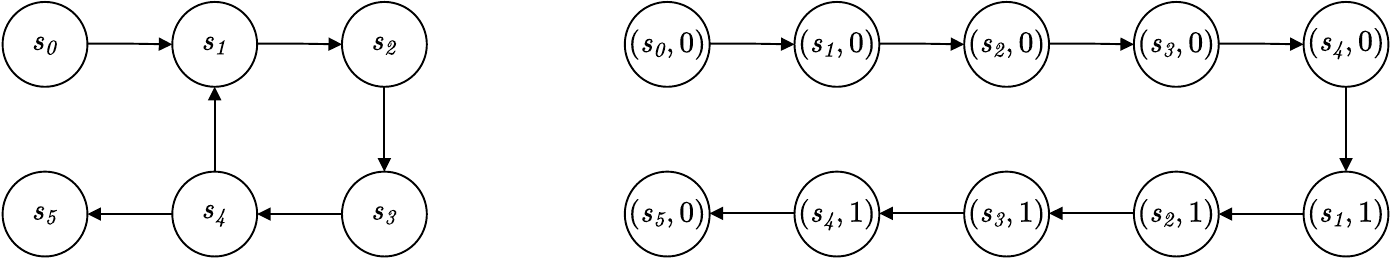}
    \caption{On the left, a fragment of an \ac{rts} containing a cycle. On the right, the corresponding unrolled fragment, in which the cycle from $s_1$ to $s_4$ is traversed exactly once in the play from $s_0$ to $s_5$. For convenience, states are represented as circles, although they may correspond to either controller or environment states. The example assumes that the controller is guaranteed to reach the goal state $s_5$ by traversing the cycle exactly once.}
    \label{fig:unrolling_example}
    \Description{RTS unrolling}
\end{figure}
\end{example}

\begin{remark}
\label{rmk:synthesys_from_extended_sg}
    A controller strategy can be synthesized from $\strategygraph^{\bigcirc}$ as detailed at the beginning of this subsection {by interpreting each}  
    edge $((s_1, n_1), \langle (s_2, n_2), a \rangle, (s_3, n_3)) \in \mathcal{E}^{\bigcirc}$, as the corresponding edge $(s_1, \langle s_2, a \rangle, s_3)$.
    The resulting strategy is determined by $\strategygraph^{\bigcirc}$ {on the finite prefixes of plays consistent with it:} at every position $i<d_k$, the controller selects a choice represented by the label of outgoing edges of $(s_i,i)$, while the environment determines the corresponding successor state.
    On other prefixes,
    $\strategyc$ can be  defined arbitrarily among the admissible moves.

    Furthermore, whenever the controller's 
     choices depend only on the current state of $\strategygraph^{\bigcirc}$, 
     the resulting strategy is positional with respect to the extended \ac{sg}.
    For instance, a controller strategy 
    {whose choices depend on}
    whether the current position is even or odd becomes positional in $\strategygraph^{\bigcirc}$. 
    This facilitates the implementation of strategy synthesis algorithms for $\Next$-conjunction \acp{tcg}. 
    Nevertheless, {non-positional strategies can also be synthesized, since the controller's choices may depend on the preceding play prefix.}
\end{remark}

\begin{figure}[tb]
\centering
\noindent
\begin{minipage}{0.68\textwidth}
\begin{algorithm}[H]
    \caption{$\Next$-conjunction for strategy synthesis}
    \label{alg:next_conjunction_and_next_strat} 
    \KwData{$\tcgshort = (\arena, \winCond := \bigwedge_{i = 0}^k \Next^{d_i} \varphi_i)$: a \ac{tcg} where each $\varphi_i$ is a pure CLTLoc formula, for $0 \le i \le k$, and $d_i < d_{i + 1}$, for $0 \le i < k$.}
    \KwResult{$\strategygraph^{\bigcirc}$: an unrolled (thus, acyclic) \ac{sg} for $\mathcal{T}$.}
    
    \Begin() { 
        $\stateSatB[0, \dots, k] \gets (\varnothing, \dots, \varnothing)$\; \label{ln:init_array_next_and_next_strat}

        \For{$i \gets 0$ \KwTo $k$}{ \label{ln:init_G_for_loop_next_and_next_strat}
            $\stateSatB[i] \gets \text{states satisfying $\varphi_i$}$\; \label{ln:init_next_and_next_strat}
        }

        $G \gets \stateSatB[k]$\; \label{ln:init_G_next_and_next_strat}

        $\mathsf{p} \gets d_k$\; \label{ln:init_pos_idx}
        


        $\mathcal{V}^{\bigcirc} \gets \varnothing$; $\mathcal{E}^{\bigcirc} \gets \varnothing$; $\mathbf{M} \gets \varnothing$\; \label{ln:init_V_E_strat_and_next_strat}

        \ForEach{$s \in G$}{ \label{ln:init_V_cycle_strat}
            $\mathcal{V}^{\bigcirc} \gets \mathcal{V}^{\bigcirc} \cup \{ (s, \mathsf{p}) \}$\; \label{ln:insert_states_into_V_init} 
        }
        

        \SetKw{KwDownTo}{downto}
        \For{$i \gets k$ \KwDownTo $0$}{ \label{ln:main_for_loop_next_and_next_strat}
            \uIf{$i > 0$}{ \label{ln:number_of_pi_app_if_next_and_next_strat}
                $\Delta \gets d_i - d_{i - 1}$\; \label{ln:get_number_of_pi_applications_and_next_strat}
            }
            \Else{
                $\Delta \gets d_0$\; \label{ln:get_number_of_pi_applications_zero_and_next_strat}
            }
        
            \For{$j \gets 1$ \KwTo $\Delta$}{ \label{ln:inner_for_loop_next_and_next_strat}



            
                \ForEach{$(s_1, s_2, a, s_3) \in \mu(G)$}{ \label{ln:foreach_cond_strat_and_next_strat}
                    
                    $\mathcal{V}^{\bigcirc} \gets \mathcal{V}^{\bigcirc} \cup \{ (s_1,\mathsf{p}-1) \}$\; \label{ln:update_V_unrolled}
                    
                    $\mathcal{E}^{\bigcirc} \gets \mathcal{E}^{\bigcirc} \cup \{ ((s_1, \mathsf{p}-1), \langle (s_2, \mathsf{p}-1), a \rangle, (s_3, \mathsf{p})) \}$\; \label{ln:update_E_unrolled}
                    
                     $\mathbf{M} \gets \mathbf{M} \cup \{ s_1 \}$\; \label{ln:update_M_and_next_strat}
                }

                $G \gets \mathbf{M}$; $\mathbf{M} \gets \varnothing$\; \label{ln:update_visited_and_next_strat}

                $\mathsf{p} \gets \mathsf{p} - 1$\; \label{ln:update_p_after_loop}
            }

            \If{$i > 0$}{ \label{ln:intersect_G_next_cond_and_next_strat}
                $G \gets G \cap \stateSatB[i - 1]$\; \label{ln:intersect_G_next_and_next_strat}

                \ForEach{$((s_1, \mathsf{p}), \langle (s_2, \mathsf{p}), a \rangle, (s_3, \mathsf{p}+1)) \in \mathcal{E}^{\bigcirc}$}{ \label{ln:foreach_trans_in_E_and_next}
                
                \If{$s_1 \notin G$}{ \label{ln:s1_not_in_G}
                    $\mathcal{V}^{\bigcirc} \gets \mathcal{V}^{\bigcirc} \setminus \{ (s_1,\mathsf{p}) \}$\; \label{ln:remove_from_V_and_next}

                    $\mathcal{E}^{\bigcirc} \gets \mathcal{E}^{\bigcirc} \setminus \{ ((s_1, \mathsf{p}), \langle (s_2, \mathsf{p}), a \rangle, (s_3, \mathsf{p}+1)) \}$\; \label{ln:remove_from_E_and_next}
                }
                }
            }
        }

       \tcp{Derive labeling function $\lambda^{\bigcirc}$ from $\mathcal{E}^{\bigcirc}$}
            \If{$(\mathcal{I} \cap G) \neq \varnothing$}
            { \label{ln:hasinitial_next_and_next_strat}
                \Return{$\strategygraph^{\bigcirc} \gets (\mathcal{V}^{\bigcirc}, \mathcal{E}^{\bigcirc}, \lambda^{\bigcirc})$}\; \label{ln:return_top_next_and_next_strat}
            }
        \Return{$\strategygraph^{\bigcirc} \gets (\varnothing, \varnothing, \lambda^{\bigcirc})$}\; \label{ln:return_bot_next_and_next_strat}
    }
\end{algorithm}
\end{minipage}
\Description{Pseudocode of the algorithms used to synthesize strategies for $\Next$-conjunction \acp{tcg}.}
    \label{fig:next_conjunction_fig_and_next_strat}
\end{figure}

To correctly construct a \ac{sg} for $\Next$-conjunction \acp{tcg}, \autoref{alg:next_conjunction} has been extended as detailed in \autoref{alg:next_conjunction_and_next_strat}.
In particular, 
in Line~\ref{ln:init_pos_idx} a variable $\mathsf{p}$ is initialized to $d_k$ and used throughout the algorithm to index the position of visited \ac{rts} states, ensuring that the resulting \ac{sg} is correctly unrolled.
Line~\ref{ln:init_V_E_strat_and_next_strat} initializes the components of the returned \ac{sg} {and the auxiliary set $\mathbf{M}$}.
{Since $\mathcal{V}^{\bigcirc}$ is initialized to the set of states satisfying $\varphi_k$
with index $\mathsf{p} = d_k$ (lines \ref{ln:init_V_cycle_strat} and \ref{ln:insert_states_into_V_init}), by \autoref{eq:mu_c} and \autoref{eq:mu_e} it is sufficient to update $\mathcal{V}^{\bigcirc}$ (Line~\ref{ln:update_V_unrolled}) by adding the source $s_1$, with index $\mathsf{p-1}$, of each tuple returned by $\mu(G)$ (Line~\ref{ln:foreach_cond_strat_and_next_strat}), thereby ensuring that the sources and targets of all edges belong to $\mathcal{V}^{\bigcirc}$.}
%
{After $\Delta$ is determined as in \autoref{alg:next_conjunction}, 
%
the sets $\mathcal{V}^{\bigcirc}$ and $\mathcal{E}^{\bigcirc}$
are updated in lines \ref{ln:update_V_unrolled} and \ref{ln:update_E_unrolled}, respectively.}
{For every $j$ from $1$ to $\Delta$, the update of $\mathbf{M}$ in Line~\ref{ln:update_M_and_next_strat} ensures that, during the $j$-th iteration, $\mathbf{M}$ contains the source states generated in that iteration. 
This is necessary to ensure that $G$ is updated with these states in Line~\ref{ln:update_visited_and_next_strat}, consistently with its assignment of the result of $\pi$ in Line~\ref{ln:update_G_next} of \autoref{alg:next_conjunction}}.
Since the algorithm restricts the states of $G$ in Line~\ref{ln:intersect_G_next_and_next_strat}, $\mathcal{V}^{\bigcirc}$ and $\mathcal{E}^{\bigcirc}$ must be updated accordingly.
Indeed, every edge $e = ((s_1, \mathsf{p}), \langle (s_2, \mathsf{p}), a \rangle, (s_3, \mathsf{p}+1)) \in \mathcal{E}^{\bigcirc}$ (Line~\ref{ln:foreach_trans_in_E_and_next}) involving states collected during the iteration in which $j = \Delta$ must be such that $s_1$---\emph{i.e.}, a state corresponding to the source of a move---is in $G$.
Otherwise, if the condition in Line~\ref{ln:s1_not_in_G} is satisfied, then $e$ is invalid.
{Consequently, the source state $(s_1, \mathsf{p})$ is removed from $\mathcal{V}^{\bigcirc}$ in Line~\ref{ln:remove_from_V_and_next}, and the edge itself is removed from $\mathcal{E}^{\bigcirc}$ in Line~\ref{ln:remove_from_E_and_next}.
Notice that we only remove $(s_1, \mathsf{p})$ since $(s_3, \mathsf{p}+1)$ may be the target of other valid moves.}
Finally, Line~\ref{ln:hasinitial_next_and_next_strat} checks whether at least one winning controller strategy exists.
The \ac{sg} from which such a strategy (and possibly other winning controller strategies) can be synthesized is returned in Line~\ref{ln:return_top_next_and_next_strat}; otherwise, an empty \ac{sg} is returned in Line~\ref{ln:return_bot_next_and_next_strat}.
%

{Since $\mu$ is applied exactly $d_k$ times, each application in $\mathcal{O}(|\mathcal{S}|^3 \cdot |\act|)$, and since the edges of each indexed layer can be examined at most once when removing invalid edges, the overall complexity of \autoref{alg:next_conjunction_and_next_strat} is $\mathcal{O}(d_k\cdot|\mathcal{S}|^3 \cdot |\act|)$.}
\begin{lemma}
\label{lem:sg_next_positions}
Let $\tcgshort = (\arena, \winCond)$ be a \ac{tcg}, where $\winCond := \bigwedge_{i = 0}^k \Next^{d_i} \varphi_i$ is such that each $\varphi_i$ is a pure \ac{cltloc} formula,  $d_i < d_{i+1}$ holds for $0 \le i < k$  and $d_k \ge 1$, and let $\strategygraph^{\bigcirc} = (\mathcal{V}^{\bigcirc}, \mathcal{E}^{\bigcirc}, \lambda^{\bigcirc})$ be the \ac{sg} returned by \autoref{alg:next_conjunction_and_next_strat} with input $\tcgshort$. 
Then, 
\emph{(i)} every edge of $\mathcal{E}^{\bigcirc}$ is of the form 
           $((s_1, \mathsf{p}), \langle (s_2, \mathsf{p}), a \rangle, (s_3, \mathsf{p}+1))$, with $0 \le \mathsf{p} < d_k$;
\emph{(ii)} every state $(s, \mathsf{p}) \in \mathcal{V}^{\bigcirc}$ with $\mathsf{p} < d_k$ is the source of at least one edge of $\mathcal{E}^{\bigcirc}$;
and \emph{(iii)} for every $0 \le i \le k$, if $(s, d_i) \in \mathcal{V}^{\bigcirc}$, then $s$ satisfies $\varphi_i$.
\end{lemma}

\begin{proof}
Property~(i) follows immediately from \autoref{alg:next_conjunction_and_next_strat}: the variable $\mathsf{p}$ is initialized to $d_k$ (Line~\ref{ln:init_pos_idx}) and decremented once per application of $\mu$ (Line~\ref{ln:update_p_after_loop}), so that it always holds the position of the states currently in $G$; hence, an edge created at Line~\ref{ln:update_E_unrolled} has source indexed by $\mathsf{p}-1$ and target indexed by $\mathsf{p}$, where $1 \le \mathsf{p} \le d_k$. 
Moreover, the total number of applications of $\mu$ is $\sum_{i=1}^{k}(d_i - d_{i-1}) + d_0 = d_k$, hence $\mathsf{p}$ ranges over $d_k, \dots, 0$ and never becomes negative.

Property~(ii) follows from the same lines of Property (i), since a state $(s, \mathsf{p})$ with $\mathsf{p} < d_k$ is added to $\mathcal{V}^{\bigcirc}$ only together with an edge.
Indeed, $(s, \mathsf{p})$ is added as the source of an edge 
$((s, \mathsf{p}), \langle (s', \mathsf{p}), a \rangle, (s'', \mathsf{p} + 1))$ in Line~\ref{ln:update_V_unrolled}.
%
%
%
If $s$ does not belong to $G$ after the intersection on Line~\ref{ln:intersect_G_next_and_next_strat}, then Line~\ref{ln:remove_from_V_and_next} removes the \Red{node itself and the edges outgoing from $(s, \mathsf{p})$.}

Property~(iii) follows from the intersection of Line~\ref{ln:intersect_G_next_and_next_strat} and from the loop of Line~\ref{ln:foreach_trans_in_E_and_next}: 
\Red{upon reaching Line~\ref{ln:intersect_G_next_cond_and_next_strat},
the set $G$ contains the states of position $d_{i-1}$; after the intersection, all remaining states in $G$ satisfy $\varphi_{i-1}$,
and every edge whose source does not belong to $G$}
is removed from $\mathcal{E}^{\bigcirc}$, together with its source state. 
%
For $i = k$, the property holds since {$\mathcal{V}^{\bigcirc}$ is initialized to the set of states satisfying $\varphi_k$ with index $\mathsf{p} = d_k$ (lines \ref{ln:init_V_cycle_strat} and \ref{ln:insert_states_into_V_init}).}
%
\end{proof}

The following theorem proves the correctness of the algorithm:

\begin{theorem}
    {Let $\tcgshort = (\arena, \winCond)$ be a \ac{tcg}, where $\winCond := \bigwedge_{i = 0}^k \Next^{d_i} \varphi_i$ is such that each $\varphi_i$ is a pure \ac{cltloc} formula and $d_i < d_{i+1}$ holds for $0 \le i < k$, and let $\strategygraph^{\bigcirc} =  ( \mathcal{V}^{\bigcirc}, \mathcal{E}^{\bigcirc}, \lambda^{\bigcirc})$ be the extended \ac{sg} returned by \autoref{alg:next_conjunction_and_next_strat} with input $\tcgshort$. 
    Then, \emph{(i)} every controller strategy synthesized from $\strategygraph^{\bigcirc}$ from an initial state $(s_0, 0) \in \mathcal{V}^{\bigcirc}$ with $s_0 \in \mathcal{I}$ is winning for $\tcgshort$; 
    and \emph{(ii)} if there exists a winning controller strategy $\strategyc$ for $\tcgshort$ from an initial state $s_0$, then $(s_0, 0) \in \mathcal{V}^{\bigcirc}$ and $\strategyc$ can be synthesized from $\strategygraph^{\bigcirc}$ from $(s_0, 0)$.}
\end{theorem}

\begin{proof}
    \Blue{For an initial state $s_0$ and a controller strategy $\strategyc$, let $\mathcal{S}_i$ be the set of states occurring at position $i$ in at least one play from $s_0$ consistent with $\strategyc$, for every $i\in\mathbb{N}$.
    In particular, $\mathcal{S}_0=\{s_0\}$.}

    
    We first prove that every controller strategy synthesized from $\strategygraph^{\bigcirc}$ from an initial state $(s_0, 0) \in \mathcal{V}^{\bigcirc}$ with $s_0 \in \mathcal{I}$ is winning for $\tcgshort$.
    Let $\strategyc$ be a controller strategy synthesized from $\strategygraph^{\bigcirc}$ from $(s_0, 0)$. 
    We show, by induction on $i$, that $\{ (s, i) \mid s \in \mathcal{S}_i \} \subseteq \mathcal{V}^{\bigcirc}$, for all $0 \le i \le d_k$. 
    
    The claim holds for $i = 0$, since $(s_0, 0) \in \mathcal{V}^{\bigcirc}$. 
    Let $0 \le i < d_k$ and 
    \Blue{let $s' \in \mathcal{S}_{i+1}$. 
    By definition of $\mathcal{S}_{i+1}$, there exists a prefix $\rho_{\le i}$ of a play from $s_0$ consistent with $\strategyc$, ending in a state $s \in \mathcal{S}_i$, such that $s'$ is a possible successor under the controller choice $\strategyc(\rho_{\le i})$.
    By the inductive hypothesis, $(s,i) \in \mathcal{V}^{\bigcirc}$.
    %
    By Property (ii) of \autoref{lem:sg_next_positions}, $(s, i)$ is the source of at least one edge of $\mathcal{E}^{\bigcirc}$.
    Since $\strategyc$ is synthesized from $\strategygraph^{\bigcirc}$, its choice at $\rho_{\le i}$ is represented by the {labels of the} 
    outgoing edges of $(s,i)$. 
    By the construction of the \ac{sg}, all possible environment responses to that choice are represented.
    Hence, for some $s_2$ and $a$, there exists an edge 
    $((s,i), \langle (s_2,i),a \rangle, (s',j)) \in  \mathcal{E}^{\bigcirc}$.
    By Property (i) of \autoref{lem:sg_next_positions}, $j=i+1$. 
    Therefore, $(s',i+1) \in \mathcal{V}^{\bigcirc}$, which completes the inductive claim.}
    %
    %
    %
    By Property (iii) of \autoref{lem:sg_next_positions}, every state $s \in \mathcal{S}_{d_i}$ is such that, if $(s, d_i) \in \mathcal{V}^{\bigcirc}$, then $s$ satisfies $\varphi_i$.
    Therefore, in every play from $s_0$ consistent with $\strategyc$, the state occurring at position $d_i$ satisfies $\varphi_i$, \emph{i.e.}, $\winCond$ is satisfied.
    It follows that $\strategyc$ is winning.
    
    We now prove that if there exists a winning controller strategy $\strategyc$ for $\tcgshort$ from an initial state $s_0$, then $(s_0, 0) \in \mathcal{V}^{\bigcirc}$ and $\strategyc$ can be synthesized from $\strategygraph^{\bigcirc}$ from $(s_0, 0)$.
    \Blue{Recall that, in the proof of \autoref{thm:next_conjunction_solving}, $G_m$ denotes the set of states from which, considering the current position \Red{of a play} to be $m$, the controller can guarantee that $\varphi_i$ holds at position $d_i$, for every $i$ such that $d_i \ge m$.
    
    Since $\mu$ extends $\pi$ (\autoref{rek:mu_remark}), \autoref{alg:next_conjunction_and_next_strat} computes the same sets $G_m$ as \autoref{alg:next_conjunction}.
    In particular, when $\mathsf{p}=m+1$, \autoref{alg:next_conjunction_and_next_strat} applies $\mu$ to $G_{m+1}$ to construct the edges between positions $m$ and $m+1$.
    We first show that $\mathcal{S}_m \subseteq G_m$ for every $0 \le m \le d_k$. Let $\rho_{\le m}$ be any prefix of a play from $s_0$ consistent with $\strategyc$, ending in a state $s$. 
    Since $\strategyc$ is winning, every play extending $\rho_{\le m}$ and consistent with $\strategyc$ satisfies $\varphi_i$ at position $d_i$, for every $i$ such that $d_i \ge m$. By the characterization of $G_m$ recalled above, $s \in G_m$. Hence, $\mathcal{S}_m \subseteq G_m$.
    Now, let $0 \le m < d_k$ and consider any prefix $\rho_{\le m}$ of a play from $s_0$ consistent with $\strategyc$, ending in a state $s \in \mathcal{S}_m$.
    Every possible successor under the controller choice $\strategyc(\rho_{\le m})$ belongs to $\mathcal{S}_{m+1} \subseteq G_{m+1}$. Therefore, by \autoref{eq:mu_c} and \autoref{eq:mu_e}, $\mu(G_{m+1})$ contains all tuples $(s,s_2,a,s_3)$ corresponding to this choice and its possible environment outcomes. 
    When $\mathsf{p}=m+1$, lines~\ref{ln:update_V_unrolled} and~\ref{ln:update_E_unrolled} therefore add $(s,m)$ to $\mathcal{V}^{\bigcirc}$ and the corresponding edge
    $
    ((s,m),\langle(s_2,m),a\rangle,(s_3,m+1))
    $
    to $\mathcal{E}^{\bigcirc}$.
    Moreover, these {states} and edges
    are never removed. 
    Indeed, if $m=d_i$ for some $0 \le i < k$, then the intersection at Line~\ref{ln:intersect_G_next_and_next_strat} retains $s$, since $s \in \mathcal{S}_m \subseteq G_m$.
    At other positions, no such removal is performed. 
    In particular, since $\mathcal{S}_0=\{s_0\}$, we obtain $(s_0,0)\in\mathcal{V}^{\bigcirc}$.
    Since the argument applies to every prefix $\rho_{\le m}$ consistent with $\strategyc$, all controller choices prescribed by $\strategyc$ up to position $d_k$ are represented in $\strategygraph^{\bigcirc}$, together with all their possible environment choices. 
    Thus, $\strategyc$ can be synthesized from $\strategygraph^{\bigcirc}$ from $(s_0,0)$.}
\end{proof}

\section{Implementation and Experimental Evaluation}
\label{sec:implementation_and_experimental_evaluationn}
This section briefly introduces \toolname{}, a \Red{a C\texttt{++} library available on GitHub~\cite{tarzan_github}, originally designed for forward and backward reachability analysis of \ac{ta} and networks of TA, which has been extended with the algorithms presented in \autoref{sec:solving_timed_CLTLoc_games} and \autoref{sec:synthesizing_winning_controller_strategies}.}
Then, it provides a detailed empirical evaluation 
of the implemented algorithms to solve and synthesize winning controller strategies for \acp{tcg}
\Red{considereing two case studies: a Production Cell and a medical Human-Robot Interaction scenario.}
Experiments were conducted on a MacBook Pro with an M3 chip (8-core CPU with 4 performance cores up to 4.06 GHz and 4 efficiency cores up to 2.75 GHz) and 24 GB of RAM.

\subsection{The \textnormal{\toolname{}} Library}
\toolname{} leverages a region-based abstraction
tracking the order in which clocks become unbounded---\emph{i.e.}, when their value exceeds the maximum constant $\cmax$ against which they are compared.
In the following we provide a brief overview of this abstraction; a detailed description is presented in~\cite{tarzan_forte}.

Let $\aut = (\act, Q, Q_{0}, \ackc, T, \invariant)$ be a \ac{ta} and let $q \in Q$.
Furthermore, let $\lidx, \ridx \in \mathbb{N}$ be two integers such that $\lidx + \ridx \le \lvert \ackc \rvert$.
We denote by $X_i \subseteq \ackc$ a subset of clocks of $\aut$, for $-\lidx \le i \le \ridx$, such that: 
(i) $\ackc = \bigcup_{i} X_i$,
and (ii) for all $-\lidx \le i,j \le \ridx$ such that $i \neq j$, $(X_i \cap X_j) = \varnothing$ holds.
We also introduce a function $h : \ackc \rightarrow \{0, \dots, \cmax \}$ assigning integer values to clocks.
A state region of $\aut$ is represented as:
\begin{equation}
\label{eq:region}
    \reg := \langle \loc, h,  X_{-\lidx}, X_{-(\lidx-1)}, \dots, X_{-1}, X_{0}, X_1, \dots, X_{(\ridx-1)}, X_{\ridx} \rangle.
\end{equation}
In particular, the sets $X_{-\lidx}, \dots, X_{-1}$ represent unbounded clocks, capturing the ordering in which these clocks become unbounded.
Conversely, the sets $X_1, \dots, X_{\ridx}$ represent bounded clocks, capturing the ordering of their fractional parts.
Set $X_0$ contains bounded clocks whose fractional part is zero.
This region abstraction, while potentially increasing the state space size, ensures that each region has at most three immediate delay predecessors (\emph{i.e.}, regions reachable via the shortest delay transition going backwards).
\Red{This benefits backward state space exploration while having a negligible impact on forward exploration.}


\toolname{} requires two separate input files containing, respectively, the specification of 
a \ac{tarena} and a winning condition following the syntax of our Domain-Specific Language called \emph{Liana}. 
\toolname{} parses the given input files
and generates a C\texttt{++} representation of both a \ac{tarena} and a winning condition, effectively defining a \ac{tcg}. 
Once parsing is successfully completed, the resulting \ac{tcg} is processed by calling suitable functions provided by the \toolname{} API.

The implementation of \toolname{} is engineered for multi-core architectures.
Indeed, by leveraging the nature of the algorithms presented in both  \autoref{sec:solving_timed_CLTLoc_games} and \autoref{sec:synthesizing_winning_controller_strategies}, we integrated OpenMP~\cite{openmp18} parallelization to distribute the computational workload across multiple hardware cores.

Let $\tcgshort = (\aut, \winCond)$ be a \ac{tcg}. 
Currently, \toolname{} does not reset clocks external to $\aut$, \emph{i.e.}, those in $\cckc \setminus \ackc$, unless the \ac{tarena}-extension of $\aut$ is used. 
{In all winning conditions considered below, external clocks are used as global elapsed-time clocks (hence, a single external clock suffices), and resetting them is never required by the winning controller strategies; hence, we do not employ the \ac{tarena}-extension construction.} 


\subsection{Application to a Production Cell Case Study}
We illustrate the modeling capabilities of \acp{tcg} through a Production Cell (henceforth referred to as Cell) case study,
which represents a standard benchmark for realistic industrial applications.
Consider the Cell depicted in \autoref{fig:production_cell}, which is a simplified \ac{tarena} adaptation of~\cite{Burns2003}.
The Cell orchestrates the sequential automated processing of metal plates.
At the beginning, an unprocessed plate is transported from a feed belt ($\mathit{feed}$) to a rotary table ($\mathit{table}$). 
A two-armed robot then manages the subsequent transit: arm $P$ ($\mathit{robot_p}$) extracts the plate from the table to load a press ($\mathit{press}$), while arm $D$ ($\mathit{robot_d}$) transfers the pressed plate to the final deposit ($\mathit{depot}$). 
To maintain a continuous cycle, a crane finally refills the feed belt with raw material taken from the deposit.
Notably, if the robot moves too quickly or holds a plate for too long, physical instability may cause the plate to fall to the ground ($\mathit{fall_p}$ and $\mathit{fall_d}$).
\Red{Since the time required to restore fallen plates to the feed belt is not under the Cell’s control, $\mathit{fall_p}$ and $\mathit{fall_d}$ represent environment locations and are therefore depicted as squares.}
The Cell also features different operational modalities,
which can be set by a human operator, including a {\emph{fast loading mode}, requiring that every loading of plates into the press happens within at most $\alpha$ time units from the beginning of the Cell’s execution.
\Red{Notably, these modalities are not explicitly modeled in the \ac{tarena} depicted in \autoref{fig:production_cell}, but can instead be directly specified in the winning condition. For instance, $\alpha$ can be used as a constant in a clock constraint of the corresponding winning condition.}
\begin{figure}
    \centering
    \includegraphics[width=0.8\linewidth]{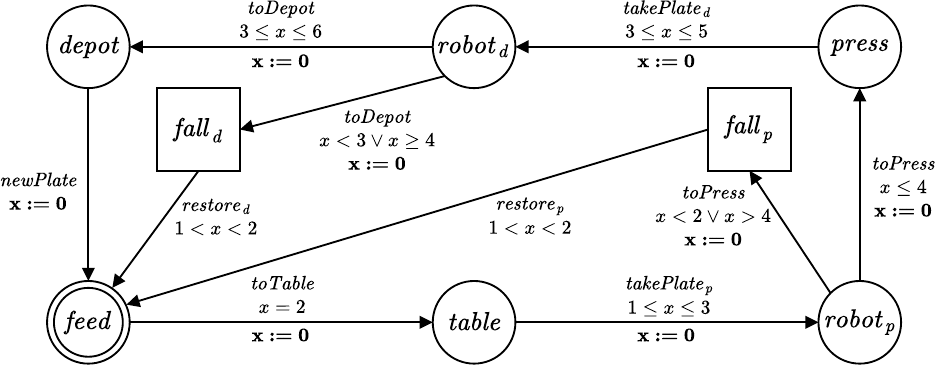}
    \caption{\ac{tarena} adaptation of the Production Cell introduced in~\cite{Burns2003}.
Environment locations, represented as squares, model the uncontrollable time, determined by the environment, required for a human operator to restore a fallen plate.}
\Description{\ac{tarena} adaptation}
\label{fig:production_cell}
\end{figure}

We evaluated various \acp{tcg} $\tcgshort_i = (\arena, \winCond_i)$, where $\arena$ is the \ac{tarena} of \autoref{fig:production_cell}, $\ackc = \{x\}$, $\cckc = \{ x, y \}$, while $\winCond_i$ varies as detailed below, 
for $i = 1, \dots, 4$.
For each $\winCond_i$, we analyze whether there exists a winning controller strategy for $\tcgshort_i$.
$\winCond_1,\dots,\winCond_4$ are described below, and three additional winning conditions are introduced in \autoref{tab:production_cell}.
\Red{We recall that our aim is to demonstrate the modeling capabilities of \acp{tcg} by exploring different winning conditions, some of which may not describe behaviors that would be desirable in a practical application. For instance, a winning condition may require a plate to fall for the Cell to win.}

\begin{enumerate}[nosep]

\item \label{ex:example_simple_bound} 
$\winCond_1 := \Next ((\neg \mathit{depot} \land  y > 0) \Until(\mathit{depot} \land 25 \le y \le 30)) $: 
a plate must reach the deposit between 25 and 30 time units after the beginning of execution.
A winning strategy for the Cell (\emph{i.e.}, the controller) is the following:
the lower bound on $y$ requires the Cell to exceed the holding time in $\mathit{robot_p}$, causing the plate to fall---an environmental interference.
The time required by a human operator, which is part of the environment, to restore the plate is beyond the control of the Cell.
Once the plate is restored, the Cell can process the plate again and reach the deposit.
Notably, if the Cell moves too quickly or exceeds the holding time in $\mathit{robot}_d$---\emph{i.e.}, the last station before the deposit---a human operator may then require $1 < \delta < 2$ time units to restore the fallen plate.

{\item \label{ex:example_bound} $ \winCond_2 :=  \Next ((\neg \mathit{depot} \land  y > 0 \land (\mathit{press} \Rightarrow y \le 10)) \Until (\mathit{depot} \land 25 \le y \le 30)) $: 
here the reasoning is similar to the previous case, \Red{\emph{i.e.}, a plate must reach the deposit between 25 and 30 time units
after the beginning of execution,} but now the Cell operates in fast loading mode, where $\alpha = 10$. 
As in the previous case, winning requires making a plate fall.
However, doing so makes it impossible for every loading of the plate into the press (even after being restored) to occur within 10 time units from the beginning of the Cell’s execution.}
\Red{It follows that a winning strategy for the Cell does not exist.}

\item $ \winCond_3 := \Box ( \neg \mathit{fall}_p \land \neg \mathit{fall}_d )$: a plate must never fall.
Winning strategies for the Cell consist in avoiding all locations representing a fallen plate, by selecting a delay $2 \le \delta \le 4$ in  $\mathit{robot}_p$ and a delay $3 \le \delta < 4$ in  $\mathit{robot}_d$.

\item \label{ex:next_reachability}
$\winCond_4 := 
\mathit{feed} \land 
\Next^5 (\mathit{depot} \land 10 \le y \le 12) \land 
\Next^6 (\mathit{feed} \land y = 14) \land 
\Next^{11} (\mathit{depot} \land 29 \le y \le 31) \land 
\Next^{12} (\mathit{feed} \land y = 32) \land \Next^{17} (\mathit{depot} \land y = 43)$: three different plates must be completed sequentially, subject to different timing constraints.
The first plate's production begins at $t = 0$ with duration $\delta_1 \in [10, 12]$.
The second begins at $t = 14$ with duration $\delta_2 \in [15,17]$.
The third begins at $t = 32$ with duration $\delta_3 = 11$.

{Notably, the \Red{winning condition implicitly encodes, in the number of moves needed to reach the deposit, the requirement that the Cell must never drop a plate.}
Indeed, completing the first plate requires exactly five moves, but any drop would reset execution to the feed belt.
The same reasoning applies to the other plates.}
\end{enumerate}

In cases~\ref{ex:example_simple_bound} and \ref{ex:example_bound}, the $\Next$ operator and the constraint $\neg \mathit{depot} \land y > 0$ guarantee that $\winCond_1$ and $\winCond_2$ target the processing of a single plate within a given time interval.
In particular, {since the constraint $y > 0$ occurs on the left-hand side of an $\Until$ formula, resetting $y$ would violate the constraint;}
this effectively forces $y$ to measure the total elapsed time (even though \toolname{} does not reset $y$ in these experiments, recall that external clocks can be reset independently of the underlying \ac{tarena}).
The operator $\Next$ skips the initial configuration where $y = 0$ holds, ensuring the $\Until$ subformula is not trivially falsified by initial clock values.

\begin{table}[tb]
\centering
\definecolor{lightgray}{gray}{0.95}
\setlength{\aboverulesep}{0pt} 
\setlength{\belowrulesep}{0pt}
\renewcommand{\arraystretch}{1.35} 
\setlength{\tabcolsep}{4.5pt}

\resizebox{0.7\textwidth}{!}{
\begin{tabular}{rr|rr|cc} 
\toprule
\multicolumn{2}{c|}{\textbf{Solving}} & \multicolumn{2}{c|}{\textbf{Synthesis}} & 
\multicolumn{1}{c}{\multirow{2}{*}{\textbf{\#Iter}}} & 
\multicolumn{1}{c}{\multirow{2}{*}{\textbf{Sat}}} \\
\cmidrule(lr){1-2} \cmidrule(lr){3-4}
\multicolumn{1}{c}{\textbf{Time} [ms]} & \multicolumn{1}{c|}{\textbf{Mem} [MB]} & \multicolumn{1}{c}{\textbf{Time} [ms]} & \multicolumn{1}{c|}{\textbf{Mem} [MB]} & & \\
\midrule

\multicolumn{6}{c}{\cellcolor{lightgray} 
$\blacktriangleright$ 
$\winCond_1 := \Next ((\neg \mathit{depot} \land y > 0) \Until(\mathit{depot} \land 25 \le y \le 30))$ 
$\blacktriangleleft$} \\
108.803 & 56.72 & 120.900 & 71.01 & 11  & \textsc{win} \\
\midrule

\multicolumn{6}{c}{\cellcolor{lightgray} 
$\blacktriangleright$ 
$\winCond_2 :=  \Next ((\neg \mathit{depot} \land  y > 0 \land (\mathit{press} \Rightarrow y \le 10)) \Until (\mathit{depot} \land 25 \le y \le 30)) $ $\blacktriangleleft$} \\
18.240 & 37.90 & --- & --- & 2  & \textsc{lose} \\
\midrule

\multicolumn{6}{c}{\cellcolor{lightgray} 
$\blacktriangleright$ 
$\winCond_3 := \Box ( \neg \mathit{fall}_p \land \neg \mathit{fall}_d )$ 
$\blacktriangleleft$} \\
3.059 & 9.90 & 3.715 & 9.99 & 2  & \textsc{win} \\
\midrule

\multicolumn{6}{c}{\cellcolor{lightgray}
$\blacktriangleright$ 
$\winCond_4 := \vcenter{\hbox{\begin{tabular}[c]{@{}l@{}}
$\mathit{feed} \land \Next^5 (\mathit{depot} \land 10 \le y \le 12) \land \Next^6 (\mathit{feed} \land y = 14)$ \\
$\land \, \Next^{11} (\mathit{depot} \land 29 \le y \le 31) \land \Next^{12} (\mathit{feed} \land y = 32) \land \Next^{17} (\mathit{depot} \land y = 43)$
\end{tabular}}}$
$\blacktriangleleft$} \\
8.415 & 9.49 & 8.746 & 11.18 & 17  & \textsc{win} \\
\midrule

\multicolumn{6}{c}{\cellcolor{lightgray} 
$\blacktriangleright$ 
$\winCond_5 := \Diamond ( \mathit{depot} \land 0 < y \le 11 )$ 
$\blacktriangleleft$} \\
17.490 & 13.44 & 17.652 & 15.90 & 7  & \textsc{win} \\
\midrule

\multicolumn{6}{c}{\cellcolor{lightgray} 
$\blacktriangleright$ 
$\winCond_6 := \Diamond ( \mathit{depot} \land 0 < y \le 9 )$ 
$\blacktriangleleft$} \\
11.384 & 11.92 & --- & --- & 5  & \textsc{lose} \\
\midrule

\multicolumn{6}{c}{\cellcolor{lightgray} 
$\blacktriangleright$ 
$\winCond_7 := \Box ( \neg \mathit{fall}_p \land \neg \mathit{fall}_d \land \neg \mathit{table})$ 
$\blacktriangleleft$} \\
3.366 & 9.64 & --- & --- & 6 & \textsc{lose} \\
\bottomrule
\end{tabular}
}

\captionsetup{skip=10pt}
\caption{Experimental results for various \acp{tcg} defined over an adaptation of the Production Cell~\cite{Burns2003} case study.
Time is measured in milliseconds, memory in Megabytes.
\textsc{win} means that a winning strategy for the Cell exists, \textsc{lose} otherwise.}
\label{tab:production_cell}
\end{table}
\autoref{tab:production_cell}
illustrates the experimental results (averaged over 100 runs) for various \acp{tcg} based on the \ac{tarena} shown in \autoref{fig:production_cell}, defined using different winning conditions.
Beyond $\winCond_1$--$\winCond_4$, we also evaluated two reachability winning conditions ($\winCond_5$ and $\winCond_6$) and a safety winning condition ($\winCond_7$).
All experiments considered both solving (\textbf{Solving}) and, whenever a winning controller strategy exists, the synthesis of the \ac{sg} (\textbf{Synthesis}).
Specifically, the table reports the total execution time (\textbf{Time}) in milliseconds, the peak physical memory consumption (\textbf{Mem}) in Megabytes,
the number of iterations required by the algorithms (\textbf{\#Iter}), 
and whether there exists a winning controller strategy (\textbf{Sat})---\textsc{win} means that a winning strategy for the Cell exists, \textsc{lose} otherwise.
The symbol --- in the \textbf{Synthesis} column indicates that a winning controller strategy cannot be synthesized since the corresponding \ac{tcg} is losing for the controller. 

\subsection{Application to a Therapeutic Human-Robot Interaction Example}

To demonstrate the applicability of \acp{tcg} to Human-Robot Interaction scenarios, we introduce the following example, which models the \emph{HumanApplicant} human-robot interaction pattern introduced in~\cite{Lestingi2024} as the \ac{tarena} shown in \autoref{fig:med_app}.
This pattern captures human agents requesting robot-assisted support in performing tasks that require close contact or tight temporal synchronisation.
In this example, these interactions are modeled as a \ac{tcg} between a controller (a therapeutic robot, called \emph{TBot}) and its environment (a nurse and a doctor).

\begin{figure}[t]
    \centering
    \includegraphics[width=\linewidth]{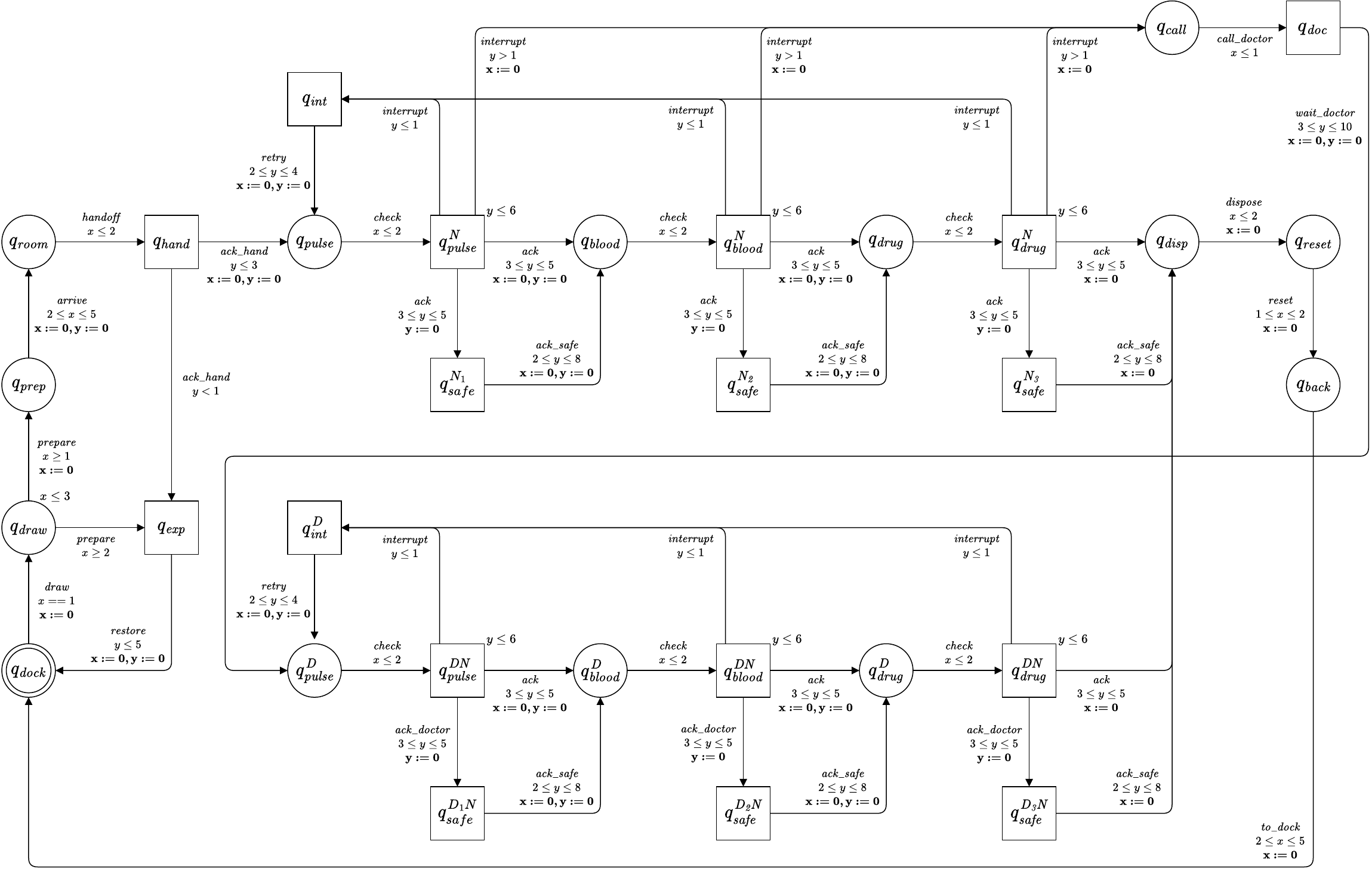}
    \caption{\ac{tarena} model of a therapeutic Human-Robot Interaction example inspired by~\cite{Lestingi2024}.}
    \label{fig:med_app}
    \Description{medapp}
\end{figure}

The main objective of TBot is to assist a nurse in administering an intravenous drug via a syringe.
TBot also assists the nurse by performing medical checks prior to administration, \emph{i.e.}, pulse assessment and blood pressure measurement.

Whenever TBot is invoked, it initiates its operation by drawing the drug ($\mathit{q_{dock}}$ to $\mathit{q_{draw}}$) from a cabinet.
The drug must then be prepared. 
If preparation takes too long ($\mathit{q_{draw}}$ to $\mathit{q_{exp}}$), the drug may expire, forcing TBot to return to the cabinet and draw a new dose ($\mathit{q_{exp}}$ to $\mathit{q_{dock}}$).
Otherwise ($\mathit{q_{draw}}$ to $\mathit{q_{prep}}$), TBot reaches the patient's room ($\mathit{q_{prep}}$ to $\mathit{q_{room}}$) and then hands the syringe to the nurse ($\mathit{q_{room}}$ to $\mathit{q_{hand}}$).
The nurse must acknowledge the handoff. 
Prior to delivery, the drug requires temperature equilibration in TBot’s gripper; if the handoff is too brief, the drug may still be at an incorrect temperature (\emph{e.g.}, too cold for injection) and thus expire ($\mathit{q_{hand}}$ to $\mathit{q_{exp}}$), requiring TBot to restart the same process again ($\mathit{q_{exp}}$ to $\mathit{q_{dock}}$).
Otherwise, the handoff is successful ($\mathit{q_{hand}}$ to $\mathit{q_{pulse}}$), and the nurse proceeds to assess the patient’s pulse.

Each clinical operation on the patient, starting with pulse measurement, consists of a preliminary TBot preparation phase ($\mathit{q_{pulse}}$ to $\mathit{q^{N}_{pulse}}$), followed by the nurse’s execution phase. 
Completion of each nurse operation must be explicitly acknowledged to TBot before the workflow can proceed to the next step.
During execution, the nurse may operate at a regular pace ($\mathit{q^{N}_{pulse}}$ to $\mathit{q_{blood}}$) or deliberately slow down to ensure additional safety ($\mathit{q^{N}_{pulse}}$ to $\mathit{q^{N_1}_{safe}}$).
If the measured parameters are outside a safe range, TBot may raise an interrupt during the operation. 
This would require restarting the clinical sequence from the pulse measurement phase ($\mathit{q^{N}_{pulse}}$ to $\mathit{q_{int}}$) and ($\mathit{q_{int}}$ to $\mathit{q_{pulse}}$), or to request assistance from a doctor ($\mathit{q^{N}_{pulse}}$ to $\mathit{q_{call}}$), which can be called at most once during the entire procedure ($\mathit{q_{call}}$ to $\mathit{q_{doc}}$).
The same interaction pattern---TBot preparation, nurse execution, and acknowledgment---applies to subsequent phases, namely blood pressure measurement and drug administration.

If the doctor is called, the entire sequence of operations must be repeated under assisted conditions. In this setting, TBot is responsible for selecting whether each phase is carried out in regular or safe mode.
After administration, TBot disposes the used syringe ($\mathit{q_{disp}}$ to $\mathit{q_{reset}}$).
It then recovers by resetting relevant operational parameters ($\mathit{q_{reset}}$ to $\mathit{q_{back}}$) and finally returns to its docking station ($\mathit{q_{back}}$ to $\mathit{q_{dock}}$).

We evaluated various \acp{tcg} $\tcgshort_i = (\arena, \winCond_i)$, where $\arena$ is the \ac{tarena} of \autoref{fig:med_app}, $\ackc = \{x, y\}$, $\cckc = \{ y, z \}$, while $\winCond_i$ varies as detailed below, 
for $i = 1, \dots, 7$.
For each $\winCond_i$, we analyze whether there exists a winning controller strategy for $\tcgshort_i$.
Experimental results (averaged over 100 runs) are summarized in \autoref{tab:med_app}.

\begin{table}[tb]
\centering
\definecolor{lightgray}{gray}{0.95}
\setlength{\aboverulesep}{0pt} 
\setlength{\belowrulesep}{0pt}
\renewcommand{\arraystretch}{1.35} 
\setlength{\tabcolsep}{4.5pt}

\resizebox{0.9\textwidth}{!}{
\begin{tabular}{rr|rr|cc} 
\toprule
\multicolumn{2}{c|}{\textbf{Solving}} & \multicolumn{2}{c|}{\textbf{Synthesis}} & 
\multicolumn{1}{c}{\multirow{2}{*}{\textbf{\#Iter}}} & 
\multicolumn{1}{c}{\multirow{2}{*}{\textbf{Sat}}} \\
\cmidrule(lr){1-2} \cmidrule(lr){3-4}
\multicolumn{1}{c}{\textbf{Time} [ms]} & \multicolumn{1}{c|}{\textbf{Mem} [MB]} & \multicolumn{1}{c}{\textbf{Time} [ms]} & \multicolumn{1}{c|}{\textbf{Mem} [MB]} & & \\
\midrule

\multicolumn{6}{c}{\cellcolor{lightgray} 
$\blacktriangleright$ 
$\winCond_1 := \Next ( ( \neg q_{\mathit{dock}} \land z > 0) \Until (q_{\mathit{dock}} \land y > 0))$ 
$\blacktriangleleft$} \\
524.507 & 227.32 & 667.064 & 319.65 & 17 & \textsc{win} \\
\midrule

\multicolumn{6}{c}{\cellcolor{lightgray} 
$\blacktriangleright$ 
$\winCond_2 := \Next ( ( \neg q_{\mathit{dock}} \land z > 0) \Until (q_{\mathit{dock}} \land y > 0 \land z \le 50))$ $\blacktriangleleft$} \\
53944.238 & 14015.61 & 76405.869 & 13881.97 & 18 & \textsc{win} \\
\midrule

\multicolumn{6}{c}{\cellcolor{lightgray} 
$\blacktriangleright$ 
$\winCond_3 := \Box (\neg q_{\mathit{exp}} \land \neg q_{\mathit{int}} \land \neg q_{\mathit{call}})$ $\blacktriangleleft$} \\
255.961 & 183.09 & 387.068 & 188.91 & 2 & \textsc{win} \\
\midrule

\multicolumn{6}{c}{\cellcolor{lightgray} 
$\blacktriangleright$ 
$\winCond_4 := q_{\mathit{dock}} \land \Next^7 (q_{\mathit{call}}) \land \Next^{19} (q_{\mathit{dock}})$ 
$\blacktriangleleft$} \\
447.708 & 68.94 & 759.950 & 253.26 & 19 & \textsc{win} \\
\midrule

\multicolumn{6}{c}{\cellcolor{lightgray}
$\blacktriangleright$ 
$\winCond_5 := q_{\mathit{dock}} \land \Next^7 (q_{\mathit{int}}) \land \Next^{10} (q_{\mathit{int}}) \land \Next^{13} (q_{\mathit{int}}) \land \Next^{16} (q_{\mathit{call}}) \land \Next^{27} (q_{\mathit{dock}})$
$\blacktriangleleft$} \\
382.352 & 79.23 & --- & --- & 27 & \textsc{lose} \\
\midrule

\multicolumn{6}{c}{\cellcolor{lightgray} 
$\blacktriangleright$ 
$\winCond_6 := q_{\mathit{dock}} \land \Next^7 (q_{\mathit{call}}) \land \Next^9 (q^D_{\mathit{pulse}}) \land \Next^{11} (q^D_{\mathit{int}}) \land \Next^{14} (q^D_{\mathit{int}}) \land \Next^{17} (q^D_{\mathit{int}}) \land \Next^{27} (q_{\mathit{dock}} \land z \le 50)$ 
$\blacktriangleleft$} \\
21244.263 & 4240.19 & 39550.433 & 10892.29 & 27 & \textsc{win} \\
\midrule

\multicolumn{6}{c}{\cellcolor{lightgray} 
$\blacktriangleright$ 
$\winCond_7 := \Diamond (q_{\mathit{int}})$ 
$\blacktriangleleft$} \\
4.941 & 13.06 & --- & --- & 1  & \textsc{lose} \\
\bottomrule
\end{tabular}
}

\captionsetup{skip=10pt}
\caption{Experimental results for various \acp{tcg} defined over the therapeutic Human-Robot Interaction example inspired by~\cite{Lestingi2024}.}
\label{tab:med_app}
\end{table}

\begin{enumerate}[nosep]
\item \label{ex:single_patient_win} 
$\winCond_1 := \Next ( ( \neg q_{\mathit{dock}} \land z > 0) \Until (q_{\mathit{dock}} \land y > 0))$: TBot is capable of assisting the nurse, and potentially the doctor, with drug administration for a single patient, \Red{as enforced by the constraint $\neg q_{\mathit{dock}} \land z > 0$}.
Notably, the constraint $y > 0$ ensures that TBot can return to its docking station only after the entire procedure has been completed. Indeed, TBot may also return to its docking station if the drug expires; in this case, $y$ is reset to zero upon entering $q_{\mathit{dock}}$.
\Red{$\winCond_1$ does not impose further constraints on TBot's behavior during the assistance procedure.}
\Red{A possible winning strategy for TBot is to complete the procedure without triggering any interrupts.}

\item \label{ex:single_patient_time_win} 
$\winCond_2 := \Next ( ( \neg q_{\mathit{dock}} \land z > 0) \Until (q_{\mathit{dock}} \land y > 0 \land z \le 50))$: the reasoning
is analogous to that of $\winCond_1$, \Red{\emph{i.e.}, TBot is capable of assisting the nurse, and potentially the doctor,
with drug administration for a single patient}, except that the entire assistance procedure must now be completed within 50 time units.
\Red{A possible winning strategy for TBot is to complete the procedure without raising any interrupts.}

\item \label{ex:patient_no_int_no_drug_win}
$\winCond_3 := \Box (\neg q_{\mathit{exp}} \land \neg q_{\mathit{int}} \land \neg q_{\mathit{call}})$: TBot is capable of assisting the nurse without causing the drug to expire or raising an interrupt. 
Notice that the doctor is not involved in this case since $\neg q_{\mathit{call}}$ must always hold.

\item \label{ex:patient_doc_in_pulse_win}
$\winCond_4 := q_{\mathit{dock}} \land \Next^7 (q_{\mathit{call}}) \land \Next^{19} (q_{\mathit{dock}})$: in this case, TBot must raise an interrupt
in $q^N_{\mathit{pulse}}$, \Red{since $\Next^7 (q_{\mathit{call}})$ requires $q_{\mathit{call}}$ to be reached exactly after 7 moves,} \emph{i.e.}, the doctor is called in exactly 7 moves from $q_{\mathit{dock}}$. 
After the doctor has been called, the entire assistance procedure must be completed in exactly twelve moves.
This additionally requires that no further interrupts are raised after the doctor has been called, and that safe mode is used exactly once during the post-call phase.

\item \label{ex:patient_doc_after_three_int_lose}
$\winCond_5 := q_{\mathit{dock}} \land \Next^7 (q_{\mathit{int}}) \land \Next^{10} (q_{\mathit{int}}) \land \Next^{13} (q_{\mathit{int}}) \land \Next^{16} (q_{\mathit{call}}) \land \Next^{27} (q_{\mathit{dock}})$: TBot is required to raise three interrupts in $q^N_{\mathit{pulse}}$
leading to $q_{\mathit{int}}$, \Red{followed by an interrupt (still raised in $q^N_{\mathit{pulse}}$) leading to $q_{\mathit{call}}$,} before proceeding with the completion of the assistance procedure. 
\Red{Interestingly}, reachability of $q_{\mathit{int}}$ cannot be enforced by TBot alone, as further discussed in $\winCond_7$; \Red{in this case, TBot has no winning strategy}.

\item \label{ex:patient_doc_atfer_three_int_time_win}
$\winCond_6 := q_{\mathit{dock}} \land \Next^7 (q_{\mathit{call}}) \land \Next^9 (q^D_{\mathit{pulse}}) \land \Next^{11} (q^D_{\mathit{int}}) \land \Next^{14} (q^D_{\mathit{int}}) \land \Next^{17} (q^D_{\mathit{int}}) \land \Next^{27} (q_{\mathit{dock}} \land z \le 50)$: \Red{TBot is required to raise three interrupts in $q^{\mathit{DN}}_{\mathit{pulse}}$ leading to $q^D_{\mathit{int}}$, before proceeding with the completion of the \Red{assistance} procedure.}
In this setting, TBot has a winning strategy, since once the doctor has been called the interrupts are fully under TBot’s control.
Furthermore, a global time bound of 50 time units is enforced.

\item \label{ex:w_int_unreachable}
$\winCond_7 := \Diamond (q_{\mathit{int}})$: TBot does not have a winning strategy to reach $q_{\mathit{int}}$, meaning that the decision to either restart the operation sequence (when the nurse is operating alone) or call a doctor may be influenced by TBot, but it is ultimately resolved by the timing of the nurse’s actions. 
In particular, the nurse’s delay determines whether the system transitions to calling the doctor or restarting the sequence. 
\Blue{This example illustrates how the model distinguishes decisions controlled by TBot from outcomes influenced by the nurse's timing, a distinction that is relevant to the discussion of human responsibility in safety-critical systems~\cite{moralcrumplezones,Strasser2022}.}
\end{enumerate}

\subsection{Discussion of Experimental Results}

The experimental results presented in this section demonstrate that \toolname{} can effectively solve \acp{tcg} and, whenever a winning controller strategy exists, synthesize a \ac{sg}, whose structure depends on the \ac{tcg} under analysis.
{In particular, the results reported in \autoref{tab:production_cell} demonstrate the feasibility of applying \toolname{} to small models.}

As illustrated by the results in \autoref{tab:med_app}, performance degrades for larger models {consistent with} the increased \ac{tarena} size, the higher number of clocks, and the larger clock constants involved, including those appearing in the winning conditions. 
This is expected, since the region abstraction requires explicit enumeration of the state space.
\Red{Nevertheless, it is worth noting that, even for the largest models considered, execution times remain within the order of seconds.}
This effect is particularly evident when comparing the winning conditions $\winCond_1$ and $\winCond_2$ in \autoref{tab:med_app}. Although these winning conditions are essentially identical, $\winCond_2$ additionally imposes the upper bound $z \le 50$. Enforcing this constraint requires precomputing all goal states satisfying it, thereby substantially increasing both time and memory consumption.

The reported execution times could be further reduced through several optimizations (some of which are discussed in~\cite{tarzan_forte}). 
For instance, in reachability winning conditions involving negated propositions (such as $\winCond_1,\winCond_2$ in both \autoref{tab:production_cell} and \autoref{tab:med_app}), \toolname{} currently constructs the entire state space excluding the states corresponding to the negated propositions. 
Supporting negation directly within the algorithms for \acp{tcg} would avoid this overhead.

It is worth emphasizing that both solving and synthesis algorithms are affected by the size of the underlying \ac{rts}, consistently with the complexity analysis in \autoref{subsec:solvingalg} and \autoref{sec:synthesizing_winning_controller_strategies}, respectively.
We also observe that strategy synthesis is generally more computationally expensive than solving, \Red{although the difference remains relatively moderate and does not amount to orders of magnitude}.

\section{Conclusion and Future Works}
\label{sec:conclusion}
This paper introduced \aclp{tcg}, a novel variant of \aclp{tg} that facilitates the modeling of reactive systems where a controller always chooses an action, while the non-determinism is resolved after the action has been chosen.
After establishing the theoretical foundations, we defined algorithms for solving and synthesizing winning controller strategies for \acp{tcg} with reachability and safety winning conditions.
The synthesis algorithms construct a \acl{sg}, from which multiple strategies can be synthesized. 
Notably, our \ac{sg} construction for reachability \acp{tcg} enables the synthesis of controller strategies that minimize the number of moves required to reach a goal state.
We demonstrated that solving reachability \acp{tcg} is EXPTIME-complete, matching the
complexity of the corresponding problem for traditional \acp{tg}.
Finally, we evaluated our theoretical results on a classical Production Cell case study and on a Human-Robot Interaction example, which we verified using our C\texttt{++} library \toolname{}, which leverages parallelization techniques.
Despite relying on the region abstraction,
the experimental evaluation demonstrates that our approach is not only applicable in practice, but also {feasible} on the considered experiments.

Future work will proceed along several directions. 
The proposed algorithms for \acp{tcg} perform backward exploration. 
Although such algorithms are straightforward to define and implement, backward exploration leads to limited scalability.
We therefore plan to develop and implement in \toolname{} forward on-the-fly algorithms for both solving \acp{tcg} and synthesizing winning controller strategies.
We also plan to generalize \acp{tcg} to networks of \acp{tcg}, thereby enabling the modeling of multiple interacting agents. 
Furthermore, we intend to broaden the class of supported \ac{cltloc} winning conditions. 
Finally, we aim to investigate hybrid approaches that complement \toolname{} with zone-based techniques, thus leveraging the strengths of both abstractions.

\bibliographystyle{ACM-Reference-Format}
\bibliography{references}


\end{document}